\documentclass[aps,preprint,amsmath,amssymb,amsfonts]{revtex4}
\usepackage{epsfig}
\usepackage{graphicx}
\usepackage{subfigure}
\usepackage{dcolumn}
\usepackage{bm}
\usepackage{amsthm}
\usepackage{enumitem}
\usepackage{slashed}
\usepackage{braket}
\usepackage{amsmath}
\usepackage{mathtools}

\usepackage{color}   
\usepackage{hyperref}
\hypersetup{
	colorlinks=true,  
	linkcolor=black,  
	citecolor=black,
	filecolor=black,
	urlcolor=black,
}

\makeatletter
\def\l@subsubsection#1#2{}
\makeatother

\newcommand{\del}{\partial}

\renewcommand{\Re}{\operatorname{Re}}
\renewcommand{\Im}{\operatorname{Im}}

\newcommand{\diag}{\operatorname{diag}}

\newcommand{\bbR}{\mathbb{R}}

\newcommand{\calD}{\mathcal{D}}

\newcommand{\calF}{\mathcal{F}}
\newcommand{\calG}{\mathcal{G}}

\newcommand{\calO}{\mathcal{O}}

\newcommand{\calT}{\mathcal{T}}

\newcommand{\dSCFT}[2]{dS\textsubscript{#1}/CFT\textsubscript{#2}}

\newtheorem{lemma}{Lemma}
\newtheorem*{radial_locality_criterion}{Radial locality criterion}
\newtheorem*{time_locality_criterion}{Time locality criterion}

\begin{document}

\title{Bulk locality and gauge invariance for boundary-bilocal cubic correlators in higher-spin gravity}

\author{Vyacheslav Lysov}
\email{vyacheslav.lysov@oist.jp}
\affiliation{Okinawa Institute of Science and Technology, 1919-1 Tancha, Onna-son, Okinawa 904-0495, Japan}
\author{Yasha Neiman}
\email{yashula@icloud.com}
\affiliation{Okinawa Institute of Science and Technology, 1919-1 Tancha, Onna-son, Okinawa 904-0495, Japan}

\date{\today}

\begin{abstract}
We consider type-A higher-spin gravity in 4 dimensions, holographically dual to a free $O(N)$ vector model. In this theory, the cubic correlators of higher-spin boundary currents are reproduced in the bulk by the Sleight-Taronna cubic vertex. We extend these cubic correlators from local boundary currents to bilocal boundary operators, which contain the tower of local currents in their Taylor expansion. In the bulk, these boundary bilocals are represented by linearized Didenko-Vasiliev (DV) ``black holes''. We argue that the cubic correlators are still described by local bulk structures, which include a new vertex coupling two higher-spin fields to the ``worldline'' of a DV solution. As an illustration of the general argument, we analyze numerically the correlator of two local scalars and one bilocal. We also prove a gauge-invariance property of the Sleight-Taronna vertex outside its original range of applicability: in the absence of sources, it is invariant not just within transverse-traceless gauge, but rather in general traceless gauge, which in particular includes the DV solution away from its ``worldline''.
\end{abstract}

\maketitle
\tableofcontents
\newpage

\section{Introduction} \label{sec:intro}

\subsection{Setup and motivation}

Higher-spin (HS) gravity \cite{Vasiliev:1990en,Vasiliev:1995dn,Vasiliev:1999ba} is the conjectured interacting theory of an infinite tower of massless gauge fields of all spins. It can be thought of as a ``smaller cousin'' of string theory. In its simplest version, the theory lacks a realistic GR limit. However, it has the virtues of being native to 4 spacetime dimensions, and consistent with both signs of the cosmological constant. We consider here the ``smallest'' version of HS gravity in 4d: the so-called minimal type-A theory, which has a single, parity-even field of every even spin. This theory admits a particularly simple holographic dual \cite{Klebanov:2002ja,Sezgin:2002rt,Sezgin:2003pt} via AdS/CFT: a free $O(N)$ vector model on the 3d boundary of AdS\textsubscript{4}, whose primary single-trace operators form a tower of conserved HS currents. A major reason to be interested in this particular duality is that it also admits a positive cosmological constant \cite{Anninos:2011ui}, providing a concrete model of \dSCFT{4}{3}. In the present paper, we stick for simplicity to AdS\textsubscript{4}, in Euclidean signature. 

The biggest outstanding question in HS theory concerns its locality properties. In general, since the theory involves infinitely many massless fields interacting at all orders in derivatives, it was always expected to be non-local in some way. Moreover, at the classical level, the only length scale in the theory is the cosmological curvature radius. Thus, the theory was expected to be non-local \emph{at the cosmological scale}. Though exotic, this still implies a positive expectation of some degree of locality: in particular, at distances \emph{much larger} than the AdS radius, one expects the couplings to vanish sufficiently fast.

This expectation was put to the test, by a research program to explicitly reproduce the theory's vertices from its holographic boundary correlators. For 3-point correlators, bulk locality is satisfied automatically: all gauge-invariant cubic vertices for given spins $(s_1,s_2,s_3)$ can be reduced to a finite set of structures with \emph{finitely many derivatives} \cite{Joung:2011ww}. Nevertheless, it seems significant that the particular cubic vertex \cite{Sleight:2016dba} found for the minimal type-A theory takes a remarkably simple form. However, at the 4-point level, disaster strikes: the spin-0 quartic bulk vertex, derived in \cite{Bekaert:2015tva}, turns out \cite{Sleight:2017pcz} to be as non-local as an exchange diagram. This result was foreshadowed some years before, in the flat-spacetime context \cite{Fotopoulos:2010ay,Taronna:2011kt}. In particular, the authors of \cite{Fotopoulos:2010ay} conjectured that some additional degrees of freedom should be added to make the theory local. A more subtle resolution is being advocated in e.g. \cite{Didenko:2019xzz,Gelfond:2019tac}: to keep the same degrees of freedom, but to extend the ordinary notion of locality to so-called ``spin-locality''. 

Our own approach to the locality problem is to try and mimic string theory: interactions that appear non-local in terms of field theory may become local when viewed in terms of more appropriate structures, such as the string worldsheet. While HS gravity (in its simplest version) doesn't give rise to strings, it does contain an analogous object -- the Didenko-Vasiliev ``BPS black hole'' solution \cite{Didenko:2008va,Didenko:2009td}. The analogy between this solution and the string is twofold. First, one can view the fundamental string (and all the other branes of string theory) as BPS solutions of supergravity \cite{Schwarz:1996bh,Blumenhagen:2013fgp}, with the Didenko-Vasiliev (henceforth, DV) solution playing the analogous role in HS gravity. Second, in AdS/CFT, one can view the string as the bulk dual of boundary Wilson lines or loops \cite{Rey:1998ik,Maldacena:1998im}, which contain as a Taylor expansion the whole tower of local single-trace boundary operators (whose bulk duals are the string's modes). Similarly, in HS holography, the DV solution is the bulk dual \cite{David:2020fea,Lysov:2022zlw} of the boundary bilocal operator \cite{Das:2003vw,Douglas:2010rc,Das:2012dt}, which contains as a Taylor expansion the tower of local boundary HS currents (whose bulk duals are the individual HS gauge fields). Due to these analogies, we believe that the key to understanding HS theory lies in the bulk dynamics of not just HS fields, but also DV solutions.

Our focus is on the \emph{linearized} DV solution \cite{Didenko:2008va}, which consists simply of linearized HS fields, sourced by a particle-like singularity located on a geodesic ``worldline'' in the AdS\textsubscript{4} bulk. This particle-like source is charged under the gauge fields of all spins, following a BPS-like proportionality pattern. In \cite{David:2020fea,Lysov:2022zlw}, we explored the bulk interaction between two such solutions, showing that it reproduces the CFT correlator of two boundary bilocals. In that case, the ``interaction'' was simply that of charged particles exchanging (an HS multiplet of) gauge fields, with no self-coupling among the gauge fields themselves. In the present paper, we extend the analysis to \emph{three} DV solutions, and ask what kind of bulk interactions can reproduce the corresponding \emph{cubic} CFT correlator. Here, the cubic self-interaction of the HS gauge fields becomes important. In fact, in an appropriate limit, the DV solutions reduce to usual boundary-bulk propagators \cite{Lysov:2022zlw}, and the boundary correlator is then captured fully by the on-shell cubic vertex found by Sleight and Taronna \cite{Sleight:2016dba}. Our goal in this paper will be to step away from this limit, and study the locality and gauge-invariance properties of the resulting bulk interactions. 

Our eventual goal is to reformulate the entirety of HS theory in terms of cubic interactions between DV solutions \cite{FeynmanRules}, entirely bypassing the need for quartic or higher vertices. It is this larger project that lends importance to the locality of such cubic interactions.

The formalism we'll employ is the same as in \cite{Sleight:2016dba}, combining Fronsdal's ``metric-like'' approach to linearized HS fields \cite{Fronsdal:1978rb,Fronsdal:1978vb} with the radial-reduction approach to bulk AdS fields \cite{Biswas:2002nk}, where we choose the scaling weights to match those of the relevant boundary-bulk propagators (as opposed to the more common choice \cite{Joung:2011ww}, which simplifies the gauge-invariance analysis for general vertices).

\subsection{Summary of locality results} \label{sec:intro:locality}

\begin{figure}%
	\centering%
	\includegraphics[scale=0.6]{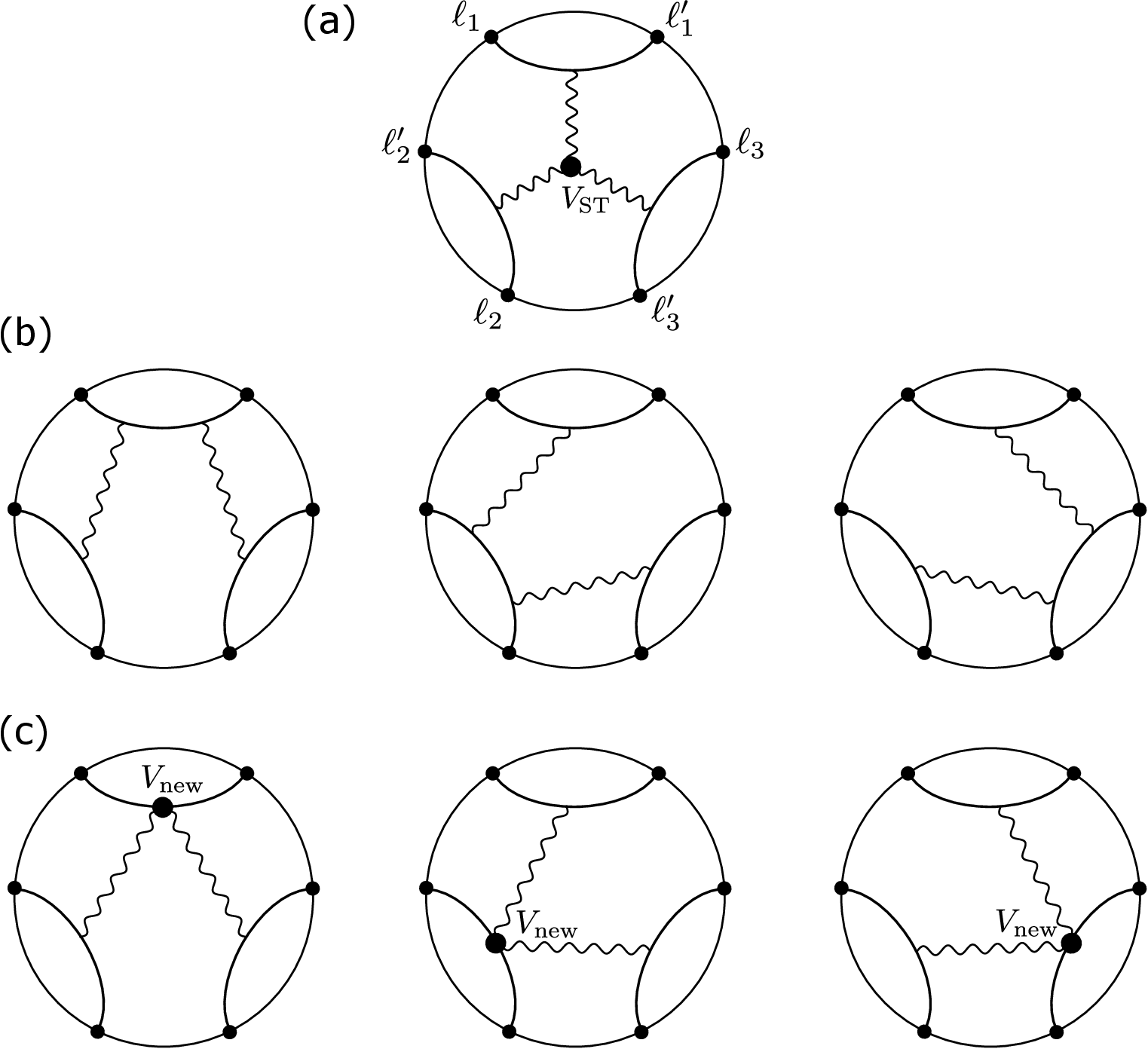} \\
	\caption{Bulk diagrams for the connected correlator $\langle\calO(\ell_1,\ell'_1)\calO(\ell_2,\ell'_2)\calO(\ell_3,\ell'_3) \rangle$ of three boundary bilocals, in terms of bulk DV solutions and their ``worldlines'': (a) the Sleight-Taronna cubic vertex; (b) double exchanges of HS fields between the worldlines; (c) a new vertex, coupling two HS fields to a worldline.}
	\label{fig:OOO} 
\end{figure}

We will argue that the cubic correlator of boundary bilocals is reproduced by a set of local Witten diagrams that couple the corresponding DV solutions and their geodesic ``worldlines''. These diagrams can be divided into three groups:
\begin{enumerate}[label=(\alph*)]
	\item The Sleight-Taronna vertex \cite{Sleight:2016dba} coupling the three DV solutions.
	\item Exchange of two HS gauge fields between the three geodesic ``worldlines''. This is just a product of two pairwise interactions between the DV solutions, of the type considered in \cite{David:2020fea,Lysov:2022zlw}. In particular, it doesn't involve self-interaction of HS fields.
	\item A new vertex, coupling the fields of two DV solutions to the ``worldline'' of the third.
\end{enumerate}
These different terms (a)-(c) comprising the correlator are depicted in figure \ref{fig:OOO}. Let us now comment on the extent to which each term is known, and the sense in which it is local.

Term (a) -- the on-shell cubic coupling of HS fields -- is known explicitly \cite{Sleight:2016dba}, and is local in the traditional sense, i.e. it involves a finite number of derivatives for each set of spins $(s_1,s_2,s_3)$. Note, however, that the DV solutions contain all spins. Therefore, the sum over spins will introduce an infinite tower of derivatives, and with it some degree of non-locality. Fortunately, as we'll argue in section \ref{sec:locality:general}, this non-locality is in fact at the scale of $\sim 1$ AdS radius, matching the original expectation for HS theory. 

Term (b) consists of simple diagrams whose only ``vertices'' are the local minimal couplings \cite{David:2020fea} between an HS-charged particle and an HS gauge field. As such, it is fully known, and manifestly local \emph{if} we agree to view the DV solutions' worldlines as HS-charged particles. If one tried instead to express these diagrams as a cubic vertex between HS fields, that vertex would of course be non-local.

Now we turn to term (c) -- a new vertex, which will be discussed at length in section \ref{sec:locality}. One may alternatively view it as an ``off-shell'' correction to the Sleight-Taronna vertex (i.e. a correction proportional to the free equations of motion), due to the DV fields not being source-free, and thus possessing Fronsdal curvature, concentrated on the corresponding ``worldlines''. Even for fixed spins, this new vertex may include an infinite tower of derivatives. The question then is whether the resulting non-locality is restricted to $\sim 1$ AdS radius. We will argue that this question can be reframed as a set of proxy criteria, involving not the vertex formula itself, but rather its contribution to the correlator in certain limits. We will then show that our criteria are indeed satisfied, once the other contributions (a)-(b) to the correlator are taken into account. We won't evaluate the new vertex explicitly, aside from a numerical study in one simple case (section \ref{sec:numeric}).

An alternative concise way of introducing the three terms (a)-(c) is as follows:
\begin{enumerate}[label=(\alph*)]
	\item We draw the most obvious cubic coupling between the three DV solutions, via the Sleight-Taronna vertex. We find that this doesn't reproduce the boundary cubic correlator of bilocals.
	\item We add the double-exchange diagrams, still constructed purely from known elements. We find that the boundary correlator is still not reproduced.
	\item We parameterize the difference between the boundary correlator and terms (a)-(b) in terms of a new vertex (or, alternatively, an off-shell correction to the Sleight-Taronna vertex). Our main result is then that this new vertex has appropriate locality properties.
\end{enumerate}
Finally, note that our terms (a)-(c) don't include any \emph{gauge} corrections to the Sleight-Taronna vertex, i.e. corrections due to the DV solutions not being in transverse-traceless gauge. The vanishing of such corrections is one of our results, derived in section \ref{sec:free_invariance} and summarized below.
 
\subsection{Plan of the paper} \label{sec:intro:plan}

The rest of the paper is structured as follows. In section \ref{sec:preliminaries}, we review the formalism of \cite{Sleight:2016dba} for HS fields in Euclidean AdS\textsubscript{4}, along with other relevant ingredients: the free vector model on the boundary, asymptotics of bulk fields, boundary-bulk propagators, the DV solution and the Sleight-Taronna vertex. 

Section \ref{sec:free_invariance} contains our gauge-invariance results for the Sleight-Taronna vertex. We show that, if one merely symmetrizes the original vertex formula from \cite{Sleight:2016dba} over permutations of its 3 legs, then the vertex's gauge invariance is extended from source-free fields in transverse-traceless gauge (as originally intended in \cite{Sleight:2016dba}) to source-free fields in \emph{general traceless gauge}. In section \ref{sec:free_invariance:proof}, we prove that this extended gauge-invariance holds up to boundary terms. Then, in section \ref{sec:free_invariance:boundary_terms}, we show that the boundary terms \emph{also} vanish under appropriate assumptions on the fields' asymptotics, which in particular are satisfied by the DV solution away from its singular worldline.

In section \ref{sec:locality}, we present our main argument vis. the locality structure of the general cubic correlator and the new vertex. In section \ref{sec:numeric}, we illustrate the locality argument by a numerical analysis in a simple case: a single DV solution coupled to a pair of spin-0 boundary-bulk propagators. In section \ref{sec:delta_gauge}, we outline an alternative technique for calculating the relevant bulk diagram, using a new non-traceless gauge \cite{Lysov:2022zlw} for the DV solution. Section \ref{sec:discuss} is devoted to discussion and outlook.

We note that section \ref{sec:free_invariance}'s gauge-invariance result for the Sleight-Taronna vertex is not essential for the abstract locality argument in section \ref{sec:locality}. However, the existence of this nice result reinforces our sense that the paper's main idea -- of combining the DV solution with the Sleight-Taronna vertex -- is on the right track.

\section{Preliminaries} \label{sec:preliminaries}

\subsection{Bulk geometry} \label{sec:preliminaries:bulk_geometry}

To write the Sleight-Taronna vertex in a simple form, one must use an embedding-space formalism, and in particular the radial reduction approach of \cite{Biswas:2002nk}. Thus, we describe Euclidean AdS\textsubscript{4} as the hyperboloid of unit timelike radius within 5d flat spacetime $\bbR^{1,4}$:
\begin{align}
 EAdS_4 = \left\{x^\mu\in\bbR^{1,4}\, |\, x_\mu x^\mu = -1, \ x^0 > 0 \right\} \ . \label{eq:EAdS}
\end{align}
Here, indices $(\mu,\nu,\dots)$ are 5-dimensional, and are raised and lowered with the Minkowski metric $\eta_{\mu\nu} = \diag(-1,1,1,1,1)$. 4d vectors at a point $x^\mu \in EAdS_4$ are simply 5d vectors $v^\mu$ that satisfy $v\cdot x\equiv v_\mu x^\mu = 0$. Covariant derivatives in $EAdS_4$ are simply flat $\bbR^{1,4}$ derivatives, followed by a projection of all indices back into the $EAdS_4$ tangent space:
\begin{align}
 \nabla_\mu v_\nu &= P_\mu^\rho(x) P_\nu^\sigma(x) \frac{\del v_\sigma}{\del x^\rho} \ ; \label{eq:nabla} \\
 P_\mu^\nu(x) &\equiv \delta_\mu^\nu - \frac{x_\mu x^\nu}{x\cdot x} \ . \label{eq:P}
\end{align}
With lowered indices, the projector $P_\mu^\nu(x)$ becomes the 4d metric of $EAdS_4$ at $x$:
\begin{align}
 g_{\mu\nu}(x) &\equiv P_{\mu\nu}(x) = \eta_{\mu\nu} - \frac{x_\mu x_\nu}{x\cdot x} \ . \label{eq:g}
\end{align}
Our use of different letters for $P_\mu^\nu$ and $g_{\mu\nu}$ is purely cosmetic.

Since HS fields carry many symmetrized tensor indices, it is convenient to package them as functions of an auxiliary ``polarization vector'' $u^\mu\in\bbR^{1,4}$. Thus, we encode a rank-$p$ symmetric tensor by a function of the form:
\begin{align}
 f(x,u) = \frac{1}{p!}\,u^{\mu_1}\!\dots u^{\mu_p}f_{\mu_1\dots\mu_p}(x) \ . \label{eq:generating_function}
\end{align}
We denote flat $\bbR^{1,4}$ derivatives w.r.t. $x^\mu$ and $u^\mu$ as $\del_x^\mu$ and $\del_u^\mu$, respectively. The tensor rank of $f_{\mu_1\dots\mu_p}$, and the fact that it's tangential to the $EAdS_4$ hyperboloid, can be expressed as constraints on $f(x,u)$: 
\begin{align}
 (u\cdot\del_u) f = pf \ ; \quad (x\cdot\del_u) f = 0 \ .
\end{align}
Tracing a pair of indices on $f_{\mu_1\dots\mu_p}$ is encoded by acting on $f(x,u)$ with the operator $\del_u\cdot\del_u$. A factor of the metric $EAdS_4$ metric \eqref{eq:g} can be encoded as:
\begin{align}
 g_{\mu\nu} u^\mu u^\nu = u\cdot u - \frac{(u\cdot x)^2}{x\cdot x} \ .
\end{align}
It is convenient to introduce a notation for the \emph{traceless part} of a symmetric $EAdS_4$ tensor $\hat t^{\mu_1}\dots\hat t^{\mu_s}$ at a point $x$. This traceless part can be encoded by the function:
\begin{align}
 \begin{split}
   \calT^{(p)}(x,\hat t,u) &\equiv \frac{1}{p!}u^{\mu_1}\!\dots u^{\mu_p}\calT_{\mu_1\dots\mu_p}(x,\hat t) = \frac{(\hat t\cdot u)^p}{p!} - \text{traces} \\
     &= \frac{1}{p!}\sum_{n=0}^{\lfloor p/2 \rfloor} \binom{p-n}{n}\left(-\frac{1}{4}(\hat t\cdot\hat t)(g_{\mu\nu}(x)u^\mu u^\nu) \right)^n (\hat t\cdot u)^{p-2n} \\
     &= \frac{1}{2^p p!}\sum_{n=0}^{\lfloor p/2 \rfloor} \binom{p+1}{2n+1} \big({-(\hat t\cdot\hat t)}(q_{\mu\nu}(x,\hat t)u^\mu u^\nu) \big)^n (\hat t\cdot u)^{p-2n} \ ,
 \end{split} \label{eq:T}
\end{align}
where, in the third line, we introduced the 3d metric $q_{\mu\nu} = g_{\mu\nu} - \frac{\hat t_\mu\hat t_\nu}{\hat t\cdot\hat t}$ of the subspace orthogonal to both $x^\mu$ and $\hat t^\mu$.

So far, everything was defined on the $EAdS_4$ hyperboloid $x\cdot x = -1$. The idea of the radial reduction approach \cite{Biswas:2002nk} is to define our functions $f(x,u)$ also away from $x\cdot x = -1$, by introducing a scaling law of the form $(x\cdot\del_x) f = -\Delta f$ with some weight $\Delta$, usually chosen to match the conformal weight of relevant boundary data. This gives meaning to the 5d flat derivative $\del_x^\mu$ in all directions, which can lead to substantial simplifications, in particular for the Sleight-Taronna vertex. Within this formalism, the $EAdS_4$ symmetrized gradient, divergence and Laplacian take the form:
\begin{align}
 u\cdot\nabla ={}& u\cdot\del_x + \frac{u\cdot x}{x\cdot x}(u\cdot\del_u - x\cdot\del_x) \ ; \label{eq:grad} \\
 \del_u\cdot\nabla ={}& \del_u\cdot\del_x + \frac{u\cdot x}{x\cdot x}(\del_u\cdot\del_u) \ ; \label{eq:div} \\
 \begin{split}
   \nabla\cdot\nabla ={}& \del_x\cdot\del_x + 2\,\frac{u\cdot x}{x\cdot x}(\del_u\cdot\del_x) + \left(\frac{u\cdot x}{x\cdot x}\right)^2 (\del_u\cdot\del_u) \\
     &- \frac{1}{x\cdot x}\left((x\cdot\del_x)^2 + 3(x\cdot\del_x) - u\cdot\del_u\right) \ .
 \end{split} \label{eq:bulk_Laplacian}
\end{align}
In these expressions, we see two kinds of correction terms: 
\begin{itemize}
	\item The $\sim u\cdot x$ terms serve to project the 5d derivatives back into $EAdS_4$. 
	\item The terms on the bottom line of \eqref{eq:bulk_Laplacian} are just a constant multiple $\Delta(\Delta-3)-p$ of the $EAdS_4$ curvature -$\frac{1}{x\cdot x}$ (which we set equal to 1 in \eqref{eq:EAdS}). The 4d Laplacian $\nabla\cdot\nabla$ is then the $EAdS_4$ projection of the 5d d'Alembertian $\del_x\cdot\del_x$, shifted by this constant.
\end{itemize}

\subsection{Fronsdal fields in the bulk} \label{sec:preliminaries:Fronsdal}

Let us review the form of Fronsdal's field equations for linearized HS fields \cite{Fronsdal:1978vb} in the above framework. In Fronsdal's formalism, a spin-$s$ field (more precisely, gauge potential) is a totally symmetric rank-$s$ tensor with vanishing \emph{double trace}. This can be encoded by a scalar function $h^{(s)}(x,u)$, as in \eqref{eq:generating_function}. For its scaling weight, we choose $\Delta = 1+s$ -- the conformal weight of the dual boundary currents. This is the choice of \cite{Sleight:2016dba}, which brings the Sleight-Taronna vertex into a simple form. Note that this weight is different from that in the general literature on HS cubic vertices \cite{Joung:2011ww}, where the dual weight choice $\Delta = 2-s$ is used. Overall, the constraints on the field $h^{(s)}(x,u)$ read:
\begin{align}
 (u\cdot \del_u)h^{(s)} &= s h^{(s)} \ ; & (x\cdot\del_u)h^{(s)} &= 0 \ ; \label{eq:h_tensor_properties_1} \\ 
 (x\cdot\del_x)h^{(s)} &= -(s+1) h^{(s)} \ ; & (\del_u\cdot\del_u)^2 h^{(s)} &= 0 \ .  \label{eq:h_tensor_properties_2}
\end{align}
Gauge transformations take the form:
\begin{align}
h^{(s)} \ \rightarrow \ h^{(s)} + (u\cdot\nabla_x)\,\Lambda^{(s)} = h^{(s)} + \left( u\cdot\del_x + (2s-1)\frac{u\cdot x}{x\cdot x} \right) \Lambda^{(s)} \ , \label{eq:gauge_transformation}
\end{align}
where $\Lambda^{(s)}$ represents a traceless gauge parameter with $s-1$ tensor indices and weight $\Delta=s$, i.e.:
\begin{align}
 (u\cdot \del_u)\Lambda^{(s)} &= (s-1) \Lambda^{(s)} \ ; & (x\cdot\del_u)\Lambda^{(s)} &= 0 \ ; \label{eq:Lambda_tensor_properties_1} \\ 
 (x\cdot\del_x)\Lambda^{(s)} &= -s \Lambda^{(s)} \ ; & (\del_u\cdot\del_u)\Lambda^{(s)} &= 0 \ . \label{eq:Lambda_tensor_properties_2}
\end{align}
Out of the field $h^{(s)}$, we can construct a gauge-invariant curvature, which generalizes the $s=2$ linearized Ricci tensor to all spins. This is the Fronsdal tensor $\calF h^{(s)}$, where the operator $\calF$ is given by:
\begin{align}
 \begin{split}
   \calF ={}& -\nabla\cdot\nabla + \frac{2+2s-s^2}{x\cdot x} + (u\cdot\nabla)(\del_u\cdot\nabla) - \left(\frac{1}{2}(u\cdot\nabla)^2 + \frac{g_{\mu\nu}u^\mu u^\nu}{x\cdot x} \right)(\del_u\cdot\del_u) \\ 
      ={}& -\del_x\cdot\del_x + \left(u\cdot\del_x + (2s-1)\frac{u\cdot x}{x\cdot x} \right)(\del_u\cdot\del_x) \\
     &- \left(\frac{u\cdot u}{x\cdot x} + \frac{1}{2}\left(u\cdot\del_x + (2s+1)\frac{u\cdot x}{x\cdot x} \right)\left(u\cdot\del_x + (2s-3)\frac{u\cdot x}{x\cdot x} \right) \right)(\del_u\cdot\del_u) \ .
 \end{split} \label{eq:Fronsdal}
\end{align}
$\calF$ is a second-order differential operator with respect to $x$. The Fronsdal tensor $\calF h^{(s)}$ has the same tensor properties as the potential $h^{(s)}$, but with scaling weight increased by 2:
\begin{align}
 (u\cdot \del_u)\calF h^{(s)} &= s\calF h^{(s)} \ ; & (x\cdot\del_u)\calF h^{(s)} &= 0 \ ; \label{eq:F_tensor_properties_1} \\ 
 (x\cdot\del_x)\calF h^{(s)} &= -(s+3)\calF h^{(s)} \ ; & (\del_u\cdot\del_u)^2\calF h^{(s)} &= 0 \ .  \label{eq:F_tensor_properties_2}
\end{align}
In analogy with GR, we can rearrange the trace of $\calF h^{(s)}$ to obtain the Einstein tensor:
\begin{align}
 \calG h^{(s)} = \left(1 - \frac{1}{4}(g_{\mu\nu}u^\mu u^\nu)(\del_u\cdot\del_u) \right) \calF h^{(s)} \ . \label{eq:Einstein}
\end{align}
This has the same tensor properties \eqref{eq:F_tensor_properties_1}-\eqref{eq:F_tensor_properties_2}, but also satisfies a conservation law of the form:
\begin{align}
 (\del_u\cdot\nabla)\calG h^{(s)} = (g_{\mu\nu}u^\mu u^\nu)(\dots) \ , \label{eq:Einstein_conservation}
\end{align}
i.e. the $EAdS_4$ divergence of $\calG h^{(s)}$ vanishes up to trace terms. This allows us to write a gauge-invariant quadratic action for linearized HS fields:
\begin{align}
 S_s = \int_{EAdS_4} d^4x\,s!\,h^{(s)}(x,\del_u)\left(\frac{1}{2}\calG h^{(s)}(x,u) - J^{(s)}(x,u) \right) \ . \label{eq:Fronsdal_action}
\end{align}
Here, $J^{(s)}(x,u)$ is an external HS current, which must be conserved in the same sense \eqref{eq:Einstein_conservation} as $\calG h^{(s)}$. The field equations for the action \eqref{eq:Fronsdal_action} read simply:
\begin{align}
 \calG h^{(s)}(x,u) = J^{(s)}(x,u) \ . \label{eq:Fronsdal_equation}
\end{align}
This formalism for HS theory is substantially simplified in a \emph{traceless gauge} (which can also be viewed as a framework in its own right \cite{Skvortsov:2007kz,Campoleoni:2012th}). In this gauge, the double-traceless condition $(\del_u\cdot\del_u)^2 h^{(s)} = 0$ is strengthened into ordinary tracelessness $(\del_u\cdot\del_u) h^{(s)} = 0$. The remaining gauge freedom is parameterized by \eqref{eq:gauge_transformation}-\eqref{eq:Lambda_tensor_properties_2}, with the further constraint:
\begin{align}
 (\del_u\cdot\nabla)\Lambda^{(s)} = 0 \ . \label{eq:div_Lambda_raw}
\end{align}
Since $\Lambda^{(s)}$ is traceless, we see from \eqref{eq:div} that its 4d divergence $(\del_u\cdot\nabla)\Lambda^{(s)}$ is equal to the 5d one $(\del_u\cdot\del_x)\Lambda^{(s)}$. Thus, the constraint \eqref{eq:div_Lambda_raw} can also be written as:
\begin{align}
 (\del_u\cdot\del_x)\Lambda^{(s)} = 0 \ . \label{eq:div_Lambda} 
\end{align}
In this gauge, the Fronsdal operator \eqref{eq:Fronsdal} simplifies into:
\begin{align}
 \begin{split}
   \calF &= -\nabla\cdot\nabla + \frac{2+2s-s^2}{x\cdot x} + (u\cdot\nabla)(\del_u\cdot\nabla) \\
     &= -\del_x\cdot\del_x + \left(u\cdot\del_x + (2s-1)\frac{u\cdot x}{x\cdot x} \right)(\del_u\cdot\del_x) \ .
 \end{split} \label{eq:Fronsdal_traceless}
\end{align}
Note also that the trace of the Fronsdal tensor now reads simply:
\begin{align}
 (\del_u\cdot\del_u)\calF h^{(s)} = 2(\del_u\cdot\nabla)^2 h^{(s)} = 2(\del_u\cdot\del_x)^2 h^{(s)} \ . \label{eq:trace_Fronsdal}
\end{align}
With the exception of section \ref{sec:delta_gauge}, we will work in traceless gauge throughout. For source-free fields, one can specialize further to transverse-traceless gauge, by imposing also the zero-divergence condition $(\del_u\cdot\nabla)h^{(s)} = 0$, or, equivalently, $(\del_u\cdot\del_x)h^{(s)} = 0$. A gauge parameter that preserves traceless gauge, i.e. that satisfies \eqref{eq:div_Lambda_raw}-\eqref{eq:div_Lambda}, will shift the divergence of $h^{(s)}$ as:
\begin{align}
 (\del_u\cdot\nabla)h^{(s)} \ \rightarrow \ (\del_u\cdot\nabla)h^{(s)} + \left(\nabla\cdot\nabla + \frac{s^2-1}{x\cdot x} \right)\Lambda^{(s)} \ , \label{eq:4d_Box_Lambda}
\end{align}
or, equivalently:
\begin{align}
 (\del_u\cdot\del_x)h^{(s)} \ \rightarrow \ (\del_u\cdot\del_x)h^{(s)} + \left(\del_x\cdot\del_x + \frac{2(2s-1)}{x\cdot x} \right)\Lambda^{(s)} \ . \label{eq:Box_Lambda}
\end{align}

\subsection{Boundary theory} \label{sec:preliminaries:CFT}

The 3d boundary of $EAdS_4$ is given by the projective lightcone in $\bbR^{1,4}$, i.e. by null vectors $\ell^\mu\in\bbR^{1,4}$, $\ell\cdot\ell = 0$, modulo rescalings $\ell^\mu \cong \rho\ell^\mu$. Boundary quantities will transform under such rescalings as $(\ell\cdot\del_\ell)f = -\Delta f$, according to their conformal weights $\Delta$. We describe 3d vectors at a boundary point $\ell^\mu$ as 5d vectors $\lambda^\mu$ that satisfy $\lambda\cdot\ell = 0$, modulo shifts $\lambda^\mu\cong \lambda^\mu + \alpha\ell^\mu$. For a boundary scalar $f(\ell)$ with weight $\Delta=\frac{1}{2}$, we can define the conformal Laplacian $\Box_\ell f$. In the embedding-space language, this is the same as the 5d d'Alambertian $(\del_\ell\cdot\del_\ell)f$, provided that $f$ is extended away from the $\ell\cdot\ell = 0$ lightcone in a way that preserves the scaling law $(\ell\cdot\del_\ell)f = -\frac{1}{2}f$. The operator $\Box_\ell$ itself has conformal weight 2.

The CFT that lives on our 3d boundary is a free $O(N)$ vector model. It is convenient to assume that $N$ is even, and package the vector model's $N$ real fields as $\frac{N}{2}$ \emph{complex} fields $\chi^I(\ell)$ with complex conjugates $\bar\chi_I(\ell)$, where $I=1,\dots,\frac{N}{2}$ is a color index. The theory then takes the form of a $U(N/2)$ vector model, whose action reads simply:
\begin{align}
 S_{\text{CFT}} = -\int d^3\ell\,\bar\chi_I(\ell)\Box_\ell\chi^I(\ell) \ , \label{eq:CFT}
\end{align}
where $\chi^I$ and $\bar\chi_I$ each have conformal weight $\Delta = \frac{1}{2}$. The propagator for these fundamental fields reads:
\begin{align}
 G_{\text{CFT}}(\ell,\ell') = \frac{1}{4\pi\sqrt{-2\ell\cdot\ell'}} \ ; \quad \Box_\ell G_{\text{CFT}}(\ell,\ell') = -\delta^{\frac{5}{2},\frac{1}{2}}(\ell,\ell') \ , \label{eq:G_CFT}
\end{align}
where the superscripts on the boundary delta function $\delta(\ell,\ell')$ denote its conformal weight with respect to each argument.

The fundamental single-trace operators in the theory \eqref{eq:CFT} are the \emph{bilocals}:
\begin{align}
 \calO(\ell,\ell') \equiv \frac{2\chi^I(\ell)\bar\chi_I(\ell')}{G(\ell,\ell')} \ . \label{eq:bilocal}
\end{align}
Here, we made an unconventional normalization choice, which makes $\calO(\ell,\ell')$ invariant under rescalings of $\ell,\ell'$. Thus, our $\calO(\ell,\ell')$ depends only on the actual choice of two boundary points, which will allow a cleaner interpretation of the bulk dual. The numerical factor in \eqref{eq:bilocal} is chosen to ensure the proper relative normalization of the first and second terms in eq. \eqref{eq:jjO} below.

\begin{figure}%
	\centering%
	\includegraphics[scale=1.0]{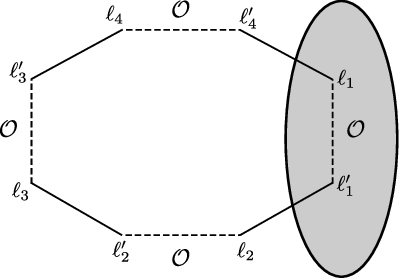} \\
	\caption{Boundary Feynman diagram for a quartic correlator of bilocals $\calO(\ell_i,\ell'_i)$. From the point of view of the operator $\calO(\ell_1,\ell'_1)$ in the shaded region, the other operators couple to it as a single bilocal, in this case as $\calO(\ell_2,\ell'_4)$.}
	\label{fig:boundary_diagram} 
\end{figure}

By Taylor-expanding the bilocals \eqref{eq:bilocal} around $\ell=\ell'$, we obtain the local single-trace primaries, i.e. the tower of HS currents \cite{Craigie:1983fb,Anselmi:1999bb,David:2020ptn} (including the honorary spin-0 ``current'' $\bar\chi_I(\ell)\chi^I(\ell)$). These local currents can be encoded conveniently by contracting their indices with a null polarization vector $\lambda^\mu$ at $\ell^\mu$, satisfying $\lambda\cdot\lambda = \lambda\cdot\ell = 0$:
\begin{align}
  j^{(s)}(\ell,\lambda) = \lambda^{\mu_1}\!\dots\lambda^{\mu_s} j_{\mu_1\dots\mu_s}(\ell) \ .
\end{align}
The currents' relation to the bilocal \eqref{eq:bilocal} is then expressed compactly via a differential operator $D^{(s)}$, as:
\begin{align}
 \begin{split}
   j^{(s)}(\ell,\lambda) &= D^{(s)}(\del_\ell,\del_{\ell'},\lambda)\big[\chi^I(\ell)\bar\chi_I(\ell')\big]\Big|_{\ell=\ell'} \\
     &= \frac{1}{2}\,D^{(s)}(\del_\ell,\del_{\ell'},\lambda)\big[G(\ell,\ell')\calO(\ell,\ell')\big]\Big|_{\ell=\ell'} \ ; 
 \end{split} \label{eq:j} \\
 D^{(s)}(\del_\ell,\del_{\ell'},\lambda) &= i^s\sum_{m=0}^s (-1)^m \binom{2s}{2m} (\lambda\cdot\del_\ell)^m (\lambda\cdot\del_{\ell'})^{s-m} \ . \label{eq:D}
\end{align}
The connected correlators of bilocals \eqref{eq:bilocal} are given by simple 1-loop Feynman diagrams composed of propagators \eqref{eq:G_CFT} (see figure \ref{fig:boundary_diagram}), with the normalization factor in \eqref{eq:bilocal} simply along for the ride:
\begin{align}
 \left<\calO(\ell_1,\ell_1')\dots \calO(\ell_n,\ell_n')\right> = \frac{2^n}{\prod_{p=1}^n G(\ell_p,\ell_p')}\times\frac{N}{2}\left(\prod_{p=1}^n G(\ell_p',\ell_{p+1}) + \text{permutations}\right) \ , \label{eq:correlators_bilocal}
\end{align}
where the product in the numerator is cyclic, i.e. $\ell_{n+1}\equiv\ell_1$, and the sum is over cyclically inequivalent permutations of $(1,\dots,n)$. From these bilocal correlators, one can derive the correlators of local currents $j^{(s)}$, via the Taylor expansion \eqref{eq:j}. 

Up to the boundary field equation $\Box_\ell\chi^I(\ell) = \Box_{\ell'}\chi_I(\ell') = 0$, the local currents \eqref{eq:j} span the full space of single-trace operators. This means in particular that, given two points $\ell,\ell'$ and a compact boundary region $B$ that includes them, the bilocal $\calO(\ell,\ell')$ is equivalent to some superposition of local currents \eqref{eq:j} inside $B$:
\begin{align}
 \calO(\ell,\ell') \cong \sum_{s=0}^\infty \int_B d^3L\,A^{(s)}_{\ell,\ell'}(L,\del_\lambda)\,j^{(s)}(L,\lambda) \ , \label{eq:local_decomposition}
\end{align}
where $A^{(s)}_{\ell,\ell'}(L,\lambda)$ is some configuration of traceless spin-$s$ sources at the boundary point $L$:
\begin{align}
 (\lambda\cdot\del_\lambda)A^{(s)} = sA^{(s)} \ ; \quad (L\cdot\del_L)A^{(s)} = (s-2)A^{(s)} \ ; \quad (\del_\lambda\cdot\del_\lambda)A^{(s)} = (L\cdot\del_\lambda)A^{(s)} = 0 \ .
\end{align}
The sense in which the equivalence \eqref{eq:local_decomposition} holds is that the LHS and RHS have the same correlators with any number of operators $\calO$ or $j^{(s)}$ \emph{with support in the complement $\bar B$ of $B$}. On the other hand, to check that some configuration of local sources $A^{(s)}$ in $B$ satisfies \eqref{eq:local_decomposition}, it is sufficient to check just the \emph{quadratic} correlators with \emph{local} currents $j^{(s)}$ in $\bar B$. This can be seen in two steps. First, in any correlator of one single-trace operator in $B$ and $(n-1)$ such operators in $\bar B$, the diagrams \eqref{eq:correlators_bilocal} are always arranged such that the operator in $B$ effectively couples to a \emph{single bilocal} in $\bar B$ (see figure \ref{fig:boundary_diagram}). Thus, it's enough to match the \emph{quadratic} correlators with bilocals in $\bar B$. But, using now the equivalence \eqref{eq:local_decomposition} for $\bar B$, we see that these can be reconstructed from the quadratic correlators with \emph{local} currents. 

Again, the theory described above is not quite the $O(N)$ vector model, but the $U(N/2)$ one. However, we can obtain the $O(N)$ model by simply truncating the single-trace operators \eqref{eq:bilocal},\eqref{eq:j} from all those invariant under $U(N/2)$ to those invariant under the larger group $O(N)$. For the bilocals \eqref{eq:bilocal}, this requires symmetrizing over $\ell\leftrightarrow\ell'$:
\begin{align}
 \calO^+(\ell,\ell') = \frac{1}{2}\big(\calO(\ell,\ell') + \calO(\ell',\ell) \big) \ , \label{eq:bilocal_even}
\end{align}
whereas for the local currents \eqref{eq:j}, it requires restricting to even spins $s$. It's easy to see that the even-spin currents $j^{(s)}$ can indeed be constructed from the symmetrized bilocal \eqref{eq:bilocal_even}. For odd $N$, the above construction starting from $U(N/2)$ doesn't directly apply. However, the end results for the correlators are the same, with $N$ simply an overall prefactor, as in \eqref{eq:correlators_bilocal}.

\subsection{Boundary asymptotics of bulk fields} \label{sec:preliminaries:asymptotics}

In this subsection, we set up a framework for discussing the \emph{asymptotic behavior} of fields in $EAdS_4$. For this purpose, it's convenient to use Poincare coordinates $(z,y^a)$ for $EAdS_4$:
\begin{align}
x^\mu(z,y^a) = \frac{1}{z}\left(\frac{1+z^2+y^2}{2}, \frac{1-z^2-y^2}{2}, y^a \right) \ ; \quad dx\cdot dx = \frac{dz^2 + dy^2}{z^2} \ , \label{eq:Poincare}
\end{align}
where $y^2\equiv \delta_{ab}y^a y^b$. The boundary of $EAdS_4$ can be similarly parameterized as:
\begin{align}
\ell^\mu(y^a) = \left(\frac{1+y^2}{2}, \frac{1-y^2}{2}, y^a \right) \ ; \quad d\ell\cdot d\ell = dy^2 \ . \label{eq:flat_section}
\end{align}
The parameterization \eqref{eq:flat_section} chooses a flat section of the $\bbR^{1,4}$ lightcone, defined by $\ell\cdot n = -\frac{1}{2}$, where $n^\mu = \left(\frac{1}{2},-\frac{1}{2},\vec{0}\right)$. The bulk and boundary coordinates \eqref{eq:Poincare}-\eqref{eq:flat_section} are related by:
\begin{align}
x^\mu(z,y^a) = \frac{1}{z}\ell^\mu(y^a) + zn^\mu \ . \label{eq:x_ell}
\end{align}
In the limit $z\rightarrow 0$, the bulk point $x(z,y^a)$ asymptotes to the boundary point $\ell(y^a)$, in the precise manner defined by \eqref{eq:x_ell}. 

To study the asymptotics of tensor fields, it is convenient to use an orthonormal basis $(e_0,e_a)$ along the $(z,y^a)$ coordinate axes:
\begin{align}
e_0^\mu(z,y^a) = -z\frac{\del x^\mu}{\del z} \ ; \quad e_a^\mu(y^a) = z\frac{\del x^\mu}{\del y^a} \ .
\end{align}
In the boundary limit $z\rightarrow 0$, the $\bbR^{1,4}$ components of the ``tangential'' basis vectors $e_a^\mu$ are $z$-independent, while those of the ``radial'' vector $e_0^\mu$ behave as:
\begin{align}
e_0^\mu(z,y^a) = x^\mu(z,y^a) + O(z) = \frac{1}{z}\ell^\mu(y^a) + O(z) \ .
\end{align}
We can now discuss the asymptotics of symmetric bulk tensor fields \eqref{eq:generating_function} by describing the $z\rightarrow 0$ scaling of their different components in the orthonormal $(e_0,e_a)$ basis. For a rank-$p$ field $f(x,u)$, we'll use the compact notation $[f]_{q,p-q}$ to refer to its components with $q$ indices along $e_0$ and $p-q$ indices along $e_a$. 

\subsection{Boundary-bulk propagator} \label{sec:preliminaries:Pi}

The boundary-bulk propagators dual to the boundary HS currents \eqref{eq:j} read \cite{Mikhailov:2002bp}:
\begin{align}
 \Pi^{(s)}(x,u;\ell,\lambda) = -\frac{(\sqrt{2})^s(m\cdot u)^s}{16\pi^2(\ell\cdot x)^{2s+1}} \ ; \quad m^\mu(x;\ell,\lambda) \equiv (\lambda\cdot x)\ell^\mu - (\ell\cdot x)\lambda^\mu \ , \label{eq:Pi}
\end{align}
where we chose a non-standard normalization for later convenience. With respect to its bulk arguments $(x,u)$, the propagator $\Pi^{(s)}$ satisfies the standard constraints \eqref{eq:h_tensor_properties_1}-\eqref{eq:h_tensor_properties_2} for a Fronsdal field, as well as the traceless and transverse gauge conditions $(\del_u\cdot\del_u)\Pi^{(s)} = (\del_u\cdot\nabla)\Pi^{(s)} = 0$. With respect to its boundary arguments $(\ell,\lambda)$, $\Pi^{(s)}$ has the same conformal weight $(\ell\cdot\del_\ell)\Pi^{(s)} = -(s+1)\Pi^{(s)}$ and tensor rank $(\lambda\cdot\del_\lambda)\Pi^{(s)} = s\Pi^{(s)}$ as the boundary currents \eqref{eq:j}, and is invariant under the shift symmetry $\lambda^\mu\rightarrow \lambda^\mu + \alpha\ell^\mu$.

The propagator \eqref{eq:Pi} is a special case $(p,w)=(s,s+1)$ of the general formula:
\begin{align}
 f(x,u;\ell,\lambda) \sim \frac{(m\cdot u)^p}{(\ell\cdot x)^{p+w}} \ , \label{eq:Pi_w}
\end{align}
which spans the solution space of the free field equations for rank-$p$ symmetric, transverse-traceless fields with arbitrary mass parameterized by $w$:
\begin{align}
 (\del_u\cdot\del_u)f = (\del_u\cdot\nabla)f = \left(\nabla\cdot\nabla - \frac{w(3-w) + p}{x\cdot x} \right) f = 0 \ . \label{eq:field_eq_w}
\end{align}
Let's now apply the formalism of section \ref{sec:preliminaries:asymptotics} to discuss the asymptotic behavior of the general propagator \eqref{eq:Pi_w}. Let us choose Poincare coordinates \eqref{eq:Poincare}-\eqref{eq:flat_section} such that the boundary source point $\ell^\mu$ in \eqref{eq:Pi_w} is at $y^a=0$, i.e. $\ell^\mu = \left(\frac{1}{2},\frac{1}{2},\vec{0}\right)$. We can also choose the polarization vector $\lambda^\mu$ as $\lambda^\mu = (0,0,\lambda^a)$, which becomes $\lambda^\mu = \lambda^a e_a^\mu$ in terms of the orthonormal basis at $y^a=0$. The ingredients of the tensor field \eqref{eq:Pi_w} at an arbitrary bulk point $x^\mu(z,y^a)$ now read:
\begin{align}
 \ell\cdot x = -\frac{z^2 + y^2}{2z} \ ; \quad m^\mu = (\lambda\cdot y)e_0^\mu + \frac{1}{z}\left(\frac{z^2 + y^2}{2}\lambda^a - (\lambda\cdot y)y^a\right) e_a^\mu \ , \label{eq:building_blocks}
\end{align}
where $\lambda\cdot y\equiv \delta_{ab}\lambda^a y^b$. Assuming $y^a\neq 0$, we see that in the small-$z$ limit $\ell\cdot x$ scales as $z^{-1}$, while $m^\mu$ has $\sim z^{-1}$ components along $e_a^\mu$ and a $\sim z^0$ component along $e_0^\mu$. Thus, at $y^a\neq 0$, the various components of $f$ scale at small $z$ as:
\begin{align}
 y^a\neq 0:\quad [f]_{q,p-q} \sim z^{w+q} \ . \label{eq:Pi_w_asymptotics}
\end{align}
Now, note that under $w\rightarrow 3-w$, the field equations \eqref{eq:field_eq_w} do not change. Therefore, the same field equations must also support the asymptotics $[f]_{q,p-q} \sim z^{3-w+q}$. In a neighborhood of the boundary, the two asymptotics $\sim z^{w+q}$ and $\sim z^{3-w+q}$ constitute a pair of independent boundary data (more precisely, within each set, it is the $q=0$ data that's independent, with the $q>0$ data determined from it). For a regular solution in all of $EAdS_4$, these two boundary data cease to be independent, i.e. one becomes linearly determined by the other. In particular, a closer inspection of the solution \eqref{eq:Pi_w} reveals that it also contains the ``other'' asymptotics $\sim z^{3-w+q}$, as a delta-function-like distribution with support at $y^a=0$. Rotational invariance and the dilatation symmetry $(z,y^a)\rightarrow (\rho z,\rho y^a)$ fix this delta-function-like piece to take the form:
\begin{align}
 y^a=0:\quad [f]_{q,p-q} \sim z^{3-w+q}(e_0\cdot u)^q(\lambda\cdot u)^{p-q}(\lambda\cdot\del_y)^q\,\delta^3(y) \ ,
\end{align}
where $\lambda\cdot\del_y \equiv \lambda^a\frac{\del}{\del y^a}$. Specializing back to $(p,w)=(s,s+1)$, we obtain, for our original propagator \eqref{eq:Pi}:
\begin{align}
 y^a\neq 0:\quad &[\Pi^{(s)}]_{q,s-q} \sim z^{s+1+q} \ ; \label{eq:Pi_away_from_delta} \\
 y^a=0:\quad &[\Pi^{(s)}]_{q,s-q} \sim z^{2-s+q}(e_0\cdot u)^q(\lambda\cdot u)^{s-q}(\lambda\cdot\del_y)^q\,\delta^3(y) \ . \label{eq:Pi_delta}
\end{align}

\subsection{Bulk geodesics} \label{sec:preliminaries:geodesics}

The Didenko-Vasiliev solution is the field of an HS-charged source concentrated on a bulk geodesic. Before describing the solution and its properties, it is useful to discuss bulk geodesics in their own right.

A geodesic in $EAdS_4$ is a hyperbola in the $\bbR^{1,4}$ embedding space. The hyperbola's asymptotes are two lightrays through the origin in $\bbR^{1,4}$, or, equivalently, two points on the conformal boundary of $EAdS_4$. In fact, (oriented) bulk geodesics are in one-to-one correspondence with (ordered) pairs of boundary points. We can parameterize a geodesic's boundary endpoints by two lightlike vectors $\ell^\mu,\ell'^\mu$, keeping in mind the usual redundancy of such vectors under rescalings. The geodesic itself can then be parameterized as:
\begin{align}
 \gamma(\ell,\ell'):\quad x^\mu(\tau;\ell,\ell') = \frac{e^\tau \ell^\mu + e^{-\tau}\ell'^\mu}{\sqrt{-2\ell\cdot\ell'}} \ , \label{eq:geodesic}
\end{align}
where $\tau$ is a proper-length parameter. If we allow rescalings of $x^\mu$ away from the $EAdS_4$ hyperboloid $x\cdot x = -1$, then the geodesic \eqref{eq:geodesic} becomes just a 2d plane in the $\bbR^{1,4}$ embedding space -- the plane spanned by $\ell^\mu,\ell'^\mu$.

The distance of a bulk point $x\in EAdS_4$ from a geodesic $\gamma(\ell,\ell')$ can be parameterized by the function:
\begin{align}
 R(x;\ell,\ell') = \sqrt{\frac{2(\ell\cdot x)(\ell'\cdot x)}{(\ell\cdot\ell')(x\cdot x)} - 1} \ . \label{eq:R}
\end{align}
This has weight 0 (i.e. is invariant) under rescalings of $\ell^\mu,\ell'^\mu$, as well as rescalings of $x^\mu$. For $x^\mu$ on the $x\cdot x = -1$ hyperboloid, $R(x;\ell,\ell')$ is just the flat $\bbR^{1,4}$ distance between $x^\mu$ and the $(\ell,\ell')$ plane. This is related to the geodesic $EAdS_4$ distance $\chi$ as $R = \sinh\chi$. 

We can define a delta function that localizes $x\in EAdS_4$ on the geodesic $\gamma(\ell,\ell')$, i.e. at $R=0$, as:
\begin{align}
 \delta^3(x;\ell,\ell') = \int_{-\infty}^\infty d\tau\,\delta^4(x,x(\tau;\ell,\ell')) \ , \label{eq:geodesic_delta}
\end{align}
where $\delta^4$ is the delta function on $EAdS_4$, and $x(\tau;\ell,\ell')$ is the proper-length parameterization \eqref{eq:geodesic} of the geodesic. The formula \eqref{eq:geodesic_delta} assumes that $x^\mu$ lies on the $x\cdot x = -1$ hyperboloid. If we allow rescalings of $x^\mu$ away from $x\cdot x = -1$, an even simpler definition becomes possible: we can define $\delta^3(x;\ell,\ell')$ as just the standard flat 3d delta function in $\bbR^{1,4}$ with support on the $(\ell,\ell')$ plane. With this definition, $\delta^3(x;\ell,\ell')$ has weight $\Delta=3$ with respect to $x^\mu$ (and weight 0 with respect to $\ell^\mu,\ell'^\mu$). 

Given a geodesic $\gamma(\ell,\ell')$ and a bulk point $x^\mu$ that doesn't necessarily lie on it, one can define at $x$ the following pair of $EAdS_4$ vectors:
\begin{align}
 t_\mu(x;\ell,\ell') &= \frac{1}{2}\left(\frac{\ell'_\mu}{\ell'\cdot x} - \frac{\ell_\mu}{\ell\cdot x} \right) \ ; \label{eq:t} \\ 
 r_\mu(x;\ell,\ell') &= -\frac{x_\mu}{x\cdot x} + \frac{1}{2}\left(\frac{\ell_\mu}{\ell\cdot x} + \frac{\ell'_\mu}{\ell'\cdot x} \right) \ , \label{eq:r}
\end{align}
Here, $r^\mu(x;\ell,\ell')$ points radially away from the $\gamma(\ell,\ell')$ geodesic, while $t^\mu(x;\ell,\ell')$ points ``parallel to'' $\gamma(\ell,\ell')$, in the sense of parallel transport along $r^\mu$. These vectors satisfy:
\begin{align}
 t\cdot x = r\cdot x = t\cdot r = 0 \ ; \quad t\cdot t = -\frac{1}{x\cdot x}\cdot\frac{1}{1+R^2} \ ; \quad r\cdot r = -\frac{1}{x\cdot x}\cdot\frac{R^2}{1+R^2} \ . \label{eq:t_r_products}
\end{align}
We can then construct a \emph{complex null} vector in the $(t,r)$ plane:
\begin{align}
 k_\mu(x;\ell,\ell') &= \frac{1}{2}\left(t^\mu + \frac{ir_\mu}{R} \right) \ ; \quad k\cdot k = 0 \ ; \quad (k\cdot\nabla)k_\mu = 0 \ . \label{eq:k}
\end{align}
In Lorentzian signature, $k^\mu$ would be a real, affine tangent to radial lightrays emanating from $\gamma(\ell,\ell')$. The distance function $R$ and the null vector $k^\mu$ will be the main ingredients of the Didenko-Vasiliev solution below.

\subsection{Linearized DV solution}

The Didenko-Vasiliev solution \cite{Didenko:2009td} is a solution of the non-linear Vasiliev equations, structurally similar to supergravity's BPS black holes. We will be interested here in the solution's linearized version \cite{Didenko:2008va}, which consists of a multiplet of Fronsdal fields (one for each spin), satisfying the Fronsdal field equation \eqref{eq:Fronsdal_equation} with a \emph{particle-like source} concentrated on a bulk geodesic $\gamma(\ell,\ell')$. 

In terms of the building blocks from section \ref{sec:preliminaries:geodesics} above, the DV solution is described by the following multiplet of Fronsdal fields:
\begin{align}
  \phi^{(s)}(x,u;\ell,\ell') = \frac{1}{\pi R\sqrt{-x\cdot x}} \times \left\{
    \begin{array}{cl}
      1 & \qquad s = 0 \\
      \displaystyle \frac{2}{s!}(i\sqrt{2})^s(u\cdot k)^s & \qquad s\geq 1
    \end{array} \right. \ . \label{eq:phi}
\end{align}
Here, the spin-dependent normalization factors come from the master-field expression in \cite{Didenko:2009td}, which was translated into canonically normalized Fronsdal fields in \cite{David:2020fea}, by matching the normalizations of 2-point functions $\left< j^{(s)}j^{(s)} \right>$ in both languages. In its bulk arguments $(x,u)$, $\phi^{(s)}$ satisfies the standard constraints \eqref{eq:h_tensor_properties_1}-\eqref{eq:h_tensor_properties_2} for a Fronsdal field, as well as the traceless gauge condition $(\del_u\cdot\del_u)\phi^{(s)} = 0$. In the minimal HS theory, we include only even spins in \eqref{eq:phi}. While the potentials \eqref{eq:phi} are complex, their gauge-invariant curvatures are always real, i.e. the imaginary part of \eqref{eq:phi} is pure gauge. For odd spins, these reality properties are reversed.

The Einstein curvature of the DV solution \eqref{eq:phi}, i.e. the bulk source in its Fronsdal equation \eqref{eq:Fronsdal_equation}, is given by a delta function at $R=0$, as:
\begin{align}
 \calG \phi^{(s)}(x;\ell,\ell') = \frac{4}{s!}\,(i\sqrt{2})^s\,\delta^3(x;\ell,\ell') \big[(u\cdot t)^s - \text{double traces} \big] \ . \label{eq:Einstein_phi}
\end{align}
Here, $\delta^3(x;\ell,\ell')$ is the geodesic delta function \eqref{eq:geodesic_delta}, with support on $R=0$; $t^\mu$ is the vector \eqref{eq:t}, which at $R=0$ becomes just the tangent to $\gamma(\ell,\ell')$, normalized as $t\cdot t = -\frac{1}{x\cdot x}$; and ``${}-\text{double traces} $'' means that we subtract $\sim (g_{\mu\nu}u^\mu u^\nu)^2$ pieces so as to satisfy the double-tracelessness condition $(\del_u\cdot\del_u)^2\calG \phi^{(s)} = 0$. Eq. \eqref{eq:Einstein_phi} shows explicitly the HS charges carried by the geodesic. In particular, the factor of $(i\sqrt{2})^s$ encodes the BPS-like proportionality between the HS charges of different spins. In terms of the traceless structure \eqref{eq:T}, the Einstein curvature \eqref{eq:Einstein_phi} and the corresponding Fronsdal curvature can be written as:
\begin{align}
 \calG \phi^{(s)} &= 4(i\sqrt{2})^s\,\delta^3(x;\ell,\ell') \left(\calT^{(s)}(x,t,u) - \frac{\theta(s-2)(g_{\mu\nu}u^\mu u^\nu)}{4s(x\cdot x)}\,\calT^{(s-2)}(x,t,u) \right) \ ; \\
 \calF \phi^{(s)} &= 4(i\sqrt{2})^s\,\delta^3(x;\ell,\ell') \left(\calT^{(s)}(x,t,u) + \frac{\theta(s-2)(g_{\mu\nu}u^\mu u^\nu)}{4s(s-1)(x\cdot x)}\,\calT^{(s-2)}(x,t,u) \right) \ , \label{eq:Fronsdal_phi}
\end{align}
where $\theta$ is the step function:
\begin{align}
 \theta(p) = \left\{
   \begin{array}{cl}
     1 & \qquad p\geq 0 \\
     0 & \qquad p<0
  \end{array} \right. \ ,
\end{align}
and we assume the convention that $\theta(p)$ for negative $p$ vanishes ``stronger than anything else'', so that e.g. $\frac{\theta(s-2)}{s(s-1)}$ is zero for $s=0$.

\begin{figure}%
	\centering%
	\includegraphics[scale=0.6]{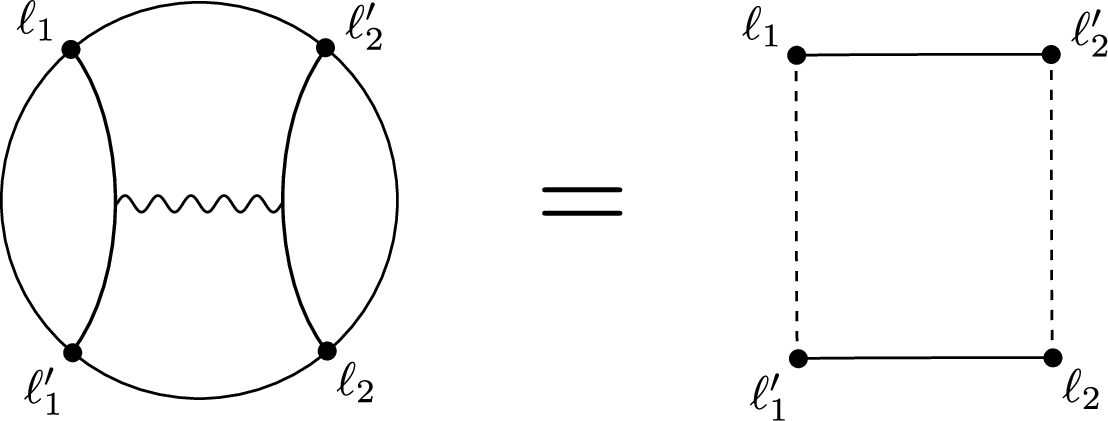} \\
	\caption{Bulk and boundary diagrams for the correlator of two boundary bilocals. On the left, each bilocal is represented in the bulk by a Didenko-Vasiliev solution. The solid lines represent each solution's central geodesic ``worldline'', where its Fronsdal curvature is concentrated. The wavy line represents the multiplet of HS gauge fields exchanged between the two worldlines. On the right, each bilocal is represented by a dashed line, while the solid lines represent propagators \eqref{eq:G_CFT} of the fundamental boundary fields $\chi^I,\bar\chi_I$. Upon restricting to even spins, one should average the boundary diagram under $\ell_1\leftrightarrow\ell'_1$.}
	\label{fig:quadratic} 
\end{figure}

It was recently understood \cite{David:2020fea,Lysov:2022zlw} that the DV solution \eqref{eq:phi} is the bulk dual of the bilocal boundary operator $\calO(\ell,\ell')$ from \eqref{eq:bilocal}, in the same way that the boundary-bulk propagators \eqref{eq:Pi} are the bulk duals of the local boundary currents \eqref{eq:j}. The main aspect of this correspondence is an agreement between the on-shell bulk action \eqref{eq:Fronsdal_action} for a pair of interacting DV solutions, and the CFT correlator of the corresponding boundary bilocals. The relevant Feynman/Witten diagrams are shown in figure \ref{fig:quadratic}.
The bulk action \eqref{eq:Fronsdal_action} in this case (for each spin channel) can be expressed as an integral $S_s[\phi^{(s)}_1,\gamma_2]$ of the first DV solution's field $\phi^{(s)}(x,u;\ell_1,\ell'_1)\equiv \phi_1$ over the second DV solution's worldline $\gamma(\ell_2,\ell'_2)\equiv \gamma_2$. The explicit formula for this action, with a general field $h^{(s)}_1(x,u)$ in place of $\phi_1^{(s)}$, reads:
\begin{align}
 S_s[h^{(s)}_1,\gamma_2] = -4(i\sqrt{2})^s s!\int_{-\infty}^\infty d\tau\,h_1^{(s)}\big(x(\tau;\ell_2,\ell'_2),\dot x(\tau;\ell_2,\ell'_2)\big) \ . \label{eq:quadratic_bulk}
\end{align}
Note that, despite the apparent asymmetry, $S_s[\phi^{(s)}_1,\gamma_2]$ is the same as $S_s[\phi^{(s)}_2,\gamma_1]$ (this is obvious from the Witten diagram in figure \ref{fig:quadratic}). 

The holographic duality between the bulk action and the boundary correlator now takes the form:
\begin{align}
  -N\sum_{s=0}^\infty S_s[\phi^{(s)}_1,\gamma_2] = \left<\calO(\ell_1,\ell'_1)\calO(\ell_2,\ell'_2)\right> \ , \label{eq:quadratic_correlator}
\end{align}
where the sum is over all spins, and the correlator on the RHS is from the $U(N/2)$ vector model. The restriction to even spins and the $O(N)$ vector model is immediate:
\begin{align}
 -N\sum_{\text{even }s} S_s[\phi^{(s)}_1,\gamma_2] = \left<\calO^+(\ell_1,\ell'_1)\calO^+(\ell_2,\ell'_2)\right> \ . \label{eq:quadratic_correlator_even}
\end{align}
The prefactor $N$ on the LHS can be thought of as an inverse Planck's constant, converting a classical bulk action into a proper quantum correlator:
\begin{align}
 N\equiv \frac{1}{\hbar} \ . \label{eq:hbar}
\end{align}
Our aim in the present paper is to study the extension of eq. \eqref{eq:quadratic_correlator_even} from the quadratic to the cubic level. 

Another aspect of the DV-solution/boundary-bilocal correspondence is that in the bilocal$\rightarrow$local limit \eqref{eq:j}, the DV solution simply reduces to the boundary-bulk propagators \cite{Lysov:2022zlw}:
\begin{align}
 \frac{1}{2}\,D^{(\tilde s)}\big[G(\ell,\ell')\phi^{(s)} \big]\Big|_{\ell,\ell'} = \delta_{s,\tilde s}\,\Pi^{(s)}(x,u;\ell,\lambda) \ , \label{eq:local_limit}
\end{align}
where, on the LHS, we act on $G(\ell,\ell')\phi^{(s)}(x,u;\ell,\ell')$ with the differential operator $D^{(\tilde s)}(\del_\ell,\del_{\ell'},\lambda)$ from \eqref{eq:D}, and then set $\ell=\ell'$. On the RHS, $\delta_{s,\tilde s}$ is a Kronecker symbol imposing $s=\tilde s$, and $\Pi^{(s)}$ is the boundary-bulk propagator \eqref{eq:Pi}. One application of the limit \eqref{eq:local_limit} is to impose it on the DV solution $\phi^{(s)}_1$ in \eqref{eq:quadratic_correlator}, making a boundary-bulk propagator $\Pi^{(s)}(x,u;\ell_1,\lambda_1)\equiv\Pi^{(s)}_1$ (with a single spin $s$ picked out of the HS multiplet). This produces a bulk calculation for the CFT correlator of a bilocal with a local current, as:
\begin{align}
 -N S_s[\Pi^{(s)}_1,\gamma_2] = \left<j^{(s)}(\ell_1,\lambda_1)\calO(\ell_2,\ell'_2)\right> \ , \label{eq:jO}
\end{align}
where, for even $s$, we can replace $\calO\rightarrow\calO^+$. An alternative bulk calculation of the same correlator is to evaluate the asymptotic electric field strength (or, in the $s=0$ case, the boundary data with weight $\Delta=1$) of the DV field $\phi^{(s)}(x,u;\ell_2,\ell'_2)$ at $\ell_1$. This calculation was carried out in \cite{Neiman:2017mel}.

\subsection{Relation to geodesic Witten diagrams}

The holographic relation \eqref{eq:quadratic_correlator} for the quadratic correlator of bilocals, as depicted in figure \ref{fig:quadratic}, is closely related to the literature on geodesic Witten diagrams \cite{Hijano:2015zsa,Dyer:2017zef}. There, the contribution of a \emph{particular OPE block} to a \emph{quartic} correlator is computed by a Witten diagram much like figure \ref{fig:quadratic}, with two geodesics exchanging a bulk field that corresponds to the conformal block in question. Our eq. \eqref{eq:quadratic_correlator} can be seen as a special case of this general relation.

To see this in detail, let us (for the sake of this discussion) lift the restriction of the boundary vector model \eqref{eq:CFT} to color-singlet operators. The fundamental colored fields $\chi^I$ then become primaries in their own right, and we can consider the quartic correlator $\left<\chi^I(\ell_1)\bar\chi_I(\ell'_1)\chi^J(\ell_2)\bar\chi_J(\ell'_2)\right>$. Expanding this in an OPE in the $(11'|22')$ channel, we find that two kinds of primaries contribute: the identity, and the tower of single-trace HS currents \eqref{eq:j}. In this decomposition, the single-trace blocks precisely describe the connected correlator $\left<\calO(\ell_1,\ell'_1)\calO(\ell_2,\ell'_2)\right>$, while the identity block describes its disconnected counterpart $\left<\calO(\ell_1,\ell'_1)\right>\!\left<\calO(\ell_2,\ell'_2)\right>$. Thus, the connected correlator $\left<\calO(\ell_1,\ell'_1)\calO(\ell_2,\ell'_2)\right>$ can be computed by summing over the single-trace blocks, which, in the language of geodesic Witten diagrams, becomes the sum over exchanged spins in figure \ref{fig:quadratic}. 

Finally, we should comment on a cosmetic difference between figure \ref{fig:quadratic} and the original construction of geodesic Witten diagrams \cite{Hijano:2015zsa}. In the original construction, there are additional boundary-bulk propagators (corresponding in our case to the boundary operators $\chi^I$), which connect the endpoints of each geodesic to the vertex that emits/absorbs the exchanged field. In figure \ref{fig:quadratic}, such propagators are absent. In fact, these propagators don't affect the mathematical structure of the diagram, because their product $\sim 1/\sqrt{(\ell\cdot x)(\ell'\cdot x)}$ for $x$ on the geodesic, i.e. at $R=0$, is just a constant (c.f. \eqref{eq:R}).

\subsection{Alternative non-traceless gauge for the DV solution}

In \cite{Lysov:2022zlw}, we found expressions for the DV solution in a set of alternative, non-traceless gauges. We will use one of these in section \ref{sec:delta_gauge}. The HS potentials in these new gauges, denoted in \cite{Lysov:2022zlw} as $\Phi^{(s)}$, $\Phi'^{(s)}$ and $\Phi_{\text{symm}}^{(s)}$, read:
\begin{align}
 &\Phi^{(s)} = \frac{(i\sqrt{2})^s}{\pi R\sqrt{-x\cdot x}} \left(\calT^{(s)}(x,t+r,u) + \frac{\theta(s-2)(g_{\mu\nu}u^\mu u^\nu)}{4s(s-1)(x\cdot x)}\,\calT^{(s-2)}(x,t+r,u) \right) \ ; \label{eq:Phi} \\
 &\Phi'^{(s)} = \frac{(i\sqrt{2})^s}{\pi R\sqrt{-x\cdot x}} \left(\calT^{(s)}(x,t-r,u) + \frac{\theta(s-2)(g_{\mu\nu}u^\mu u^\nu)}{4s(s-1)(x\cdot x)}\,\calT^{(s-2)}(x,t-r,u) \right) \ ; \label{eq:Phi'} \\
 &\Phi_{\text{symm}}^{(s)} = \frac{1}{2}\left( \Phi^{(s)} + \Phi'^{(s)} \right) \ , \label{eq:Phi_symm}
\end{align}
featuring the same tensor structure as the Fronsdal curvature \eqref{eq:Fronsdal_phi}. The virtue of the gauges \eqref{eq:Phi}-\eqref{eq:Phi_symm} is their simple behavior when applying the boundary field equation, i.e. the boundary conformal Laplacian, at one or both of $\ell$ and $\ell'$:
\begin{align}
 \Box_\ell\frac{\Phi^{(s)}}{\sqrt{-\ell\cdot\ell'}} &= -\frac{(\ell'\cdot x)^2}{(-\ell\cdot\ell')^{5/2}}\,\calF\phi^{(s)} \ ; \label{eq:Box_Phi} \\
 \Box_{\ell'}\frac{\Phi'^{(s)}}{\sqrt{-\ell\cdot\ell'}} &= -\frac{(\ell\cdot x)^2}{(-\ell\cdot\ell')^{5/2}}\,\calF\phi^{(s)} \ ; \label{eq:Box_Phi'} \\
 \Box_\ell\Box_{\ell'}\frac{\Phi^{(s)}_{\text{symm}}}{\sqrt{-\ell\cdot\ell'}} &= -\frac{(i\sqrt{2})^s(x\cdot x)^2}{(-\ell\cdot\ell')^{5/2}} \left[ Q^{(s)} + \frac{\theta(s-2)(g_{\mu\nu}u^\mu u^\nu)}{4s(s-1)(x\cdot x)}\,Q^{(s-2)} \right] \ . \label{eq:double_Box} 
\end{align}
Here, $\calF\phi^{(s)}$ are the gauge-independent Fronsdal tensors \eqref{eq:Fronsdal_phi}, proportional to the geodesic delta function, while $Q^{(p)}$ is a traceless tensor involving the geodesic delta function and its bulk Laplacian:
\begin{align}
 Q^{(p)}(x,u;\ell,\ell') \equiv \frac{1}{p!}u^{\mu_1}\dots u^{\mu_p}Q_{\mu_1\dots\mu_p} = \calT^{(p)}(x,t,u)\left(\nabla\cdot\nabla + \frac{p(p-1)}{x\cdot x} \right)\delta^3(x;\ell,\ell') \ . \label{eq:Q}
\end{align}
We see that the RHS of \eqref{eq:Box_Phi}-\eqref{eq:double_Box} are all delta-function-like distributions which vanish away from the geodesic $\gamma(\ell,\ell')$. This can be viewed as a bulk version of the free field equation $\Box_\ell\chi^I(\ell) = \Box_{\ell'}\bar\chi_I(\ell') = 0$ on the boundary, which becomes $\Box_\ell\frac{\calO(\ell,\ell')}{\sqrt{-\ell\cdot\ell'}} = \Box_{\ell'}\frac{\calO(\ell,\ell')}{\sqrt{-\ell\cdot\ell'}} = 0$ in terms of bilocals.

\subsection{Sleight-Taronna on-shell cubic vertex} \label{sec:preliminaries:ST}

Let us now review the Sleight-Taronna cubic vertex \cite{Sleight:2016dba} for on-shell HS fields. In general, a cubic vertex is a symmetric scalar function of three HS fields $h^{(s_i)}_i$ ($i=1,2,3$) and their spacetime derivatives. To keep track of which field the derivatives act on, it's convenient to use a ``point-split'' formalism. This means that the three fields are temporarily associated with different spacetime points $x^\mu_i$, which we set equal \emph{after} acting as needed with derivatives $\del^\mu_{x_i}$. Similarly, the vertex's tensor structure can be encoded by using a different polarization vector $u_i^\mu$ to package each field's indices as in \eqref{eq:generating_function}. The vertex will then contain derivatives $\del^\mu_{u_i}$, which ``expose'' the fields' tensor indices before contracting them appropriately into a scalar. Thus, a general cubic vertex is a differential operator $V^{(s_1,s_2,s_3)}(\del_{x_1},\del_{u_1};\del_{x_2},\del_{u_2};\del_{x_2},\del_{u_3})$, which must contain $s_i$ factors of $\del_{u_i}$ for each $i=1,2,3$. Overall, the bulk action from coupling the three HS fields $h^{(s_i)}_i$ via the vertex $V^{(s_1,s_2,s_3)}$ evaluates to:
\begin{align}
 \begin{split}
   S_{s_1,s_2,s_3}[V;h_1,h_2,h_3] = -\int_{EAdS_4}&d^4x\,V^{(s_1,s_2,s_3)}(\del_{x_1},\del_{u_1};\del_{x_2},\del_{u_2};\del_{x_3},\del_{u_3}) \\
     &\times h_1^{(s_1)}(x_1,u_1)h_2^{(s_2)}(x_2,u_2)h_3^{(s_3)}(x_3,u_3)\Big|_{x_1=x_2=x_3=x} \ .
 \end{split} \label{eq:vertex_integral}
\end{align}
The specific on-shell vertex discovered in \cite{Sleight:2016dba} is given by the simple formula:
\begin{align}
 \begin{split}
   &V_{\text{ST}}^{(s_1,s_2,s_3)}(\del_{x_1},\del_{u_1};\del_{x_2},\del_{u_2};\del_{x_3},\del_{u_3}) = \frac{8\!\left(i\sqrt{2}\right)^{s_1+s_2+s_3}}{\Gamma(s_1+s_2+s_3)} \\
   &\quad \times \big[(\del_{u_1}\cdot\del_{x_2})^{s_1}(\del_{u_2}\cdot\del_{x_3})^{s_2}(\del_{u_3}\cdot\del_{x_1})^{s_3} + (\del_{u_1}\cdot\del_{x_3})^{s_1}(\del_{u_2}\cdot\del_{x_1})^{s_2}(\del_{u_3}\cdot\del_{x_2})^{s_3}\big]  \ .
 \end{split} \label{eq:V_ST}
\end{align}
We wrote the vertex \eqref{eq:V_ST} as a sum of two tensor structures, each corresponding to a cyclic ordering of the 3 legs. Taking the average over both orderings makes \eqref{eq:V_ST} completely symmetric under permutations. The 3-point function calculation of \cite{Sleight:2016dba} did not require this averaging, but it will prove important for gauge invariance beyond transverse-traceless gauge. Note that the vertex \eqref{eq:V_ST} doesn't carry on overall factor of $\sim \frac{1}{\sqrt{N}}$, due to our normalization choices \eqref{eq:bilocal}-\eqref{eq:D} for the boundary operators and our decision in \eqref{eq:quadratic_correlator}-\eqref{eq:hbar} to separate a factor of $N$ from the $N$-independent ``classical'' action. The factor of $i^{s_1+s_2+s_3}$ in \eqref{eq:V_ST} does not appear in \cite{Sleight:2016dba}, and is due to the factor of $(-i)^s$ in our definition \eqref{eq:j}-\eqref{eq:D} of the boundary currents. 

The cubic-scalar case $s_1=s_2=s_3=0$ has a well-known singularity: the coupling in \eqref{eq:V_ST} vanishes, but the bulk integral in \eqref{eq:vertex_integral} diverges. Through dimensional regularization, upon inserting the appropriate dimension-dependence in \eqref{eq:V_ST}, one can show that the answer is given by a \emph{boundary} integral:
\begin{align}
 \begin{split}
   S_{0,0,0}[V_{\text{ST}};h_1,h_2,h_3] &= -\lim_{D\rightarrow 4} V_{\text{ST}}^{(0,0,0)}\! \int_{EAdS_D} d^Dx\,h_1^{(0)}(x)h_2^{(0)}(x)h_3^{(0)}(x) \\
     &= -8\int d^3\ell\,h_1^{(0)}(\ell)h_2^{(0)}(\ell)h_3^{(0)}(\ell) \ , \label{eq:V_000}
 \end{split}
\end{align}
where $h^{(0)}(\ell)$ is the analytic continuation of the bulk field $h^{(0)}(x)$ onto the $\bbR^{1,4}$ lightcone. Since $h^{(0)}(x)$ has scaling weight $\Delta=1$, this is the same as evaluating its weight-1 boundary data.

Now, the main result of \cite{Sleight:2016dba} is that the simple vertex formula \eqref{eq:V_ST}, acting on three boundary-bulk propagators $\Pi^{(s_i)}(x,u;\ell_i,\lambda_i)\equiv \Pi_i$, reproduces the CFT correlator of the corresponding boundary HS currents $j^{(s_i)}(\ell_i,\lambda_i)\equiv j^{(s_i)}_i$:
\begin{align}
 -N S_{s_1,s_2,s_3}[V_{\text{ST}};\Pi_1,\Pi_2,\Pi_3] = \left< j^{(s_1)}_1 j^{(s_2)}_2 j^{(s_3)}_3 \right> \ , \label{eq:on_shell_correlator}
\end{align}
where $N$ again plays the role of an inverse Planck constant, as in \eqref{eq:quadratic_correlator}-\eqref{eq:hbar}. Abstractly, eq. \eqref{eq:on_shell_correlator} defines the action of the vertex $V_{\text{ST}}$ on a certain class of field configurations, spanned by the boundary-bulk propagators. This class of field configurations is defined by three constraints:
\begin{itemize}
	\item Source-free, i.e. vanishing Fronsdal curvature.
	\item Transverse-traceless, i.e. vanishing divergence and trace.
	\item Decaying with weight $\Delta = s+1$ as $x$ approaches the boundary, except near the insertion points $\ell_i$.
\end{itemize}

\section{Gauge invariance of Sleight-Taronna vertex for traceless source-free fields} \label{sec:free_invariance}

In this section, we prove that the Sleight-Taronna vertex \eqref{eq:V_ST} is gauge-invariant up to boundary terms, when restricted to \emph{source-free, traceless} fields. This extends the original statement in \cite{Sleight:2016dba}, which was that \emph{each of the two cyclic terms} in \eqref{eq:V_ST} is gauge-invariant when further restricted to \emph{source-free, transverse-traceless} fields. We will use the techniques of \cite{Joung:2011ww} for manipulating a cubic vertex in the radial-reduction formalism (see also \cite{Buchbinder:2006eq}), while adjusting for the fact that our bulk fields have scaling weight $\Delta=s+1$ rather than $\Delta=2-s$. Finally, in section \ref{sec:free_invariance:boundary_terms}, we identify a class of field asymptotics for which the gauge invariance is complete, i.e. the boundary terms in the gauge transformation also vanish.

\subsection{Notations and method}

First, we introduce compact notations for various contracted derivatives (note that the field labels $i,j=1,2,3$ aren't subject to the Einstein summation convention):
\begin{align}
 \Box_i \equiv \del_{x_i}\cdot\del_{x_i} \ ; \quad \calD_i \equiv \del_{u_i}\cdot\del_{x_i} \ ; \quad Y_{ij} \equiv \del_{u_i}\cdot\del_{x_j} \ ; \quad Z_{ij} = \del_{u_i}\cdot\del_{u_j} \ .
\end{align}
With this notation, the Sleight-Taronna vertex \eqref{eq:V_ST} becomes:
\begin{align}
 V_{\text{ST}}^{(s_1,s_2,s_3)} = \frac{8\!\left(i\sqrt{2}\right)^{s_1+s_2+s_3}}{\Gamma(s_1+s_2+s_3)}\big[Y_{12}^{s_1}Y_{23}^{s_2}Y_{31}^{s_3} + Y_{13}^{s_1}Y_{21}^{s_2}Y_{32}^{s_3} \big] \ .
\end{align}
Now, consider a gauge transformation \eqref{eq:gauge_transformation} of e.g. the field $h_3$ (where we suppress the spin superscripts to reduce clutter):
\begin{align}
 \delta h_3(x_3,u_3) = \left( u_3\cdot\del_{x_3} + (2s-1)\frac{u_3\cdot x_3}{x_3\cdot x_3} \right)\Lambda_3(x_3,u_3) \ .
\end{align}
Our statement is that, for source-free traceless fields, the cubic action \eqref{eq:vertex_integral} changes under this transformation by at most boundary terms. To make the calculation tractable, we follow \cite{Joung:2011ww} in writing the bulk integral \eqref{eq:vertex_integral} as a 5d integral over $\bbR^{1,4}$ with a delta function inserted: 
\begin{align}
 \begin{split}
   S_{s_1,s_2,s_3} = 2\int&d^5x\,\delta(x\cdot x+1) \\
     &\times V(\del_{x_1},\del_{u_1};\del_{x_2},\del_{u_2};\del_{x_3},\del_{u_3})h_1(x_1,u_1)h_2(x_2,u_2)h_3(x_3,u_3)\Big|_{x_i=x} \ .
 \end{split} \label{eq:5d_integral}
\end{align}
Thus, the gauge-invariance statement that we wish to prove takes the form:
\begin{align}
   &\int d^5x\,\delta(x\cdot x+1)\left.\big[Y_{12}^{s_1}Y_{23}^{s_2}Y_{31}^{s_3} + Y_{13}^{s_1}Y_{21}^{s_2}Y_{32}^{s_3} \big] \left( u_3\cdot\del_{x_3} + (2s_3-1)\frac{u_3\cdot x_3}{x_3\cdot x_3} \right) h_1 h_2 \Lambda_3 \right|_{x_i=x} \nonumber \\
     &\qquad = \text{boundary terms} \ , \label{eq:gauge_invariance}
\end{align}
where $h_1,h_2$ are subject to the constraints for traceless Fronsdal fields on $EAdS_4$ with vanishing Fronsdal tensor:
\begin{gather}
 (x_i\cdot\del_{u_i})h_i = (\del_{u_i}\cdot\del_{u_i})h_i = 0 \ ; \quad (x_i\cdot\del_{x_i})h_i = -(s+1)h_i \ ; \label{eq:h_i_properties_1} \\
 (\del_{x_i}\cdot\del_{x_i})h_i = \left(u_i\cdot\del_{x_i} + (2s-1)\frac{u_i\cdot x_i}{x_i\cdot x_i} \right)\calD_i h_i \ , \label{eq:h_i_properties_2}
\end{gather}
and $\Lambda_3$ is subject to the constraints for a traceless, divergence-free gauge parameter:
\begin{align}
 (x_3\cdot\del_{u_3})\Lambda_3 = (\del_{u_3}\cdot\del_{u_3})\Lambda_3 = 0 \ ; \quad (x_3\cdot\del_{x_3})\Lambda_3 = -s\Lambda_3 \ ; \quad \calD_3\Lambda_3 = 0 \ . \label{eq:Lambda_3_properties}
\end{align}
Our method of proof will be to manipulate the differential operator inserted between $\delta(x\cdot x+1)$ and $h_1 h_2 \Lambda_3$ in \eqref{eq:gauge_invariance}. We will use the ``weak equality'' sign ``$\approx$'' to denote that two operators are equal when sandwiched between $\delta(x\cdot x+1)$ and $h_1 h_2 \Lambda_3$ and integrated as in \eqref{eq:gauge_invariance}, up to boundary terms. The main strategy is to commute various factors within the operator to the left or to the right, where they can vanish or simplify. When on the right, we can use the fields' properties \eqref{eq:h_i_properties_1}-\eqref{eq:Lambda_3_properties} as:
\begin{gather}
 (\dots)(x_i\cdot\del_{u_i}) \approx (\dots)(\del_{u_i}\cdot\del_{u_i}) \approx 0 \ ; \label{eq:tangential_traceless_right} \\
 (\dots)(x_i\cdot\del_{x_i}) \approx -(s_i+1)(\dots) \quad [\text{for }i=1,2] \ ; \quad (\dots)(x_3\cdot\del_{x_3}) \approx -s_3(\dots) \ ; \label{eq:scaling_right} \\
 (\dots)(\del_{x_i}\cdot\del_{x_i}) = (\dots)\left(u_i\cdot\del_{x_i} + (2s-1)\frac{u_i\cdot x_i}{x_i\cdot x_i} \right)\calD_i \quad [\text{for }i=1,2] \ ; \label{eq:Fronsdal_right} \\
 (\dots)\calD_i^2 = 0 \quad [\text{for }i=1,2] \ ; \quad (\dots)\calD_3 = 0 \ , \label{eq:div_right}
\end{gather}
where the $\calD_i^2$ identity comes from the Fronsdal tensor's trace \eqref{eq:trace_Fronsdal}. 

When on the left, we can use the coincidence relation $x_i^\mu = x^\mu$ and the $EAdS_4$ condition $x\cdot x = -1$:
\begin{align}
 x_i\cdot(\dots) \approx x\cdot(\dots) \ ; \quad (x\cdot x)(\dots) \approx -(\dots) \ .
\end{align}
Also, a factor of $u_i^\mu$ on the left always vanishes, because it implies that there are more $\del_{u_i}$ derivatives than factors of $u_i$ to its right:
\begin{align}
 u_i\cdot(\dots) = 0 \ .
\end{align}
Finally, a total derivative $\del^\mu_x = \del^\mu_{x_1}+\del^\mu_{x_2}+\del^\mu_{x_3}$ on the left can be integrated by parts, as:
\begin{align}
 \del_x\cdot(\dots) \approx -(3 + x\cdot\del_x)\,x\cdot(\dots) \ . \label{eq:integration_by_parts}
\end{align}
This arises from acting with $\del_x$ on the delta function $\delta(x\cdot x+1)$ that always implicitly stands to the left of our operator. In more detail, for any vector $f^\mu$, we have:
\begin{align}
 \int d^5x\,\delta(x\cdot x+1)(\del_x\cdot f) = -2\int d^5x\,\delta'(x\cdot x+1)(x\cdot f) + \text{boundary terms} \ . \label{eq:delta_by_parts}
\end{align}
Denoting $\rho\equiv\sqrt{-x\cdot x}$, the radial part of the integral \eqref{eq:delta_by_parts} can now be written as:
\begin{align}
 2\int \rho^4 d\rho\,\delta'(\rho^2-1)(x\cdot f) = -\int d\rho\,\delta(\rho^2-1)\frac{d}{d\rho}\left(\rho^3(x\cdot f)\right) \ .
\end{align}
Identifying $\frac{d}{d\rho}$ with $x\cdot\del_x$, this yields the desired prescription \eqref{eq:integration_by_parts}.

\subsection{Two Lemmas} \label{sec:free_invariance:lemmas}

Before proving \eqref{eq:gauge_invariance}, let us establish two useful identities, or Lemmas. The first one concerns the commutation of a factor of $x_i\cdot x_i + 1$ from the right of a differential operator to the left (where it becomes simply zero).
\begin{lemma}
  Assuming only the tangential and traceless properties \eqref{eq:tangential_traceless_right}, the following identity holds:
  \begin{align}
   Y_{12}^{p_1}Y_{23}^{p_2}Y_{31}^{p_3}\calD_1^{n_1}\calD_2^{n_2}\calD_3^{n_3}(x_3\cdot x_3 + 1) \approx -2p_1 p_2 Z_{12}Y_{12}^{p_1-1}Y_{23}^{p_2-1}Y_{31}^{p_3}\calD_1^{n_1}\calD_2^{n_2}\calD_3^{n_3} \ . \label{eq:Lemma_1}
  \end{align}
\end{lemma}
To prove this, let us start from the LHS of \eqref{eq:Lemma_1}, and commute one of the $x_3^\mu$ factors to the left:
\begin{align}
 \begin{split}
   &Y_{12}^{p_1}Y_{23}^{p_2}Y_{31}^{p_3}\calD_1^{n_1}\calD_2^{n_2}\calD_3^{n_3}(x_3\cdot x_3) = x_{3\mu}Y_{12}^{p_1}Y_{23}^{p_2}Y_{31}^{p_3}\calD_1^{n_1}\calD_2^{n_2}\calD_3^{n_3}x_3^\mu \\
   &\quad + n_3 Y_{12}^{p_1}Y_{23}^{p_2}Y_{31}^{p_3}\calD_1^{n_1}\calD_2^{n_2}\calD_3^{n_3-1}(x_3\cdot\del_{u_3}) + p_2 Y_{12}^{p_1}Y_{23}^{p_2-1}Y_{31}^{p_3}\calD_1^{n_1}\calD_2^{n_2}\calD_3^{n_3}(x_3\cdot\del_{u_2}) \ .
 \end{split}
\end{align}
The second term vanishes due to \eqref{eq:tangential_traceless_right}. In the first and third terms, we commute $x_3^\mu$ to the left again (omitting a vanishing term $\sim \del_{u_2}\cdot\del_{u_2}$):
\begin{align}
 \begin{split}
   &Y_{12}^{p_1}Y_{23}^{p_2}Y_{31}^{p_3}\calD_1^{n_1}\calD_2^{n_2}\calD_3^{n_3}(x_3\cdot x_3 + 1) = n_3(x_3\cdot\del_{u_3})Y_{12}^{p_1}Y_{23}^{p_2}Y_{31}^{p_3}\calD_1^{n_1}\calD_2^{n_2}\calD_3^{n_3-1} \\
   &\quad + 2p_2(x_3\cdot\del_{u_2})Y_{12}^{p_1}Y_{23}^{p_2-1}Y_{31}^{p_3}\calD_1^{n_1}\calD_2^{n_2}\calD_3^{n_3} + p_2 n_3 Z_{23}Y_{12}^{p_1}Y_{23}^{p_2-1}Y_{31}^{p_3}\calD_1^{n_1}\calD_2^{n_2}\calD_3^{n_3-1} \ .
 \end{split} \label{eq:both_x_commuted}
\end{align}
Now, in the first term of \eqref{eq:both_x_commuted}, we commute $x_3\cdot\del_{u_3}$ to the right, where it vanishes. The commutator with $\calD_3^{n_3-1}$ gives $\del_{u_3}\cdot\del_{u_3}$ which vanishes, while the commutator with $Y_{23}^{p_2}$ cancels the third term in \eqref{eq:both_x_commuted}. We are thus left with only the second term, in which we can trade the $x_3$ on the left for $x_2$:
\begin{align}
 Y_{12}^{p_1}Y_{23}^{p_2}Y_{31}^{p_3}\calD_1^{n_1}\calD_2^{n_2}\calD_3^{n_3}(x_3\cdot x_3 + 1) = 2p_2(x_2\cdot\del_{u_2})Y_{12}^{p_1}Y_{23}^{p_2}Y_{31}^{p_3-1}\calD_1^{n_1}\calD_2^{n_2}\calD_3^{n_3} \ .
\end{align}
We now commute $x_2\cdot\del_{u_2}$ to the right, where it vanishes. The only non-vanishing contribution comes from commuting with $Y_{12}^{p_1}$, which yields the desired result \eqref{eq:Lemma_1}.

Our second Lemma presents a particular situation in which integration by parts works just like in flat spacetime, where total-derivative terms of the form $\del_x\cdot f$ can be simply discarded.
\begin{lemma}
  Assuming only the tangential and traceless properties \eqref{eq:tangential_traceless_right}, a scaling property of the form \eqref{eq:scaling_right} with arbitrary scaling weights $(\dots)(x_i\cdot\del_{x_i}) = -\Delta_i(\dots)$, and the integration-by-parts property \eqref{eq:integration_by_parts}, the following identity holds:
  \begin{align}
   (Y_{12} - \del_{u_1}\cdot\del_x)^{p_1}(Y_{23} - \del_{u_2}\cdot\del_x)^{p_2}(Y_{31} - \del_{u_3}\cdot\del_x)^{p_3} \approx Y_{12}^{p_1}Y_{23}^{p_2}Y_{31}^{p_3} \ . \label{eq:Lemma_2}
  \end{align}
  Equivalently (expanding $\del^\mu_x = \del^\mu_{x_1}+\del^\mu_{x_2}+\del^\mu_{x_3}$ in the parentheses and reshuffling the field labels):
  \begin{align}
   Y_{13}^{p_1}Y_{21}^{p_2}Y_{32}^{p_3} \approx (-1)^{p_1+p_2+p_3}(Y_{12} + \calD_1)^{p_1}(Y_{23} + \calD_2)^{p_2}(Y_{31} + \calD_3)^{p_3} \ . \label{eq:Lemma_2'}
  \end{align}
\end{lemma}
As an aside, eq. \eqref{eq:Lemma_2'} is closely related to the fact that for boundary-bulk propagators in transverse-traceless gauge, the two terms in the vertex \eqref{eq:V_ST} yield the same result (i.e. that in this gauge, there's no need to write both terms).

Let us now prove the Lemma's statement, in the form \eqref{eq:Lemma_2}. First, we apply the integration-by-parts prescription \eqref{eq:integration_by_parts} to all the factors of $\del_{u_1}\cdot\del_x$. This yields factors of $x\cdot\del_x$ and $x\cdot\del_{u_1}$. The former simply yield some multiplicative constants due to the scaling weights; the latter can be written as $x_1\cdot\del_{u_1}$, and then commuted from the left to the right, where it vanishes. The commutation yields:
\begin{itemize} 
	\item Zero from commuting with $Y_{12}$, $Y_{23}$ or $Y_{31} - \del_{u_3}\cdot\del_x$.
	\item $\del_{u_1}\cdot\del_{u_1}\approx 0$ from commuting with $\del_{u_1}\cdot\del_x$.
	\item $Z_{12}$ from commuting with $\del_{u_2}\cdot\del_x$.
\end{itemize}
After these manipulations, we are left with a polynomial in $Y_{12}$, $Y_{23}$, $\del_{u_2}\cdot\del_x$, $Y_{31} - \del_{u_3}\cdot\del_x$ and $Z_{12}$. The next step is then to integrate by parts all the factors of $\del_{u_2}\cdot\del_x$. Analogously to the previous step, this yields factors of $x_2\cdot\del_{u_2}$, which we proceed to commute from the left to the right. The commutation yields:
\begin{itemize} 
	\item Zero from commuting with $Y_{23}$, $Y_{31}$ or $Z_{12}$.
	\item $\del_{u_2}\cdot\del_{u_2}\approx 0$ from commuting with $\del_{u_2}\cdot\del_x$.
	\item $Z_{12}$ from commuting with $Y_{12}$.
	\item $Z_{23}$ from commuting with $\del_{u_3}\cdot\del_x$.
\end{itemize}
We are now left with a polynomial in $Y_{12}$, $Y_{23}$, $Y_{31}$, $\del_{u_3}\cdot\del_x$, $Z_{12}$ and $Z_{23}$. Finally, we integrate by parts the factors of $\del_{u_3}\cdot\del_x$. Commuting the resulting factors of $x_3\cdot\del_{u_3}$ from left to right, we get:
\begin{itemize}
   \item Zero from commuting with $Y_{12}$, $Y_{31}$, $Z_{12}$ or $Z_{23}$.
   \item $\del_{u_3}\cdot\del_{u_3}\approx 0$ from commuting with $\del_{u_3}\cdot\del_x$.
   \item $Z_{23}$ from commuting with $Y_{23}$.
\end{itemize}
We finally end up with a polynomial in $Y_{12},Y_{23},Y_{31},Z_{12},Z_{23}$. But this is an artifact of the particular order $1\rightarrow2\rightarrow3$ in which we chose to integrate by parts the factors of $\del_{u_i}\cdot\del_x$. By choosing $2\rightarrow3\rightarrow1$ or $3\rightarrow1\rightarrow2$ instead, we'd end up with polynomials in $Y_{12},Y_{23},Y_{31},Z_{23},Z_{31}$ or $Y_{12},Y_{23},Y_{31},Z_{31},Z_{12}$, respectively. This is consistent only if the answer doesn't depend on the $Z_{ij}$'s at all, i.e. if the nonzero $\sim Z_{ij}$ commutators in our manipulations above all cancel. Therefore, the answer simply consists of the original factors of $Y_{12},Y_{23},Y_{31}$, as claimed in \eqref{eq:Lemma_2}.

\subsection{Proof of gauge invariance up to boundary terms} \label{sec:free_invariance:proof}

We are now ready to prove eq. \eqref{eq:gauge_invariance}, i.e.:
\begin{align}
 \big[Y_{12}^{s_1}Y_{23}^{s_2}Y_{31}^{s_3} + Y_{13}^{s_1}Y_{21}^{s_2}Y_{32}^{s_3} \big] \left( u_3\cdot\del_{x_3} + (2s_3-1)\frac{u_3\cdot x_3}{x_3\cdot x_3} \right) \approx 0 \ . \label{eq:gauge_invariance_again}
\end{align} 
We begin by manipulating the first term in \eqref{eq:gauge_invariance_again}, namely $Y_{12}^{s_1}Y_{23}^{s_2}Y_{31}^{s_3} \left( u_3\cdot\del_{x_3} + (2s_3-1)\frac{u_3\cdot x_3}{x_3\cdot x_3} \right)$. The calculation is lengthy, and consists of iterating the following steps:
\begin{itemize}
	\item Commute any factors of $u_i^\mu$ to the left, where they vanish.
	\item Rewrite any factor of $\del_{x_i}\cdot\del_{x_j}$ with $i\neq j$ as e.g. $\del_{x_1}\cdot\del_{x_2} = \frac{1}{2}\big(\del_x\cdot(\del_{x_1} + \del_{x_2} - \del_{x_3}) - \Box_1 - \Box_2 + \Box_3\big)$, and integrate the first term by parts.
	\item Evaluate any factor of $x\cdot\del_x$ or $x_i\cdot\del_{x_i}$ according to the scaling weight of the expression to its right.
	\item Rewrite any factor of $x\cdot\del_{x_i}$ on the left as $x_i\cdot\del_{x_i}$, so it can be evaluated as above.
	\item Commute any factor of $x_i\cdot\del_{x_j}$ with $i\neq j$ to the left, where it can become $x_j\cdot\del_{x_j}$ and be evaluated as above.
	\item Rewrite any factor of $x\cdot\del_{u_i}$ on the left as $x_i\cdot\del_{u_i}$, and commute it to the right, where it vanishes.
	\item Convert any factor of $Y_{13},Y_{21},Y_{32}$ back into factors of $Y_{12},Y_{23},Y_{31}$ by writing e.g. $Y_{13} = \del_{u_1}\cdot\del_x - \calD_1 - Y_{12}$, and integrate the first term by parts.
	\item Use eq. \eqref{eq:Lemma_1} (Lemma 1) to convert any term with a factor of $Z_{ij}$ into terms without it.
	\item Rewrite any factor of $\Box_1$ or $\Box_2$ on the right using the source-free condition \eqref{eq:Fronsdal_right}, \emph{unless} it occurs in the combination $\calD_1\Box_1$ or $\calD_2\Box_2$, in which case the rewriting results in a closed loop.
	\item Use eq. \eqref{eq:div_right} to discard any terms with $\calD_1^2$ or $\calD_3$ on the right.
\end{itemize}
The result of this procedure reads:
\begin{align}
 &Y_{12}^{s_1}Y_{23}^{s_2}Y_{31}^{s_3} \left( u_3\cdot\del_{x_3} + (2s_3-1)\frac{u_3\cdot x_3}{x_3\cdot x_3} \right) \approx -s_3 Y_{12}^{s_1}Y_{23}^{s_2}Y_{31}^{s_3-1} \left(\frac{\Box_3}{2} + \frac{2s_3-1}{x_3\cdot x_3}\right) \nonumber \\
 &\quad + s_1s_3 Y_{12}^{s_1-1}Y_{23}^{s_2}Y_{31}^{s_3-1}\calD_1 \left(s_1+s_2+s_3-1 + \frac{\Box_1-\Box_3}{4}\right) \label{eq:first_cyclic} \\
 &\quad - s_2s_3 Y_{12}^{s_1}Y_{23}^{s_2-1}Y_{31}^{s_3-1}\calD_2 \left(s_1+s_2+s_3-1 + \frac{\Box_2+\Box_3}{4} + \frac{2s_3-1}{x_3\cdot x_3}\right) \nonumber \\
 &\quad - s_1s_2s_3 Y_{12}^{s_1-1}Y_{23}^{s_2-1}Y_{31}^{s_3-1}\calD_1\calD_2 \left(s_1+s_2+s_3-1 + \frac{1}{4}\left(\Box_2 + \frac{2s_1-1}{x_1\cdot x_1} + \frac{2s_2-1}{x_2\cdot x_2} \right) \right) \ . \nonumber
\end{align}
In transverse-traceless gauge, the fields $h_1,h_2$ and the gauge parameter $\Lambda_3$ would satisfy $\calD_1\approx\calD_2\approx\Box_3 + \frac{2(2s_3-1)}{x_3\cdot x_3}\approx 0$ (c.f. \eqref{eq:Box_Lambda}), making the variation \eqref{eq:first_cyclic} simply vanish. In general traceless gauge, we must work a bit harder. To proceed, let us apply analogous manipulations to the second term in \eqref{eq:gauge_invariance_again}, namely to $Y_{13}^{s_1}Y_{21}^{s_2}Y_{32}^{s_3} \left( u_3\cdot\del_{x_3} + (2s_3-1)\frac{u_3\cdot x_3}{x_3\cdot x_3} \right)$. The result can be directly read off from \eqref{eq:first_cyclic}, by interchanging the field labels $1\leftrightarrow 2$:
\begin{align}
 &Y_{13}^{s_1}Y_{21}^{s_2}Y_{32}^{s_3} \left( u_3\cdot\del_{x_3} + (2s_3-1)\frac{u_3\cdot x_3}{x_3\cdot x_3} \right) \approx -s_3 Y_{13}^{s_1}Y_{21}^{s_2}Y_{32}^{s_3-1} \left(\frac{\Box_3}{2} + \frac{2s_3-1}{x_3\cdot x_3}\right) \nonumber \\
 &\quad - s_1s_3 Y_{13}^{s_1-1}Y_{21}^{s_2}Y_{32}^{s_3-1}\calD_1 \left(s_1+s_2+s_3-1 + \frac{\Box_1+\Box_3}{4} + \frac{2s_3-1}{x_3\cdot x_3}\right) \label{eq:second_cyclic} \\
 &\quad + s_2s_3 Y_{13}^{s_1}Y_{21}^{s_2-1}Y_{32}^{s_3-1}\calD_2 \left(s_1+s_2+s_3-1 + \frac{\Box_2-\Box_3}{4}\right) \nonumber \\
 &\quad - s_1s_2s_3 Y_{13}^{s_1-1}Y_{21}^{s_2-1}Y_{32}^{s_3-1}\calD_1\calD_2 \left(s_1+s_2+s_3-1 + \frac{1}{4}\left(\Box_1 + \frac{2s_1-1}{x_1\cdot x_1} + \frac{2s_2-1}{x_2\cdot x_2} \right) \right) \ . \nonumber
\end{align}
Now, let us apply eq. \eqref{eq:Lemma_2'} (Lemma 2) to each term on the RHS of \eqref{eq:second_cyclic}. We get:
\begin{align}
 &Y_{13}^{s_1}Y_{21}^{s_2}Y_{32}^{s_3} \left( u_3\cdot\del_{x_3} + (2s_3-1)\frac{u_3\cdot x_3}{x_3\cdot x_3} \right) \nonumber \\
 &\approx s_3 (Y_{12}+\calD_1)^{s_1}(Y_{23}+\calD_2)^{s_2}Y_{31}^{s_3-1} \left(\frac{\Box_3}{2} + \frac{2s_3-1}{x_3\cdot x_3}\right) \label{eq:second_cyclic_rearranged} \\
 &\quad - s_1s_3 Y_{12}^{s_1-1}(Y_{23}+\calD_2)^{s_2}Y_{31}^{s_3-1}\calD_1 \left(s_1+s_2+s_3-1 + \frac{\Box_1+\Box_3}{4} + \frac{2s_3-1}{x_3\cdot x_3}\right) \nonumber \\
 &\quad + s_2s_3 (Y_{12}+\calD_1)^{s_1}Y_{23}^{s_2-1}Y_{31}^{s_3-1}\calD_2 \left(s_1+s_2+s_3-1 + \frac{\Box_2-\Box_3}{4}\right) \nonumber \\
 &\quad + s_1s_2s_3 Y_{12}^{s_1-1}Y_{23}^{s_2-1}Y_{31}^{s_3-1}\calD_1\calD_2 \left(s_1+s_2+s_3-1 + \frac{1}{4}\left(\Box_1 + \frac{2s_1-1}{x_1\cdot x_1} + \frac{2s_2-1}{x_2\cdot x_2} \right) \right) \ , \nonumber
\end{align}
where we fixed the sign factors in \eqref{eq:Lemma_2'} using the fact that $s_1+s_2+s_3$ is even, and used \eqref{eq:div_right} to discard any terms proportional to $\calD_1^2$, $\calD_2^2$ or $\calD_3$. The last step is to expand the RHS of \eqref{eq:second_cyclic_rearranged} in powers of $\calD_1,\calD_2$, again discarding terms proportional to $\calD_1^2$ or $\calD_2^2$. The result is precisely minus the RHS of \eqref{eq:first_cyclic}, thus proving the desired relation \eqref{eq:gauge_invariance_again}.

\subsection{Constraining the boundary contribution} \label{sec:free_invariance:boundary_terms}

So far in this section, we've been evaluating gauge variations \emph{up to boundary terms}. Let us now tackle the question of boundary terms, under a certain assumption on the fields' asymptotics. Specifically, consider a traceless (not necessarily transverse) spin-$s$ pure-gauge field, whose components in an orthonormal Poincare basis (see section \ref{sec:preliminaries:asymptotics}) decay towards the boundary as $z^{s+1}$ or faster:
\begin{gather}
 \tilde h^{(s)}(x,u) = (u\cdot\nabla)\Lambda^{(s)}(x,u) \ ; \quad (\del_u\cdot\del_u)\tilde h^{(s)} = 0 \ ; \label{eq:tilde_h} \\ 
 [\tilde h^{(s)}]_{q,s-q} = O(z^{s+1}) \ . \label{eq:tilde_h_asymptotics}
\end{gather}
Our claim is that the on-shell cubic correlator formula \eqref{eq:on_shell_correlator} continues to hold when the boundary-bulk propagators $\Pi^{(s)}$ are shifted by such pure-gauge fields:
\begin{align}
 -NS_{s_1,s_2,s_3}[V_{\text{ST}};\Pi_1 + \tilde h_1,\Pi_2 + \tilde h_2,\Pi_3 + \tilde h_3] = \left< j^{(s_1)}_1 j^{(s_2)}_2 j^{(s_3)}_3 \right> \ . \label{eq:shifted_correlator}
\end{align}
This is equivalent to saying that a gauge transformation of the form \eqref{eq:tilde_h}-\eqref{eq:tilde_h_asymptotics} has no effect on correlators of the form \eqref{eq:shifted_correlator}:
\begin{align}
 S_{s_1,s_2,s_3}[V_{\text{ST}};\Pi_1 + \tilde h_1,\Pi_2 + \tilde h_2,\tilde h_3] = 0 \ . \label{eq:vanishing_shift}
\end{align}
From our previous result \eqref{eq:gauge_invariance}, we already know that \eqref{eq:vanishing_shift} is true \emph{up to boundary terms}. Our goal now is to show that the boundary terms also vanish. Unfortunately, it's difficult to track all the specific boundary terms that arise from the various integrations by parts in sections \ref{sec:free_invariance:lemmas}-\ref{sec:free_invariance:proof}, especially the ones that occur in the proof of Lemma 2. Instead, we will simply consider \emph{all possible} boundary terms, and show that they all vanish by power counting.

To perform this asymptotic power counting, we invoke the formalism of Poincare coordinates with a normalized basis from section \ref{sec:preliminaries:asymptotics}. Near the boundary $z\rightarrow0$, derivatives with respect to the ``radial'' coordinate $z$ and the ``tangential'' coordinates $y^a$ scale as:
\begin{align}
  \frac{\del}{\del z} = O(z^{-1}) \ ; \quad \frac{\del}{\del y^a} = O(1) \ . \label{eq:derivs_scaling}
\end{align}
Switching to \emph{normalized} derivatives, i.e. derivatives along unit vectors, this becomes:
\begin{align}
 e_0\cdot\nabla = O(1) \ ; \quad e_a\cdot\nabla = O(z) \ . \label{eq:normalized_derivs_scaling}
\end{align}
Now, a key difficulty in our analysis is that the boundary terms in the gauge transformation \eqref{eq:vanishing_shift} involve not the pure-gauge field $\tilde h_3$ itself, but rather its gauge parameter $\Lambda_3$. We therefore need to understand how the condition \eqref{eq:tilde_h_asymptotics} on $\tilde h^{(s)}$ constrains the asymptotics of $\Lambda^{(s)}$. To do this, we note that $\Lambda^{(s)}$ satisfies (c.f. \eqref{eq:4d_Box_Lambda}):
\begin{gather}
 (\del_u\cdot\del_u)\Lambda^{(s)} = (\del_u\cdot\nabla)\Lambda^{(s)} = 0 \ ; \label{eq:Lambda_field_equation_1} \\ 
 \left(\nabla\cdot\nabla + \frac{s^2-1}{x\cdot x} \right)\Lambda^{(s)} = (\del_u\cdot\nabla)\tilde h^{(s)} \ . \label{eq:Lambda_field_equation_2}
\end{gather}
This is nothing but an inhomogeneous version of the transverse-traceless field equations \eqref{eq:field_eq_w} for the rank-$(s-1)$ ``field'' $\Lambda^{(s)}$, with weight $w=s+2$ (or, equivalently, $w=1-s$), and with the divergence $(\del_u\cdot\nabla)\tilde h^{(s)}$ in the role of a source term. We quickly see from \eqref{eq:normalized_derivs_scaling} that the $z$ scaling of this source term is the same as that of $\tilde h^{(s)}$ itself, namely:
\begin{align}
 \big[(\del_u\cdot\nabla)\tilde h^{(s)}\big]_{q,s-1-q} = O(z^{s+1}) \ . \label{eq:div_h_scaling}
\end{align}
Note that $(\del_u\cdot\nabla)\tilde h^{(s)}$ is a divergence-free symmetric rank-$(s-1)$ tensor (the second divergence of $\tilde h^{(s)}$ vanishes due to \eqref{eq:trace_Fronsdal}), and that \eqref{eq:div_h_scaling} is the natural scaling for such divergence-free (i.e. conserved) quantities.

Now, eqs. \eqref{eq:Lambda_field_equation_1}-\eqref{eq:Lambda_field_equation_2} determine the gauge parameter $\Lambda^{(s)}$ up to boundary conditions, which are governed in turn by the source-free version of \eqref{eq:Lambda_field_equation_1}-\eqref{eq:Lambda_field_equation_2}. As we saw in section \ref{sec:preliminaries:Pi}, these boundary conditions are associated with two possible $z$ scalings for the normalized Poincare components $[\Lambda^{(s)}]_{q,s-1-q}$, namely $\sim z^{s+2+q}$ and $\sim z^{1-s+q}$. Our claim is then that \emph{the correct solution of eqs. \eqref{eq:Lambda_field_equation_1}-\eqref{eq:Lambda_field_equation_2} is the one with the $\sim z^{1-s+q}$ boundary data vanishing}. To see that this is the case, note that the dominant $z$ scaling of this solution is:
\begin{align}
 [\Lambda^{(s)}]_{q,s-1-q} = O(z^{s+1}) \ , \label{eq:Lambda_scaling}
\end{align}
since the remaining $\sim z^{s+2+q}$ boundary data is dominated by the $O(z^{s+1})$ source term. This then implies the desired scaling \eqref{eq:tilde_h_asymptotics} for the pure-gauge field $\tilde h^{(s)} = (u\cdot\nabla)\Lambda^{(s)}$ itself. Any other solution of \eqref{eq:Lambda_field_equation_1}-\eqref{eq:Lambda_field_equation_2} will differ from this one by a solution $\Lambda'^{(s)}$ to the homogeneous equations, corresponding to a transverse-traceless pure-gauge (and thus source-free) field $\tilde h'^{(s)}$. But, by the analysis of section \ref{sec:preliminaries:Pi}, any such nonzero field would contain $[\tilde h'^{(s)}]_{q,s-q}\sim z^{2-s+q}$ boundary data, in contradiction with our assumption \eqref{eq:tilde_h_asymptotics}. And if $\tilde h'^{(s)}$ \emph{is} zero, then we can simply throw away the contribution $\Lambda'^{(s)}$ to the gauge parameter, and return to the original solution $\Lambda^{(s)}$ with vanishing $\sim z^{1-s+q}$ boundary data. The upshot of this analysis is that our pure-gauge field $\tilde h^{(s)}$ can be described by a gauge parameter $\Lambda^{(s)}$ that scales near the boundary as \eqref{eq:Lambda_scaling}.

We are now ready to assemble the subsection's main claim \eqref{eq:vanishing_shift}. The most general boundary contribution from turning on the pure-gauge field $\tilde h_3$ is a boundary integral over some function of the fields $\Pi_1 + \tilde h_1$ and $\Pi_2 + \tilde h_2$, the gauge parameter $\Lambda_3$, and their $EAdS_4$ derivatives. Since volume measure scales as $\sim z^{-3}$, the integral will vanish if the integrand vanishes faster than $z^3$. Let us now show that this is the case. Away from the source points $\ell_1$ and $\ell_2$, we see from \eqref{eq:Pi_away_from_delta},\eqref{eq:normalized_derivs_scaling},\eqref{eq:Lambda_scaling} that the fields and the gauge parameter scale as $O(z^{s_1+1})$, $O(z^{s_2+1})$ and $O(z^{s_3+1})$ respectively, while the $EAdS_4$ derivatives scale as $O(1)$. Since at least $s_3$ is greater than zero (otherwise, there's no gauge transformation to speak of), we conclude that the overall power of $z$ is greater than 3, as required. 

It remains to consider the contributions from the source points $\ell_1$ and $\ell_2$, where $\Pi_1$ and $\Pi_2$ have the delta-function-like contributions \eqref{eq:Pi_delta}. Let us focus e.g. on the contribution from $\ell_1$. We can integrate by parts to remove any boundary derivatives $e_a\cdot\nabla$ from $\Pi_1$, moving them onto $\Pi_2+\tilde h_2$ and $\Lambda_3$. Now, consider separately the different components $[\Pi_1]_{q_1,s_1-q_1}$ of $\Pi_1$. These scale as $\sim z^{2-s_1+q_1}$, while $\Pi_2+\tilde h_2$ and $\Lambda_3$ still scale as $O(z^{s_2+1})$ and $O(z^{s_3+1})$ respectively. The overall power of $z$ thus appears to be $4-s_1+q_1+s_2+s_3$, which is a problem if $s_1-q_1 > s_2+s_3$. However, in that case, a new consideration comes into play. Recall that $s_1-q_1$ is the number of indices on $\Pi_1$ that are tangential to the boundary. By rotational invariance, these must be contracted with indices on $\Pi_2+\tilde h_2$, $\Lambda_3$, or derivatives. But $\Pi_2+\tilde h_2$ and $\Lambda_3$ have only $s_2$ and $s_3-1$ indices respectively, which implies that at least $s_1-q_1-s_2-s_3+1$ indices must be contracted with tangential derivatives $e_a\cdot\nabla$, each of which contributes an extra power of $z$, according to \eqref{eq:normalized_derivs_scaling}. Overall, we conclude that the delta-function-like contributions to the boundary integrand scale as $O(z^5)$, and thus their integral also vanishes.

This concludes our derivation of the invariance relation \eqref{eq:vanishing_shift}. We've thus shown that the Sleight-Taronna vertex correctly computes the cubic correlator \eqref{eq:shifted_correlator} in a general traceless gauge with the asymptotic behavior \eqref{eq:tilde_h_asymptotics}.

\section{Bulk locality structure of general cubic correlator} \label{sec:locality}

In this section, we state and argue our main claims vis. the bulk locality structure of the cubic correlator $\langle\calO^+(\ell_1,\ell'_1)\calO^+(\ell_2,\ell'_2)\calO^+(\ell_3,\ell'_3)\rangle$ of boundary bilocals. We begin in section \ref{sec:locality:jjO} by laying out the structure of the bilocal-local-local correlator $\langle j^{(s_1)}(\ell_1,\lambda_1)j^{(s_2)}(\ell_2,\lambda_2)\calO^+(\ell_3,\ell'_3)\rangle$, which involves a new interaction vertex between the DV geodesic ``worldline'' $\gamma_3$ and the fields $h_1,h_2$. In section \ref{sec:locality:new_vertex}, we describe a general ansatz for this new vertex. In sections \ref{sec:locality:radial} and \ref{sec:locality:time}, we state and verify locality criteria for the new vertex, in the directions perpendicular and parallel to $\gamma_3$, respectively. In section \ref{sec:locality:gauge}, we extend the new vertex beyond transverse-traceless gauge. Finally, in section \ref{sec:locality:general}, we show how the bulk diagrams for the general bilocal\textsuperscript{3} correlator can be ``stitched together'' from bilocal-local-local ones.

\subsection{Bulk structure of (local,local,bilocal) correlator} \label{sec:locality:jjO}

Consider the cubic correlator between two local currents $j^{(s_1)}(\ell_1,\lambda_1)\equiv j^{(s_1)}_1$ and $j^{(s_2)}(\ell_2,\lambda_2)\equiv j^{(s_2)}_2$, and one bilocal operator $\calO^+(\ell_3,\ell'_3)$. For even $s_1$ and $s_2$, the CFT correlator is automatically symmetric under $\ell_3\leftrightarrow\ell'_3$. This allows us to replace the symmetrized bilocal $\calO^+(\ell_3,\ell'_3)$ by the unsymmetrized one $\calO(\ell_3,\ell'_3)\equiv\calO_3$, which will slightly simplify the analysis. 

At the linearized level, the operators $j^{(s_1)}_1,j^{(s_2)}_2,\calO_3$ are dual in the bulk to a pair of boundary-bulk propagators $\Pi^{(s_1)}(x,u;\ell_1,\lambda_1)\equiv\Pi_1$ and $\Pi^{(s_2)}(x,u;\ell_2,\lambda_2)\equiv\Pi_2$, and a DV solution $\phi^{(s)}(x,u;\ell_3,\ell'_3)\equiv \phi_3$ associated with a worldline geodesic $\gamma(\ell_3,\ell'_3)\equiv \gamma_3$. Our statement is that the cubic correlator can be constructed from these bulk objects as:
\begin{align}
 \begin{split}
   \left< j^{(s_1)}_1 j^{(s_2)}_2 \calO_3 \right> = -N\bigg(&\sum_{s_3}S_{s_1,s_2,s_3}[V_{\text{ST}};\Pi_1,\Pi_2,\phi_3] \\
     &\quad - S_{s_1}[\Pi_1,\gamma_3]\,S_{s_2}[\Pi_2,\gamma_3] + S_{s_1,s_2}[V_{\text{new,TT}};\Pi_1,\Pi_2,\gamma_3] \bigg) \ .
 \end{split} \label{eq:jjO}
\end{align}
We will also consider the case where $\Pi_1,\Pi_2$ are shifted by traceless pure-gauge fields $\tilde h_1,\tilde h_2$, as in section \ref{sec:free_invariance:boundary_terms}, subject to the asymptotic condition \eqref{eq:tilde_h_asymptotics}. For this case, we claim that a relation of the form \eqref{eq:jjO} will hold again, as:
\begin{align}
 \begin{split}
   &\left< j^{(s_1)}_1 j^{(s_2)}_2 \calO_3 \right> = -N\bigg(\sum_{s_3}S_{s_1,s_2,s_3}[V_{\text{ST}};\Pi_1+\tilde h_1,\Pi_2+\tilde h_2,\phi_3] \\
   &\qquad - S_{s_1}[\Pi_1+\tilde h_1,\gamma_3]\,S_{s_2}[\Pi_2+\tilde h_2,\gamma_3] + S_{s_1,s_2}[V_{\text{new}};\Pi_1+\tilde h_1,\Pi_2+\tilde h_2,\gamma_3] \bigg) \ .
 \end{split} \label{eq:jjO_shifted}
\end{align}
\begin{figure}%
	\centering%
	\includegraphics[scale=0.6]{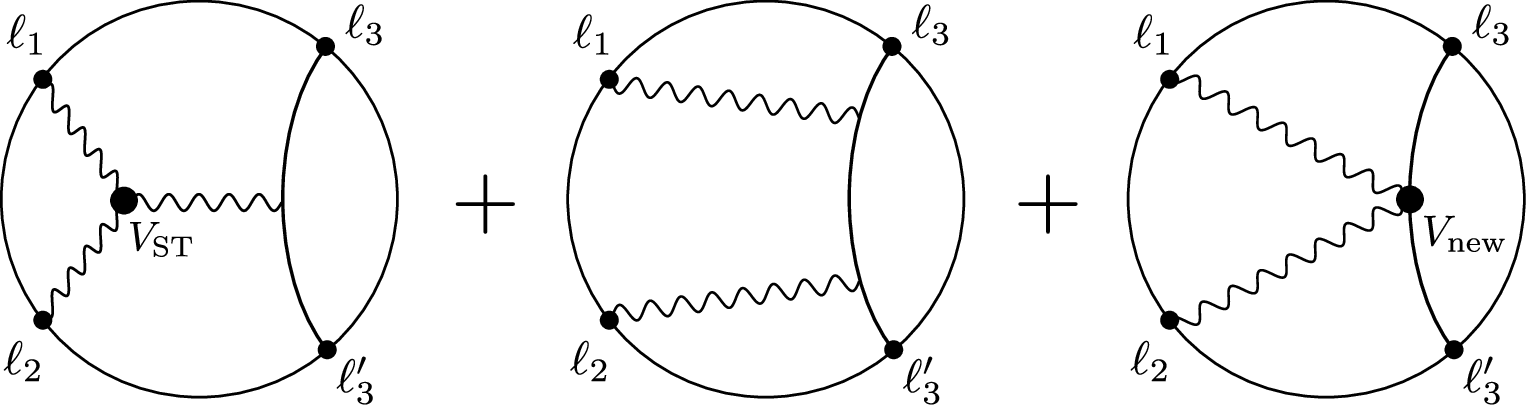} \\
	\caption{The bulk diagrams that describe the correlator $\langle j_1 j_2\calO_3\rangle$ of two local boundary currents and one bilocal. $V_{\text{ST}}$ is the Sleight-Taronna cubic vertex, while $V_{\text{new}}$ is a new vertex that couples two HS fields to a DV particle's worldline. We argue that the non-localities in $V_{\text{new}}$ are confined to $\sim 1$ AdS radius.}
	\label{fig:jjO} 
\end{figure}%
Each term in \eqref{eq:jjO}-\eqref{eq:jjO_shifted} describes a different bulk diagram, as depicted in figure \ref{fig:jjO}. The meaning of each term is as follows (referring to the input fields $\Pi_i$ or $\Pi_i + \tilde h_i$ with $i=1,2$ as simply $h_i$): 
\begin{itemize}
	\item The $S_{s_1,s_2,s_3}[V_{\text{ST}};h_1,h_2,\phi_3]$ term describes the three fields $h_1,h_2,\phi_3$ coupled by the Sleight-Taronna cubic vertex, just like in the standard $\langle jjj\rangle$ correlator \eqref{eq:on_shell_correlator}. To support our replacement of the symmetrized $\calO^+(\ell_3,\ell'_3)$ by the unsymmetrized $\calO(\ell_3,\ell'_3)$, we simply define $V_{\text{ST}}$ to vanish for odd $s_3$.
	\item The $S_{s_1}[h_1,\gamma_3]\,S_{s_2}[h_2,\gamma_3]$ term is a product of two quadratic actions of the form \eqref{eq:quadratic_bulk},\eqref{eq:jO}. It describes a diagram where each of the fields $h_1,h_2$ couples independently to the geodesic $\gamma_3$. Such a term is natural if we consider $\gamma_3$ as not just a source for the DV solution $\phi_3$, but as the physical worldline of a (infinitely heavy) particle.
	\item Finally, the $S_{s_1,s_2}[V_{\text{new}};h_1,h_2,\gamma_3]$ term describes a new cubic vertex coupling both fields $h_1,h_2$ to the $\gamma_3$ worldline. The additional ``TT'' subscript in \eqref{eq:jjO} refers to the fact that the vertex in that formula couples transverse-traceless fields, as opposed to \eqref{eq:jjO_shifted}, where transversality is dropped.
\end{itemize}

The new interaction term $S_{s_1,s_2}[V_{\text{new}};h_1,h_2,\gamma_3]$ can be written a bit more explicitly as:
\begin{align}
 \begin{split}
   S_{s_1,s_2}[V_{\text{new}};h_1,h_2,\gamma_3] = -\int_{-\infty}^\infty&d\tau\,V^{(s_1,s_2)}_{\text{new}}\big(\del_{x_1},\del_{u_1};\del_{x_2},\del_{u_2};\dot x(\tau;\ell_3,\ell'_3)\big) \\
     &\times h_1^{(s_1)}(x_1,u_1)h_2^{(s_2)}(x_2,u_2)\Big|_{x_1=x_2=x(\tau;\ell_3,\ell'_3)} \ .
 \end{split} \label{eq:new_worldline_integral}
\end{align}
This is similar to a usual cubic diagram formula \eqref{eq:vertex_integral}, except the integral is over $\gamma_3$ instead of the entire $EAdS_4$, and the vertex $V^{(s_1,s_2)}_{\text{new}}$ is allowed to depend on the geodesic's tangent vector $\dot x^\mu$. The different powers of $\dot x^\mu$ in the vertex can be viewed as couplings to the different spins $s_3$ of the HS multiplet carried by the DV ``particle'' on $\gamma_3$. It is worth emphasizing that \emph{any} cubic quantity can be reproduced by an action \eqref{eq:new_worldline_integral} with a sufficiently general vertex $V^{(s_1,s_2)}_{\text{new}}$. The non-trivial part of our statement is that this vertex \emph{satisfies appropriate locality criteria}, which we'll describe below. 

\subsection{Ansatz for $V_{\text{new,TT}}$} \label{sec:locality:new_vertex}

Let us now describe a general ansatz for $V_{\text{new,TT}}$ -- the new vertex that reproduces the correct cubic correlator as in \eqref{eq:jjO}, when coupling two boundary-bulk propagators $\Pi_1,\Pi_2$ to a geodesic worldline $\gamma_3$. These propagators span the space of source-free, transverse-traceless fields $h^{(s)}$, and we'll consider the vertex as acting on such fields.

A source-free field $h^{(s)}$ in transverse-traceless gauge is completely determined by boundary data -- for instance, in the language of sections \ref{sec:preliminaries:asymptotics}-\ref{sec:preliminaries:Pi}, by the coefficient of $z^{2-s}$ in its tangential components $[h^{(s)}]_{0,s}$ in the asymptotic limit $z\to 0$. Assuming analyticity, one can equally well formulate such boundary data on a geodesic $\gamma$, via a tower of spatial derivatives at each proper ``time'' $\tau$. To construct a basis of such derivatives, we decompose the field into components along the geodesic's ``time'' direction $\dot x^\mu = t^\mu$ vs. the ``spatial'' directions perpendicular to it, spanned by the 3d metric $q_{\mu\nu} \equiv g_{\mu\nu} - t_\mu t_\nu$. We then take either zero or one 3d curls, followed by an arbitrary number of 3d gradients, and extract the totally symmetric \& traceless part with respect to the 3d metric $q_{\mu\nu}$. Thus, a basis of boundary data on a geodesic $\gamma$ for a source-free, transverse-traceless field $h^{(s)}(x,u)$ is given by the following 3d tensors, encoded as usual through a ``polarization vector'' $u^\mu$, at each point $x^\mu(\tau)$ on $\gamma$:
\begin{align}
 \big\{h^{(s)}(\tau,u)\big\}^n_{l,+} ={}& (q_{\mu\nu}u^\mu\nabla^\nu)^{l-s+n}(q_{\mu\nu}u^\mu\del_u^\nu)^{s-n}(t\cdot\del_u)^n h^{(s)}(x,u) \Big|_{x = x(\tau)} - \text{3d traces} \ ; \label{eq:geodesic_data_even} \\
 \begin{split}
   \big\{h^{(s)}(\tau,u)\big\}^n_{l,-} ={}& (q_{\mu\nu}u^\mu\nabla^\nu)^{l-s+n}(\epsilon_{\mu\nu\rho}u^\mu\nabla^\nu\del_u^\rho)(q_{\mu\nu}u^\mu\del_u^\nu)^{s-n}(t\cdot\del_u)^n h^{(s)}(x,u) \Big|_{x = x(\tau)} \\
      &- \text{3d traces} \ .
 \end{split} \label{eq:geodesic_data_odd}
\end{align}
Here, $l$ denotes the tensors' 3d rank (i.e. their angular momentum number), and the $\pm$ superscript denotes their spatial parity. Tensors with the same 3d structure $(l,\pm)$ are distinguished by the superscript $n$, which denotes the number of indices on $h^{(s)}$ taken along the time direction. $\epsilon^{\mu\nu\rho} \equiv \epsilon^{\mu\nu\rho\sigma\lambda}t_\sigma x_\lambda$ is the 3d ``spatial'' Levi-Civita tensor, and ``$\!{}- \text{3d traces}$'' means subtracting $\sim q_{\mu\nu}u^\mu u^\nu$ terms so as to make the result traceless. $n$ runs from $0$ to $s$ for the even tensors \eqref{eq:geodesic_data_even}, and from $0$ to $s-1$ for the odd tensors \eqref{eq:geodesic_data_odd}. $l$ runs from $s-n$ to $\infty$ in both cases.

The general ansatz for the vertex $V_{\text{new,TT}}$ can now be assembled by constructing the data \eqref{eq:geodesic_data_even}-\eqref{eq:geodesic_data_odd} for the fields $h_1,h_2$ on the worldline $\gamma_3$, and then coupling the pieces with matching parity $\eta=\pm$ and angular momentum $l$:
\begin{align}
 S_{s_1,s_2}[V_{\text{new,TT}};h_1,h_2,\gamma_3] = -\int_{-\infty}^\infty&d\tau \sum_{l,\eta}\sum_{n_1,n_2} \big\{h_1^{(s_1)}(\tau,\del_u)\big\}^{n_1}_{l,\eta}\,K^{n_1,n_2}_{s_1,s_2,l,\eta}(\del_\tau)\,\big\{h_2^{(s_2)}(\tau,u)\big\}^{n_2}_{l,\eta} \ ,
 \label{eq:V_TT_ansatz}
\end{align}
where $\big\{h_1^{(s_1)}(\tau,\del_u)\big\}^{n_1}_{l,\eta}$ refers to computing $\big\{h_1^{(s_1)}(\tau,u)\big\}^{n_1}_{l,\eta}$ as in \eqref{eq:geodesic_data_even}-\eqref{eq:geodesic_data_odd} and then substituting $u^\mu\rightarrow\del_u^\mu$, in order to contract the tensor indices with those of $\big\{h_2^{(s_2)}(\tau,u)\big\}^{n_2}_{l,\eta}$.

The non-trivial information about the vertex is now contained in the kernel $K^{n_1,n_2}_{s_1,s_2,l,\eta}(\del_\tau)$. Once again, a sufficiently general $K$ can describe \emph{any} cubic quantity with the prescribed spacetime symmetries. In particular, there exists a $K$ that reproduces the cubic CFT correlator as in \eqref{eq:jjO}. Our task will be to show that this $K$ is \emph{sufficiently local}, i.e. that its non-locality is constrained to $\sim 1$ AdS curvature radius. With respect to the geodesic $\gamma_3$, this locality statement can be split into two parts. First, we can speak of ``radial locality'', transverse to $\gamma_3$. This amounts to $K^{n_1,n_2}_{s_1,s_2,l,\eta}(\del_\tau)$ vanishing fast enough as the numbers $l-s_1+n_1,l-s_2+n_2$ of ``spatial'' derivatives increase. Second, we can speak of ``time locality'', along $\gamma_3$. This amounts to $K^{n_1,n_2}_{s_1,s_2,l,\eta}(\del_\tau)$ being analytic in time derivatives $\del_\tau$, and its Taylor coefficients vanishing fast enough with increasing powers of $\del_\tau$. In this paper, we will not calculate $K^{n_1,n_2}_{s_1,s_2,l,\eta}(\del_\tau)$, and thus we won't be able to check these locality properties directly. Instead, we will formulate proxy criteria for them in terms of the behavior of the diagram $S_{s_1,s_2}[V_{\text{new,TT}};\Pi_1,\Pi_2,\gamma_3]$ in certain limits, and then demonstrate that these criteria hold.

\subsection{Radial locality of $V_{\text{new,TT}}$} \label{sec:locality:radial}

\subsubsection{Formulating the criterion}

\begin{figure}%
	\centering%
	\includegraphics[scale=0.6]{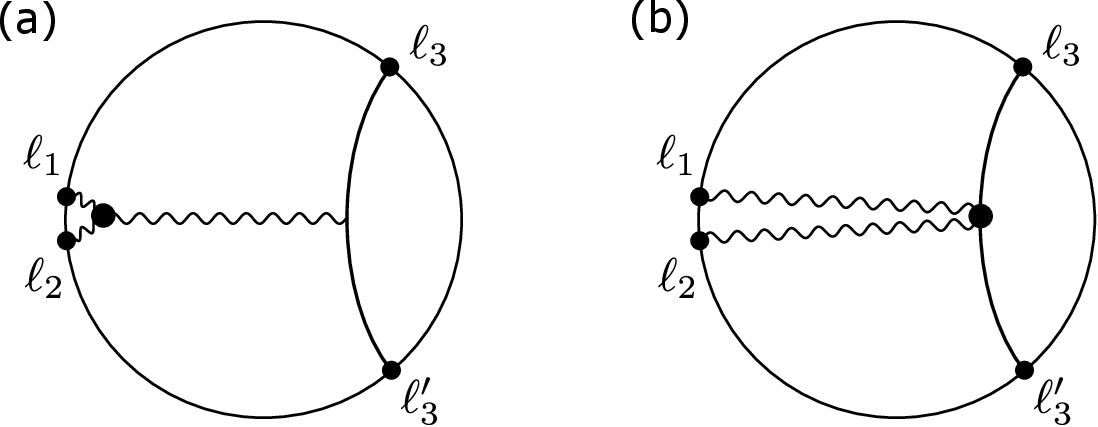} \\
	\caption{Illustration of our radial locality criterion. In panel (a), we see an interaction that is \emph{not} localized near the geodesic, i.e. that involves the fields $\Pi_1,\Pi_2$ arbitrarily far from it. In the limit of nearby source points $\ell_1,\ell_2$, this creates contributions with ``short'' propagators, which become singular at $\ell_1=\ell_2$. In panel (b), we see an interaction that \emph{is} localized near the geodesic. The propagators from $\ell_1,\ell_2$ are now ``long'', and the $\ell_1=\ell_2$ limit has no singularities.}
	\label{fig:radial_locality} 
\end{figure}

Our proxy criterion for radial locality is as follows.
\begin{radial_locality_criterion}
 A vertex $V_{\text{new,TT}}$ coupling two boundary-bulk propagators $\Pi_1,\Pi_2$ to a geodesic worldline $\gamma_3$ is radially local, if its action $S_{s_1,s_2}[V_{\text{new,TT}};\Pi_1,\Pi_2,\gamma_3]$ as a function of the source points $\ell_1,\ell_2$ is analytic at $\ell_1=\ell_2$.
\end{radial_locality_criterion}
The motivation for this criterion is depicted in figure \ref{fig:radial_locality}. A radially local vertex should only involve the fields $\Pi_1,\Pi_2$ near (i.e. within $\sim 1$ AdS radius from) the $\gamma_3$ worldline. In that situation, depicted in figure \ref{fig:radial_locality}(b), the diagram is analytic near $\ell_1=\ell_2$, because it never involves ``short'' propagators that would go singular in the limit. In contrast, in figure \ref{fig:radial_locality}(a), we see a ``vertex'' that couples $\Pi_1$ and $\Pi_2$ far from $\gamma_3$. This allows for ``short'' propagators from $\ell_1,\ell_2$, which cause a singularity at $\ell_1=\ell_2$, i.e. an infinity in the diagram itself or in its derivatives with respect to $\ell_1,\ell_2$. There is no third possibility, in the sense that the vertex cannot depend on \emph{only one} of $\Pi_1,\Pi_2$ at points distant from the geodesic. This is clear from the ansatz \eqref{eq:V_TT_ansatz}, where the number of ``spatial'' derivatives acting on $\Pi_1,\Pi_2$ can grow only together, governed by the angular momentum number $l$. 

Note the similarity between figure \ref{fig:radial_locality}(a) and the $V_{\text{ST}}$ diagram from figure \ref{fig:jjO}. Indeed, if we were to foolishly express the ``field-field-field'' diagram $\sum_{s_3}S_{s_1,s_2,s_3}[V_{\text{ST}};\Pi_1,\Pi_2,\phi_3]$ as a ``field-field-worldline'' diagram $S_{s_1,s_2}[\tilde V_{\text{ST}};\Pi_1,\Pi_2,\gamma_3]$, then $\tilde V_{\text{ST}}$ would constitute an \emph{example} of a radially non-local vertex. It's easy to see that this is consistent with our criterion above, by noting e.g. that the diagram diverges at $\ell_1=\ell_2$. To see this in detail, note that the $\ell_1\rightarrow\ell_2$ limit is conformal to the $\ell_3\rightarrow\ell'_3$ limit, where the dominant contribution to the DV field $\phi_3$ is a spin-0 boundary-bulk propagator, $\phi_3^{(0)} \sim \sqrt{-\ell_3\cdot\ell'_3}\,\Pi_3^{(0)}$. Thus, the dominant piece of $\sum_{s_3}S_{s_1,s_2,s_3}[V_{\text{ST}};\Pi_1,\Pi_2,\phi_3]$ behaves at $\ell_3\rightarrow\ell'_3$ like a standard cubic diagram $\sim \sqrt{-\ell_3\cdot\ell'_3}\,S_{s_1,s_2,0}[V_{\text{ST}};\Pi_1,\Pi_2,\Pi_3]$ computing the cubic correlator $\sim \sqrt{-\ell_3\cdot\ell'_3}\,\left<j^{(s_1)}_1 j^{(s_2)}_2 j^{(0)}_3 \right>$, which diverges at $\ell_1=\ell_2$. Since we were careful to keep track of the conformal weights, it's clear that the divergence at $\ell_1=\ell_2$ holds also in the original conformal frame, where $\ell_3,\ell'_3$ are not necessarily close.

Moreover, the radial non-locality depicted in figure \ref{fig:radial_locality}(a) is similar in nature to the infamous non-locality of HS theory's quartic scalar vertex in \cite{Sleight:2017pcz}. Indeed, the problem with the quartic vertex is that it hides within it the structure of a bulk-bulk propagator, giving the would-be contact diagram the structure of an exchange diagram. Again consistently with our criterion, this diagram is indeed singular at $\ell_1=\ell_2$, reproducing (up to a numerical coefficient) the short-distance singularity of the quartic CFT correlator.

\subsubsection{Verifying that the criterion holds} \label{sec:locality:radial:verifying}

\begin{figure}%
	\centering%
	\includegraphics[scale=0.6]{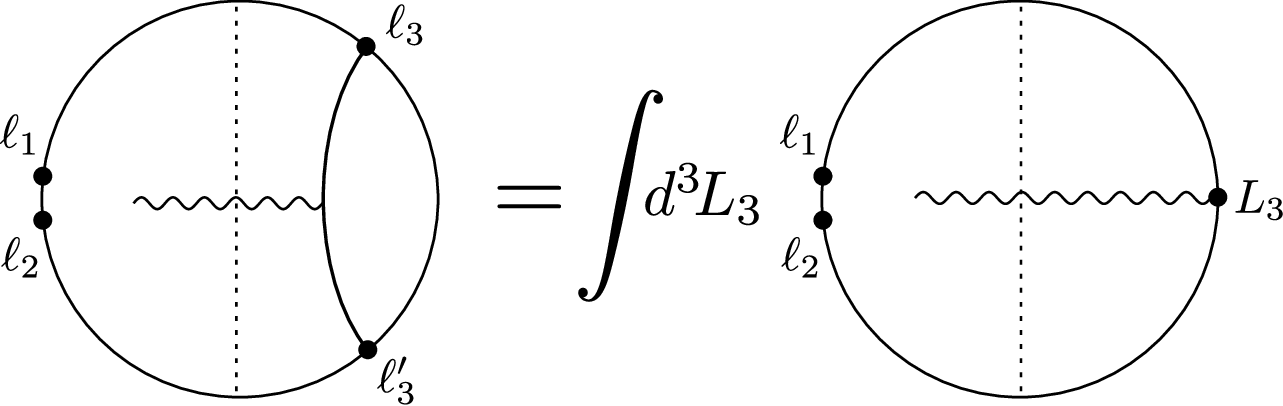} \\
	\caption{An intermediate step in the radial locality argument. To the left of the dashed hypersurface $\Sigma$, the DV field of $\gamma_3$ is source-free, and can be written (up to gauge) as a superposition of boundary-bulk propagators.}
	\label{fig:radial_split} 
\end{figure}

Having established and motivated our radial locality criterion, let us now demonstrate that it holds for the vertex $V_{\text{new,TT}}$ that satisfies eq. \eqref{eq:jjO}. First, let us notice that the $\ell_1\rightarrow\ell_2$ limit can be characterized as the limit of \emph{large bulk distance} between the geodesic $\gamma(\ell_1,\ell_2)$ and the geodesic worldline $\gamma_3$. Now, let us draw a bulk hypersurface $\Sigma$ that splits $EAdS_4$ into two regions: a region $\Omega_{12}$ containing $\gamma(\ell_1,\ell_2)$, and a region $\Omega_3$ containing $\gamma_3$. This splitting of $EAdS_4$ is depicted as a dashed line in figure \ref{fig:radial_split}. The asymptotic boundary is also split into two regions by $\Sigma$,  which we'll denote as $B_{12}$ and $B_3$. Crucially, we assume that $\Sigma$, like $\gamma_3$, is very far from $\gamma(\ell_1,\ell_2)$. 

Now, consider the restriction to $\Omega_{12}$ of the DV field $\phi_3$. Within this region, $\phi_3$ is a solution to the \emph{source-free} Fronsdal equation. From \cite{Neiman:2017mel}, we know the following about its Weyl field strength at boundary points $L_{12}$ belonging to the region $B_{12}$:
\begin{itemize}
	\item The magnetic field strength (in the spin-0 case, the boundary data with weight $\Delta=2$) vanishes.
	\item The electric field strength (in the spin-0 case, the boundary data with weight $\Delta=1$) matches the bilocal-local correlators $\left<\calO(\ell_3,\ell'_3)\,j^{(s)}(L_{12},\lambda_{12})\right>$.
\end{itemize}
Now, since it is source-free with vanishing magnetic boundary data on $B_{12}$, the restriction of $\phi_3$ to $\Omega_{12}$ must be, up to gauge, a superposition of boundary-bulk propagators $\Pi_3$ with source points in $B_3$ (see figure \ref{fig:radial_split}):
\begin{align}
 \phi^{(s_3)}(x,u;\ell_3,\ell'_3)\Big|_{x\in \Omega_{12}} = \int_{B_3} d^3L_3\,A^{(s_3)}_{\ell_3,\ell'_3}(L_3,\del_\lambda)\,\Pi^{(s_3)}(x,u;L_3,\lambda) + \tilde h^{(s_3)}_3(x,u) \ . \label{eq:local_decomposition_bulk}
\end{align} 
Here, the coefficients $A^{(s_3)}_{\ell_3,\ell'_3}$ describe some traceless boundary sources as in \eqref{eq:local_decomposition}, while $\tilde h_3$ is a pure-gauge field. Furthermore, since the RHS of \eqref{eq:local_decomposition_bulk} has the same electric field strength on $B_{12}$ as the original field $\phi_3$, we conclude that the corresponding boundary currents in $B_3$ have the same quadratic correlators with currents in $B_{12}$ as the original bilocal $\calO(\ell_3,\ell'_3)$:
\begin{gather}
 \int_{B_3} d^3L_3\,A^{(s)}_{\ell_3,\ell'_3}(L_3,\del_\lambda)\left<j^{(s)}(L_3,\lambda)\,j^{(s)}(L_{12},\lambda_{12}) \right> = \left<\calO(\ell_3,\ell'_3)\,j^{(s)}(L_{12},\lambda_{12}) \right> \\
 \text{for all }L_{12}\in B_{12} \ . \nonumber
\end{gather}
From the discussion in section \ref{sec:preliminaries:CFT}, it then follows that $\int_{B_3} d^3L_3\,A^{(s)}_{\ell_3,\ell'_3}(L_3,\del_\lambda)\,j^{(s)}(L_3,\lambda)$ and $\calO(\ell_3,\ell'_3)$ have the same correlators with \emph{any} operators in $B_{12}$. In particular, they have the same cubic correlators with our original local currents $j^{(s_1)}_1$ and $j^{(s_2)}_2$:
\begin{align}
 \int_{B_3} d^3L_3 \sum_{s_3} A^{(s_3)}_{\ell_3,\ell'_3}(L_3,\del_\lambda)\left<j^{(s_1)}_1 j^{(s_2)}_2 j^{(s_3)}(L_3,\lambda) \right> = \left<j^{(s_1)}_1 j^{(s_2)}_2 \calO_3 \right> \ . \label{eq:local_decomposition_cubic}
\end{align}
Now, consider the behavior of $\phi_3$ at the asymptotic boundary $B_{12}$, by examining the formula \eqref{eq:k}-\eqref{eq:phi} for the DV solution. Since we're away from the worldline endpoints $\ell_3,\ell'_3$, the asymptotic boundary is a large-$R$ regime.  $R$ itself scales asymptotically as $R\sim z^{-1}$, implying that the norms \eqref{eq:t_r_products} of $t^\mu$ and $r^\mu$ scale as $\sqrt{t\cdot t} \sim z$ and $\sqrt{r\cdot r} \sim 1$. It is now easy to see that $\phi^{(s_3)}_3$ satisfies the condition \eqref{eq:tilde_h_asymptotics} at $B_{12}$, i.e. its components in a normalized Poincare basis scale as $O(z^{s_3+1})$. Since this is true of the propagators $\Pi_3$ in \eqref{eq:local_decomposition_bulk}, we conclude that it must be true of the pure-gauge field $\tilde h_3$ as well.

We are now ready for the main part of the radial-locality argument. Consider the field $\hat\phi_3$, defined by the RHS of \eqref{eq:local_decomposition_bulk} throughout the bulk, i.e. in $\Omega_3$ as well as $\Omega_{12}$. Thus, $\hat\phi_3$ agrees with $\phi_3$ in $\Omega_{12}$, but is source-free in the entire bulk. We assume that the pure-gauge field $\tilde h_3$ is extended in such a way that it continues to satisfy the scaling condition \eqref{eq:tilde_h_asymptotics} at $B_3$ as well as $B_{12}$. This is easy to arrange: by the logic of section \ref{sec:free_invariance:boundary_terms}, it is sufficient to ensure that the divergence $(\del_u\cdot\nabla)\tilde h_3$ satisfies \eqref{eq:div_h_scaling} -- the natural scaling for a divergence-free symmetric tensor -- and then choose the solution of eqs. \eqref{eq:Lambda_field_equation_1}-\eqref{eq:Lambda_field_equation_2} with vanishing $\sim z^{1-s+q}$ boundary conditions.

\begin{figure}%
	\centering%
	\includegraphics[scale=0.6]{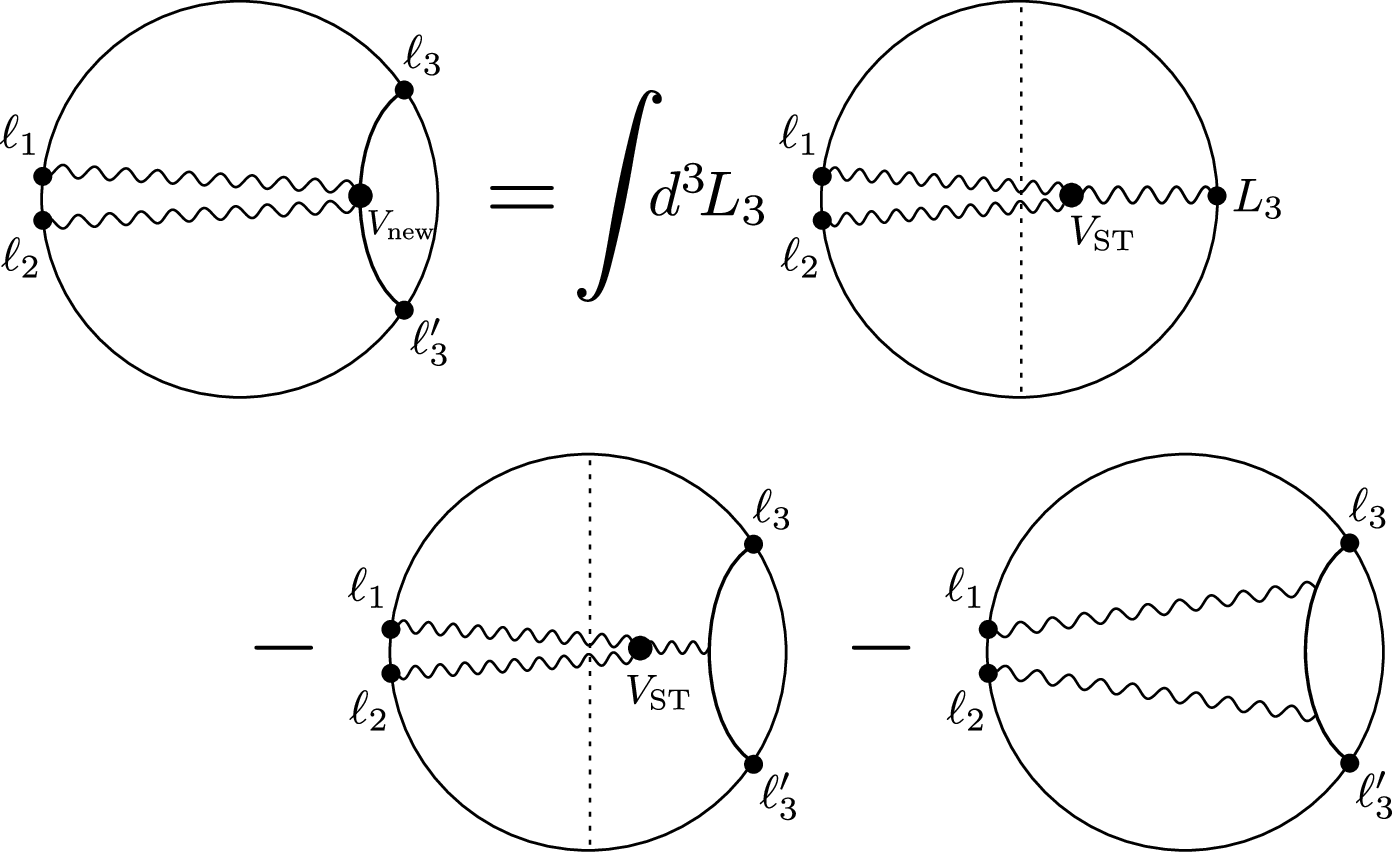} \\
	\caption{The main step in the radial locality argument. The $V_{\text{new,TT}}$ diagram is expressed as a combination of terms that are manifestly analytic at $\ell_1=\ell_2$.}
	\label{fig:radial_subtraction} 
\end{figure}

Now, consider the bulk analogue of the correlator equation \eqref{eq:local_decomposition_cubic}. The RHS of \eqref{eq:local_decomposition_cubic} is calculated by the three diagrams of \eqref{eq:jjO}, whereas the LHS is calculated by the standard Sleight-Taronna cubic diagram $S_{s_1,s_2,s_3}[V_{\text{ST}};\Pi_1,\Pi_2,\Pi_3]$, with the appropriate sum over $s_3$ and integral over $L_3$. By the results of section \ref{sec:free_invariance}, this diagram stays unchanged when we shift the propagators $\Pi_3$ by the pure-gauge field $\tilde h_3$ as in \eqref{eq:local_decomposition_bulk}. Thus, the bulk analogue of \eqref{eq:local_decomposition_cubic} can be written as:
\begin{align}
 \begin{split}
   \sum_{s_3}S_{s_1,s_2,s_3}[V_{\text{ST}};\Pi_1,\Pi_2,\hat\phi_3] ={}& \sum_{s_3}S_{s_1,s_2,s_3}[V_{\text{ST}};\Pi_1,\Pi_2,\phi_3] - S_{s_1}[\Pi_1,\gamma_3]\,S_{s_2}[\Pi_2,\gamma_3] \\
     &+ S_{s_1,s_2}[V_{\text{new,TT}};\Pi_1,\Pi_2,\gamma_3] \ .
 \end{split} \label{eq:radial_locality_argument}
\end{align}
Each of the two $V_{\text{ST}}$ diagrams in \eqref{eq:radial_locality_argument} contains a bulk integral over the position $x$ of the Sleight-Taronna vertex. The $\Omega_{12}$ portion of this integral cancels between the LHS and RHS, because $\phi_3$ and $\hat\phi_3$ are equal there. We conclude that $S_{s_1,s_2}[V_{\text{new,TT}};\Pi_1,\Pi_2,\gamma_3]$ is given by the difference between the $\Omega_3$ portions of the two $V_{\text{ST}}$ diagrams, plus the double-exchange term $S_{s_1}[\Pi_1,\gamma_3]\,S_{s_2}[\Pi_2,\gamma_3]$; this situation is depicted in figure \ref{fig:radial_subtraction}. Now, notice that all of these terms involve ``long'' propagators stretching from $\ell_1$ and $\ell_2$ into the distant region $\Omega_3$. Thus, the three terms are all analytic at $\ell_1=\ell_2$, and therefore so is the $V_{\text{new,TT}}$ diagram. This concludes our argument for the radial locality of $V_{\text{new,TT}}$.

\subsection{Time locality of $V_{\text{new,TT}}$} \label{sec:locality:time}

\subsubsection{Formulating the criterion} \label{sec:locality:time:formulating}

\begin{figure}%
	\centering%
	\includegraphics[scale=0.6]{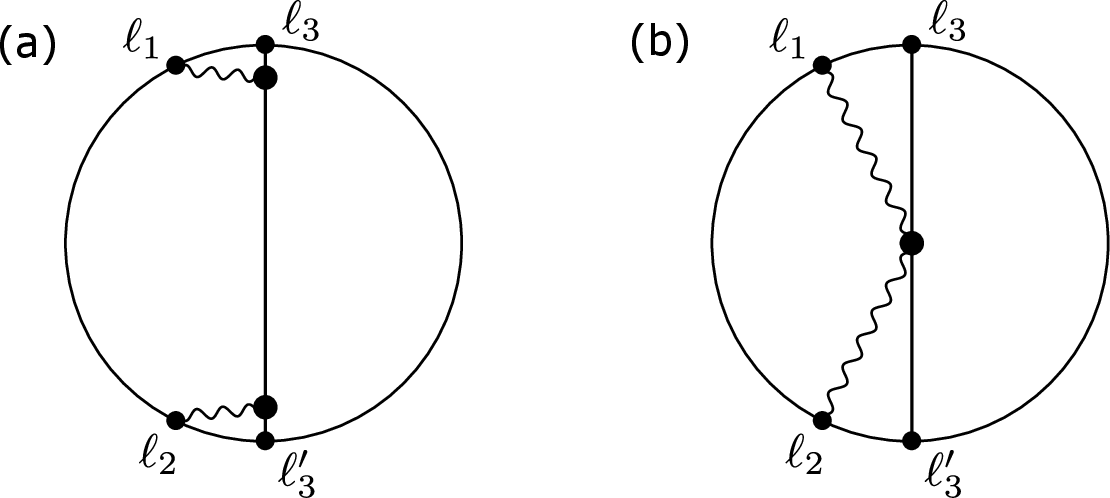} \\
	\caption{Illustration of our ``time'' locality criterion. We consider a field-field-worldline interaction in the limit of large ``time'' separation $|\tau_1-\tau_2|$ between $\ell_1$ and $\ell_2$. If the interaction is local, i.e. decays exponentially with the distance $\Delta\tau$ along the geodesic, the diagram will exponentially decay with $|\tau_1-\tau_2|$. This can happen through some combination of the scenarios in panels (a),(b). In panel (a), the diagram is dominated by contributions with $\Delta\tau\approx|\tau_1-\tau_2|$, and is governed by the interaction's decay with $\Delta\tau$. In panel (b), the diagram is dominated by contributions with $\Delta\tau=O(1)$, and its exponential decay in $|\tau_1-\tau_2|$ is due to the ``long'' boundary-bulk propagators. If the interaction is \emph{not} time-local, the dominant contribution will always be panel (a), and its failure to decay exponentially in $|\tau_1-\tau_2|$ will be governed by the interaction's failure to decay in $\Delta\tau$.}
	\label{fig:time_locality} 
\end{figure}

We now turn to our proxy criterion for ``time'' locality of the new vertex. First, let us notice that the geodesic $\gamma_3$ induces a coordinate system on $EAdS_4$ and its boundary. Setting $\ell_3^\mu = (\frac{1}{2},\frac{1}{2},\vec 0)$ and $\ell'^\mu_3 = (\frac{1}{2},-\frac{1}{2},\vec 0)$, this coordinate system reads:
\begin{align}
 x^\mu(\tau,R,\mathbf{n}) &= \sqrt{1+R^2}\,(\cosh\tau,\sinh\tau,\vec 0) + R\,(0,0,\mathbf{n}) \ ; \label{eq:x_cylindric} \\
 \ell^\mu(\tau,\mathbf{n}) &= (\cosh\tau,\sinh\tau,\mathbf{n}) \ , \label{eq:ell_cylindric}
\end{align}
where $R$ is the distance function \eqref{eq:R} from $\gamma_3$, and $\mathbf{n}\in S_2$ is a 3d unit vector. In particular, the length parameter $\tau$ along $\gamma_3$ extends into a ``time'' coordinate $\tau$ throughout the bulk and boundary, with ``time translations'' $\tau\rightarrow\tau+c$ being a spacetime symmetry (in embedding space, these are just boosts in the $(\ell_3,\ell'_3)$ plane). Our time locality criterion now reads:
\begin{time_locality_criterion}
 A vertex $V_{\text{new,TT}}$ coupling two boundary-bulk propagators $\Pi_1,\Pi_2$ to a geodesic worldline $\gamma_3$ is time-local, if its action $S_{s_1,s_2}[V_{\text{new,TT}};\Pi_1,\Pi_2,\gamma_3]$ vanishes exponentially at large time difference $|\tau_1-\tau_2|$ between the source points $\ell_1$ and $\ell_2$. 
\end{time_locality_criterion}
Let us explain the reasoning behind this criterion. We assume that radial locality is satisfied, so that $V_{\text{new,TT}}$ couples the fields $\Pi_1,\Pi_2$ only in the vicinity of $\gamma_3$. Then, our desired time-locality property is for this coupling to vanish exponentially for points separated by large distances $\Delta\tau$ \emph{along} $\gamma_3$. The premise of our criterion is that exponential decay at large $|\tau_1-\tau_2|$ \emph{on the boundary} is a good proxy for the desired exponential decay in $\Delta\tau$ on the geodesic. To become convinced of this, let us consider in detail the diagram $S_{s_1,s_2}[V_{\text{new,TT}};\Pi_1,\Pi_2,\gamma_3]$ at large $|\tau_1-\tau_2|$ (see figure \ref{fig:time_locality}). 

If the vertex couples $\Pi_1$ and $\Pi_2$ at approximately the same point $x(\tau)$ on the geodesic with $\Delta\tau=O(1)$, the diagram will appear as in figure \ref{fig:time_locality}(b). This features boundary-bulk propagators that stretch across long intervals $|\tau_1-\tau|$ and $|\tau-\tau_2|$. Let us examine the behavior of such ``long'' propagators. We focus on e.g. the $\Pi_1$ propagator, with source point $\ell_1^\mu = (\cosh\tau_1,\sinh\tau_1,\mathbf{n_1})$ at $\tau_1\gg 1$, and assume that the polarization vector $\lambda_1^\mu$ has $O(1)$ components $(\lambda_1^\tau,\boldsymbol{\lambda_1})$ along the $\tau$ axis and the 2-sphere:
\begin{align}
 \lambda_1^\mu = (\lambda_1^\tau\sinh\tau_1,\lambda_1^\tau\cosh\tau_1,\boldsymbol{\lambda_1}) \ .
\end{align}
The building blocks of the propagator \eqref{eq:Pi} then read:
\begin{align}
 \begin{split}
   \ell_1\cdot x &= -\cosh(\tau_1 -\tau )\approx -\frac{e^{|\tau_1-\tau|}}{2} \ ; \\
   m_1^\mu &= \big(0,\lambda_1^\tau,\boldsymbol{\lambda_1}\cosh(\tau_1-\tau)- \lambda_1^\tau\mathbf{n_1}\sinh(\tau_1-\tau) \big) = O\big(e^{|\tau_1-\tau|}\big) \ .
 \end{split} \label{eq:Pi_1_building_blocks}
\end{align}
We conclude that the ``long'' propagator $\Pi_1$ scales as $O\big(e^{-(s_1+1)|\tau_1-\tau|}\big)$, and similarly for $\Pi_2$. The product of the two propagators at the geodesic therefore scales as:
\begin{align}
 \Pi_1\Pi_2 = O\big(e^{-(\min(s_1,s_2)+1)|\tau_1-\tau_2|}\big) = O\big(e^{-|\tau_1-\tau_2|}\big) \ . \label{eq:Pi_product}
\end{align}
Thus, if the vertex couples $\Pi_1$ and $\Pi_2$ at distances $\Delta\tau=O(1)$, the diagram decays exponentially at large $|\tau_1-\tau_2|$, consistently with our criterion. Now, consider the complementary situation, depicted in figure \ref{fig:time_locality}(a): ``short'' $O(1)$ boundary-bulk propagators, followed by a coupling of fields at distance $\Delta\tau\approx|\tau_1-\tau_2|$ along the geodesic. In this case, the large-$|\tau_1-\tau_2|$ behavior of the diagram is directly dictated by the large-$\Delta\tau$ behavior of the vertex, again in agreement with our criterion. For a non-local vertex, the interaction of figure \ref{fig:time_locality}(a) will always dominate; for a local vertex, the interaction may be dominated by figure \ref{fig:time_locality}(a) \emph{or} \ref{fig:time_locality}(b), or some combination of the two. In any case, we see that exponential decay of the diagram as a function of $|\tau_1-\tau_2|$ on the boundary is a faithful proxy for exponential decay of the vertex as a function of $\Delta\tau$ on the geodesic.

As with radial locality, it is easy to find an example of a vertex that \emph{isn't} time-local. Such a vertex can be obtained by foolishly writing the product term $S_{s_1}[\Pi_1,\gamma_3]\,S_{s_2}[\Pi_2,\gamma_3]$ in \eqref{eq:jjO} in terms of a single field-field-worldline vertex, as $S_{s_1,s_2}[V_{\text{prod}};\Pi_1,\Pi_2,\gamma_3]$. This is immediately non-local by our criterion, since the diagram doesn't depend on $\tau_1-\tau_2$ at all. 

Finally, note that our radial and time locality criteria have different relationships with the holographic UV/IR inversion. In the bulk, both criteria are concerned with the vertex's IR behavior. In the case of radial locality, this translates into the UV limit $\ell_1=\ell_2$ on the boundary: as expected, the radial direction behaves holographically. On the other hand, for time locality, IR in the bulk stays IR on the boundary: the ``time'' coordinate $\tau$ is common to both, and does not get inverted.

\subsubsection{Verifying that the criterion holds}

Having established and motivated our ``time'' locality criterion, let us now demonstrate that it holds for the vertex $V_{\text{new,TT}}$ that satisfies eq. \eqref{eq:jjO}. As in section \ref{sec:locality:time:formulating}, we set:
\begin{gather}
 \ell_i^\mu = (\cosh\tau_i,\sinh\tau_i,\mathbf{n}_i) \ ; \quad \lambda_i^\mu = (\lambda_i^\tau\sinh\tau_i,\lambda_i^\tau\cosh\tau_i,\boldsymbol{\lambda}_i) \ ; \\
 \ell_3^\mu = \left(\frac{1}{2},\frac{1}{2},\vec 0 \right) \ ; \quad \ell'^\mu_3 = \left(\frac{1}{2},-\frac{1}{2},\vec 0 \right) \ ,
\end{gather}
with $i=1,2$. Again, we are interested in the limit of large $|\tau_1-\tau_2|$, and assume that the polarization components $\lambda_i^\tau,\boldsymbol{\lambda}_i$ are $O(1)$.

We begin by examining the CFT correlator $\left< j^{(s_1)}_1 j^{(s_2)}_2 \calO_3 \right>$ in the large $|\tau_1-\tau_2|$ limit. To simplify the analysis, we point-split the currents $j^{(s_1)}_1$ and $j^{(s_2)}_2$ into bilocals $\calO(\ell_1,\ell'_1)$ and $\calO(\ell_2,\ell'_2)$, where:
\begin{align}
 \ell'^\mu_i = (\cosh\tau'_i,\sinh\tau'_i,\mathbf{n'}_i) \ ; \quad \tau'_i-\tau_i = O(1) \ ,
\end{align}
again with $i=1,2$. We can revert back to the local currents by taking derivatives at $\ell'^\mu_i = \ell^\mu_i$, as in \eqref{eq:j}. These translate simply into derivatives (with $O(1)$ coefficients) with respect to the coordinates $(\tau_i,\mathbf{n}_i)$ and $(\tau'_i,\mathbf{n'}_i)$ at $(\tau'_i,\mathbf{n'}_i) = (\tau_i,\mathbf{n}_i)$. Thus, we consider the CFT correlator:
\begin{align}
 \langle\calO_1\calO_2\calO_3\rangle = 4N\times\frac{G(\ell'_3,\ell_1)G(\ell'_1,\ell_2)G(\ell'_2,\ell_3) + G(\ell'_3,\ell_2)G(\ell'_2,\ell_1)G(\ell'_1,\ell_3)}{G(\ell_1,\ell'_1)G(\ell_2,\ell'_2)G(\ell_3,\ell'_3)} \ , \label{eq:cubic_for_time_locality_raw}
\end{align}
where $G(\ell,\ell')$ is the boundary propagator \eqref{eq:G_CFT}. The factor of $G(\ell_1,\ell'_1)G(\ell_2,\ell'_2)$ is just an artifact of the normalization in our point-splitting procedure $j^{(s_i)}_i\rightarrow\calO_i$, and we leave it as-is. The other boundary propagators in \eqref{eq:cubic_for_time_locality_raw} can be constructed from the scalar products:
\begin{align}
 \ell_3\cdot\ell'_3 = -\frac{1}{2} \ ; \quad \ell_3\cdot\ell_i = -\frac{1}{2}e^{-\tau_i} \ ; \quad \ell'_3\cdot\ell_i = -\frac{1}{2}e^{\tau_i} \ ; \quad \ell_1\cdot\ell_2 = -\cosh(\tau_1-\tau_2) + \mathbf{n_1}\cdot\mathbf{n_2} \ ,
\end{align}
and similarly for $\ell_1\rightarrow\ell'_1$ and/or $\ell_2\rightarrow\ell'_2$. For $\tau_1-\tau_2$ large and positive (negative), the second (first) term in \eqref{eq:cubic_for_time_locality_raw} dominates. Overall, the result is an $O(1)$ term with an $O\big(e^{-|\tau_1-\tau_2|}\big)$ correction:
\begin{align}
 \langle\calO_1\calO_2\calO_3\rangle = \frac{N}{4\pi^2G(\ell_1,\ell'_1)G(\ell_2,\ell'_2)}\left( e^{(\tau'_1-\tau_1 + \tau'_2-\tau_2)/2} + O\big(e^{-|\tau_1-\tau_2|}\big) \right) \ . \label{eq:cubic_for_time_locality}
\end{align}
Now, the key observation is that the $O(1)$ term in \eqref{eq:cubic_for_time_locality} is precisely reproduced by the double-exchange term in our bulk formula \eqref{eq:jjO}. Indeed, upon extending the point-splitting procedure $j^{(s_i)}_i\rightarrow\calO_i$ to the bulk fields $\Pi_i\rightarrow\phi_i$, the double-exchange term becomes:
\begin{align}
 \begin{split}
   N\sum_{s_1,s_2}S_{s_1}[\phi_1,\gamma_3]\,S_{s_2}[\phi_2,\gamma_3] &= \frac{1}{N}\langle\calO_1\calO_3\rangle\langle\calO_2\calO_3\rangle \\
     &= 4N\times \frac{G(\ell'_3,\ell_1)G(\ell'_1,\ell_3)}{G(\ell_1,\ell'_1)G(\ell_3,\ell'_3)}\times  \frac{G(\ell'_3,\ell_2)G(\ell'_2,\ell_3)}{G(\ell_2,\ell'_2)G(\ell_3,\ell'_3)} \\
     &= \frac{Ne^{(\tau'_1-\tau_1 + \tau'_2-\tau_2)/2}}{4\pi^2G(\ell_1,\ell'_1)G(\ell_2,\ell'_2)} \ .
 \end{split} \label{eq:double_exchange}
\end{align}
Thus, the difference between \eqref{eq:cubic_for_time_locality} and \eqref{eq:double_exchange} is $O\big(e^{-|\tau_1-\tau_2|}\big)$. Reverting back to local currents $j^{(s_i)}_i$, this becomes: 
\begin{align}
 \left<j^{(s_1)}_1 j^{(s_2)}_2 \calO_3\right> - N S_{s_1}[\Pi_1,\gamma_3]S_{s_2}[\Pi_2,\gamma_3] = O\big(e^{-|\tau_1-\tau_2|}\big) \ . \label{eq:large_tau_residue}
\end{align}
The Sleight-Taronna contribution $\sum_{s_3}S_{s_1,s_2,s_3}[V_{\text{ST}},\Pi_1,\Pi_2,\phi_3]$ also decays at large ``time'' separation as $e ^{-|\tau_1-\tau_2|}$. This is easy to see by extending our analysis of $\Pi_1,\Pi_2$ in section \ref{sec:locality:time:formulating} above, away from the $\gamma_3$ geodesic. Setting the bulk position $x$ of the Sleight-Taronna vertex at an arbitrary point \eqref{eq:x_cylindric}, we see that the building blocks of e.g. $\Pi_1$ have essentially the same large-$|\tau_1-\tau|$ behavior as in \eqref{eq:Pi_1_building_blocks}:
\begin{align}
 \begin{split}
  \ell_1\cdot x ={}& -\sqrt{1+R^2}\,\cosh(\tau_1 -\tau) + R\,(\mathbf{n_1}\cdot\mathbf{n}) \approx -\frac{e^{|\tau_1-\tau|}}{2}\sqrt{1+R^2} \ ; \\
  m_1^\mu ={}& \sqrt{1+R^2}\,\big(0,\lambda_1^\tau,\boldsymbol{\lambda_1}\cosh(\tau_1-\tau)- \lambda_1^\tau\mathbf{n_1}\sinh(\tau_1-\tau) \big) \\
    &+ R\,\Big( (\boldsymbol{\lambda_1}\cdot\mathbf{n}) \big(\cosh(\tau_1-\tau),\sinh(\tau_1-\tau),\mathbf{n_1} \big) \\
    &\qquad - (\mathbf{n_1}\cdot\mathbf{n}) \big(\lambda_1^\tau\sinh(\tau_1-\tau), \lambda_1^\tau\cosh(\tau_1-\tau), \boldsymbol{\lambda_1} \big) \Big) \\
   ={}& O\big(e^{|\tau_1-\tau_2|}\big) \ .
 \end{split}
\end{align}
Therefore, the $\bbR^{1,4}$ components of $\Pi_1$ scale as $O\big(e^{-(s_1+1)|\tau_1-\tau|}\big)$, and likewise for $\Pi_2$. As a result, similarly to \eqref{eq:Pi_product}, the Sleight-Taronna diagram vanishes as $O\big(e^{-|\tau_1-\tau_2|}\big)$ at large time separation. Together with \eqref{eq:large_tau_residue}, this implies that the $S_{s_1,s_2}[V_{\text{new,TT}};\Pi_1,\Pi_2,\gamma_3]$ contribution to the correlator \eqref{eq:jjO} also vanishes as $O\big(e^{-|\tau_1-\tau_2|}\big)$, i.e. that $V_{\text{new},TT}$ satisfies our time locality criterion.

\subsection{$V_{\text{new}}$ beyond transverse-traceless gauge} \label{sec:locality:gauge}

Let's now consider shifting the boundary-bulk propagators $\Pi_i$ ($i=1,2$) by traceless pure-gauge fields $\tilde h_i$ subject to the asymptotic condition \eqref{eq:tilde_h_asymptotics}. The field-field-worldline vertex $V_{\text{new,TT}}$ from \eqref{eq:jjO} must then be generalized into the vertex $V_{\text{new}}$ from \eqref{eq:jjO_shifted}. Let us discuss the necessary corrections $V_{\text{new}}-V_{\text{new,TT}}$ to the vertex, and show that they preserve the locality properties established above for $V_{\text{new,TT}}$. Following section \ref{sec:free_invariance:boundary_terms}, we denote the gauge parameters corresponding to $\tilde h_i$ as $\Lambda_i$, recalling that these can be chosen so that their components in an orthonormal Poincare basis vanish asymptotically as \eqref{eq:Lambda_scaling}. 

We now proceed in two steps. First, we will show that under the gauge shift $\Pi_i\rightarrow\Pi_i+\tilde h_i$, the variation of the bulk diagrams in \eqref{eq:jjO} is a local functional of the fields $\Pi_i$ and gauge parameters $\Lambda_i$ in the vicinity of the worldline $\gamma_3$. Second, we'll show that this variation can be subsumed into a local vertex correction $V_{\text{new}}-V_{\text{new,TT}}$.

\subsubsection{Gauge variation of uncorrected bulk diagrams}

Let's now go over the bulk diagrams \eqref{eq:jjO}, and discuss their variation under the gauge shift $\Pi_i\rightarrow\Pi_i+\tilde h_i$. For the $V_{\text{new,TT}}$ diagram, we already established the ansatz \eqref{eq:V_TT_ansatz}, and argued that it's local on $\gamma_3$. Thus, $S_{s_1,s_2}[V_{\text{new,TT}};\Pi_1,\Pi_2,\gamma_3]$ is a local functional of $\Pi_i$ on $\gamma_3$, and similarly $S_{s_1,s_2}[V_{\text{new,TT}};\Pi_1+\tilde h_1,\Pi_2+\tilde h_2,\gamma_3]$ is a local functional of $\Pi_i+\tilde h_i$ on $\gamma_3$. Therefore, the difference between the two is also a local functional of $\Pi_i$ and $\tilde h_i$ on $\gamma_3$. 

The double-exchange diagram $S_{s_1}[\Pi_1,\gamma_3]\,S_{s_2}[\Pi_2,\gamma_3]$ is not affected by the gauge shift at all. Indeed, the effect of a gauge transformation on the field-worldline action \eqref{eq:quadratic_bulk} consists of evaluating the gauge parameter at the worldline's endpoints, its indices contracted with the worldline's unit tangent \cite{Lysov:2022zlw}:
\begin{align}
S_{s_i}[\tilde h_i,\gamma_3] = -4(i\sqrt{2})^{s_i} s_i!\left.\Lambda_i^{(s_i)}\big(x(\tau;\ell_3,\ell'_3),\dot x(\tau;\ell_3,\ell'_3)\big)\right|_{\tau=-\infty}^\infty \ . \label{eq:quadratic_action_gauge_transformation}
\end{align}
For each of the endpoints, we can choose a Poincare frame such that $\dot x^\mu$ becomes the unit vector $e_0^\mu$ in the $z$ direction at $z\rightarrow 0$. The scaling \eqref{eq:Lambda_scaling} of $\Lambda_i$ then tells us that the gauge transformation \eqref{eq:quadratic_action_gauge_transformation} indeed vanishes.

Finally, the Sleight-Taronna diagram $\sum_{s_3}S_{s_1,s_2,s_3}[V_{\text{ST}};\Pi_1,\Pi_2,\gamma_3]$ \emph{will} be affected by the gauge shift, but in a controlled way. In section \ref{sec:free_invariance}, we showed that $V_{\text{ST}}$ is invariant under such gauge transformations, but that was in the absence of a $\gamma_3$ worldline carrying Fronsdal curvature. Thus, in the present setup, $V_{\text{ST}}$ will get a gauge variation proportional to the Fronsdal curvatures $\calF\phi^{(s_3)}_3$, i.e. \emph{localized on $\gamma_3$}. The local nature of this gauge variation is somewhat disrupted by the sum over $s_3$. However, we can show that it remains local within $\sim 1$ AdS radius. Indeed, the potential source of non-locality is in derivatives of $\Pi_i$ or $\Lambda_i$ contracted with the indices of $\calF\phi^{(s_3)}_3$. The question is then how the coefficients of such derivatives behave with increasing spin. The spin-dependence \eqref{eq:Fronsdal_phi} of $\calF\phi^{(s_3)}_3$ itself is $\sim (\sqrt{2})^{s_3}$, while the coupling constant in \eqref{eq:V_ST} goes as $\sim (\sqrt{2})^{s_3}/\Gamma(s_1+s_2+s_3) = O\big((\sqrt{2})^{s_3}/s_3!\big)$ (remembering that gauge transformations require $s_1$ or $s_2$ to be greater than 0). Thus, derivatives of order $s_3$ come with $O(2^{s_3}/s_3!)$ coefficients. This is a special case $a=2$ of the scaling $a^{s_3}/s_3!$, which governs the Taylor expansion $\sum_n\frac{a^n}{n!}\nabla^n$ of a shift by distance $a$. Therefore, the point at which $\Pi_i$ or $\Lambda_i$ are evaluated is effectively shifted by $O(1)$ AdS radii, as desired.

\subsubsection{Locality of the vertex corrections}

So far, we established that the gauge shift $\Pi_i\rightarrow\Pi_i+\tilde h_i$ induces variations in the bulk diagrams of \eqref{eq:jjO} that are local, in the sense that they involve the fields $\Pi_1,\Pi_2$ and gauge parameters $\Lambda_1,\Lambda_2$ within $\sim 1$ AdS radius of each other and of the worldline $\gamma_3$. What remains is to show that these variations can be incorporated as new local terms in the vertex $V_{\text{new}}$, which only sees the fields $\Pi_i+\tilde h_i$ and not the gauge parameters $\Lambda_i$. To do this, we can follow the same logic as with ordinary cubic vertices: we'll first show that the variation strictly vanishes for transverse-traceless $\tilde h_i$, and then conclude that in the general traceless case, it's local not only in $\Lambda_i$, but in the $\tilde h_i$ themselves.

We thus begin by considering \emph{transverse-traceless} pure-gauge fields $\tilde h_i$ (for this purpose, we lift the asymptotic condition \eqref{eq:tilde_h_asymptotics}, which would have forced such fields to vanish). For such pure-gauge fields, the asymptotic value $\lim_{z\rightarrow 0} z^{s_i-2}[\tilde h_i]_{0,s_i}$ defines a pure-gauge field on the boundary, derived from the gauge parameter $\lim_{z\rightarrow 0} z^{s_i-1}[\Lambda_i]_{0,s_i-1}$. The shifted bulk fields $\Pi_i+\tilde h_i$ in this setup remain in the space spanned by boundary-bulk propagators $\Pi^{(s_i)}$, with coefficients shifted by this boundary gauge transformation. We assume that the boundary gauge shift $\lim_{z\rightarrow 0} z^{s_1-2}[\tilde h_1]_{0,s_1}$ vanishes at the points $\ell_2,\ell_3,\ell'_3$, and likewise for $1\leftrightarrow 2$. Such a gauge shift leaves us within the domain of applicability of eq. \eqref{eq:jjO}, with the CFT correlator unchanged. Therefore, the gauge variation of the sum of bulk diagrams in this case must also vanish. Since we already established that this gauge variation is \emph{local}, we conclude that it vanishes for \emph{any} transverse-traceless shift $\tilde h_i$, regardless of its asymptotic behavior.

The upshot of the preceding paragraph is that, in our original context of traceless shifts $\tilde h_i$ subject to the asymptotic condition \eqref{eq:tilde_h_asymptotics}, the gauge variation of the bulk diagrams must be proportional to the \emph{deviation} \eqref{eq:Lambda_field_equation_2} from transverse-traceless gauge. This makes the gauge variation local not only in $\Lambda_i$, but in the fields $\tilde h_i$ themselves, specifically through their divergences $(\del_u\cdot\nabla)\tilde h_i$. This variation can then be canceled by adding to the vertex $V_{\text{new,TT}}$ corrections proportional to $(\del_u\cdot\nabla)\tilde h_i$. In this way, we are able to construct a local vertex $V_{\text{new}}$ that satisfies the correlator formula \eqref{eq:jjO_shifted} in the more general gauge defined by $\Pi_i+\tilde h_i$.

\subsection{Stitching together the correlator of three bilocals} \label{sec:locality:general}

We are now ready to graduate from the bilocal-local-local correlator $\left< j^{(s_1)}_1 j^{(s_2)}_2 \calO_3 \right>$ to the general correlator $\left<\calO^+_1\calO^+_2\calO^+_3\right>$ of three (even-spin) bilocals. Our claim is that this can be expressed in the bulk as a straightforward sum of interactions between the three DV fields $\phi_i$ and their worldlines $\gamma_i$ ($i=1,2,3$), constructed from the same building blocks that we established in \eqref{eq:jjO}-\eqref{eq:jjO_shifted} (see figure \ref{fig:OOO} for the corresponding diagrams):
\begin{align}
 &\langle\calO^+_1\calO^+_2\calO^+_3\rangle = -N\bigg(\sum_{s_1,s_2,s_3}S_{s_1,s_2,s_3}[V_{\text{ST}};\phi_1,\phi_2,\phi_3] \label{eq:OOO} \\
   &\quad - \sum_{s_1,s_2}S_{s_1}[\phi_1,\gamma_3]\,S_{s_2}[\phi_2,\gamma_3] - \sum_{s_2,s_3}S_{s_2}[\phi_2,\gamma_1]\,S_{s_3}[\phi_3,\gamma_1] - \sum_{s_3,s_1}S_{s_3}[\phi_3,\gamma_2]\,S_{s_1}[\phi_1,\gamma_2] \nonumber \\
   &\quad + \sum_{s_1,s_2}S_{s_1,s_2}[V_{\text{new}};\phi_1,\phi_2,\gamma_3] + \sum_{s_2,s_3}S_{s_2,s_3}[V_{\text{new}};\phi_2,\phi_3,\gamma_1] +\sum_{s_3,s_1} S_{s_3,s_1}[V_{\text{new}};\phi_3,\phi_1,\gamma_2] \bigg) \ . \nonumber
\end{align}
Similarly, we claim that local-bilocal-bilocal correlators are given by:
\begin{align}
 \begin{split}
   \left< j^{(s_1)}_1\calO^+_2\calO^+_3\right> = -N\bigg(&\sum_{s_2,s_3}S_{s_1,s_2,s_3}[V_{\text{ST}};\Pi_1,\phi_2,\phi_3] \\
     &- S_{s_1}[\Pi_1,\gamma_3]\sum_{s_2}S_{s_2}[\phi_2,\gamma_3] - S_{s_1}[\Pi_1,\gamma_2]\sum_{s_3}S_{s_3}[\phi_3,\gamma_2] \\
     &+ \sum_{s_2}S_{s_1,s_2}[V_{\text{new}};\Pi_1,\phi_2,\gamma_3] + \sum_{s_3} S_{s_3}[V_{\text{new}};\Pi_1,\phi_3,\gamma_2] \bigg) \ .
 \end{split} \label{eq:jOO}
\end{align}
We will focus below on the more general case \eqref{eq:OOO}; the arguments can be adapted trivially to \eqref{eq:jOO} as well.

To demonstrate the relation \eqref{eq:OOO}, we divide the bulk into regions, much like we did in section \ref{sec:locality:radial:verifying}; see figure \ref{fig:stitching}. Each region contains one of the geodesics $\gamma_i$. We denote the regions as $\Omega_i$, and their asymptotic boundaries as $B_i$. By the same logic as in section \ref{sec:locality:radial:verifying}, the DV field $\phi_1$ in the regions $\Omega_2,\Omega_3$ can be expressed as a superposition of boundary-bulk propagators $\Pi_1$ with boundary sources on $B_1$, shifted by a pure-gauge field $\tilde h_1$ that satisfies the asymptotic condition \eqref{eq:tilde_h_asymptotics}. Again as in section \ref{sec:locality:radial:verifying}, we can continue this expression for $\phi_1$ back into region $\Omega_1$, making a bulk field $\hat\phi_1$ which is everywhere a superposition of $\Pi_1$'s gauge-shifted by $\tilde h_1$, which agrees with $\phi_1$ in $\Omega_1,\Omega_2$, and whose boundary data is a superposition of local currents on $B_1$ that have the same correlators with anything supported on $B_2,B_3$ as the original bilocal $\calO_1$. In the same way, we can construct source-free fields $\hat\phi_2,\hat\phi_3$ out of the other DV fields $\phi_2,\phi_3$. We can then use the already established $\langle jjj\rangle$ and $\langle jj\calO\rangle$ formulas \eqref{eq:shifted_correlator},\eqref{eq:jjO_shifted} to write $\langle\calO_1\calO_2\calO_3\rangle$ in four different ways:
\begin{figure}%
	\centering%
	\includegraphics[scale=0.6]{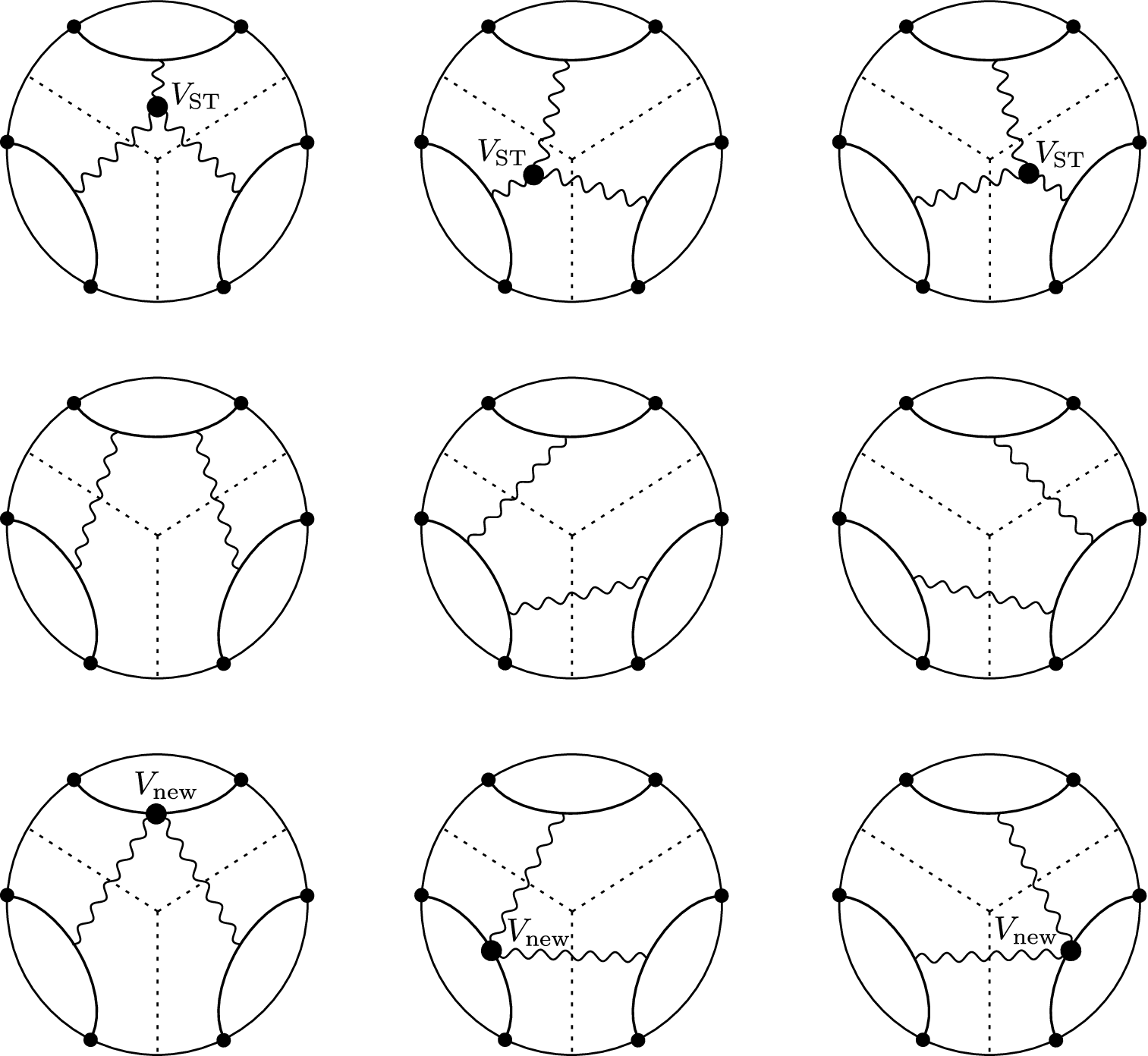} \\
	\caption{The different terms in our procedure to ``stitch together'' the correlator of three bilocals out of bilocal-local-local correlators. In the regions that do not contain its worldline, each DV field can be viewed as a gauge-transformed superposition of boundary-bulk propagators.}
	\label{fig:stitching} 
\end{figure}%
\begin{align}
 &\langle\calO^+_1\calO^+_2\calO^+_3\rangle = -N\sum_{s_1,s_2,s_3}S_{s_1,s_2,s_3}[V_{\text{ST}};\hat\phi_1,\hat\phi_2,\hat\phi_3] \label{eq:OOO_rearrangements} \\
 &= -N\left(\sum_{s_1,s_2,s_3}S_{s_1,s_2,s_3}[V_{\text{ST}};\hat\phi_1,\hat\phi_2,\phi_3] + \sum_{s_1,s_2}\left(S_{s_1,s_2}[V_{\text{new}};\hat\phi_1,\hat\phi_2,\gamma_3] - S_{s_1}[\hat\phi_1,\gamma_3]\,S_{s_2}[\hat\phi_2,\gamma_3] \right) \right) \nonumber \\
 &= -N\left(\sum_{s_1,s_2,s_3}S_{s_1,s_2,s_3}[V_{\text{ST}};\phi_1,\hat\phi_2,\hat\phi_3] + \sum_{s_2,s_3}\left(S_{s_2,s_3}[V_{\text{new}};\hat\phi_2,\hat\phi_3,\gamma_1] - S_{s_2}[\hat\phi_2,\gamma_1]\,S_{s_3}[\hat\phi_3,\gamma_1] \right) \right) \nonumber \\
 &= -N\left(\sum_{s_1,s_2,s_3}S_{s_1,s_2,s_3}[V_{\text{ST}};\hat\phi_1,\phi_2,\hat\phi_3] + \sum_{s_3,s_1}\left(S_{s_3,s_1}[V_{\text{new}};\hat\phi_3,\hat\phi_1,\gamma_2] - S_{s_3}[\hat\phi_3,\gamma_2]\,S_{s_1}[\hat\phi_1,\gamma_2] \right) \right) \ . \nonumber
\end{align}
Now, recall that the $V_{\text{ST}}$ diagrams consist of a standard local integral \eqref{eq:vertex_integral} over $EAdS_4$, which can be decomposed into a sum of integrals over the regions $\Omega_i$. We can then use the fact that $\phi_i$ and $\hat\phi_i$ are equal outside of $\Omega_i$ to write e.g. $S_{s_1,s_2,s_3}[V_{\text{ST}};\hat\phi_1,\hat\phi_2,\phi_3]$ as the $\Omega_1\cup\Omega_2$ portion of $S_{s_1,s_2,s_3}[V_{\text{ST}};\hat\phi_1,\hat\phi_2,\hat\phi_3]$, plus the $\Omega_3$ portion of $S_{s_1,s_2,s_3}[V_{\text{ST}};\phi_1,\phi_2,\phi_3]$. Similarly, we can replace e.g. $S_{s_1,s_2}[V_{\text{new}};\hat\phi_1,\hat\phi_2,\gamma_3]$ with $S_{s_1,s_2}[V_{\text{new}};\phi_1,\phi_2,\gamma_3]$, and $S_{s_1}[\hat\phi_1,\gamma_3]$ with $S_{s_1}[\phi_1,\gamma_3]$, since the value and derivatives of $\hat\phi_1,\hat\phi_2$ on $\gamma_3$ are the same as those of $\phi_1,\phi_2$. With these substitutions, when we add the last three lines of \eqref{eq:OOO_rearrangements} and subtract twice the first line, we obtain the desired formula \eqref{eq:OOO}.

There remain two subtleties worth addressing. First, are e.g. $\phi_1$ and $\hat\phi_1$ \emph{really} interchangeable inside $S_{s_1,s_2}[V_{\text{new}};\phi_1,\phi_2,\gamma_3]$, or in the $\Omega_2\cup\Omega_3$ portion of $S_{s_1,s_2,s_3}[V_{\text{ST}};\phi_1,\phi_2,\phi_3]$? One may worry that the answer is sensitive to the order of operations. For each spin, and at each order in derivatives, $\phi_1$ and $\hat\phi_1$ are indeed the same within $\Omega_2\cup\Omega_3$, and in particular on the worldlines $\gamma_2,\gamma_3$. Thus, if we evaluate the derivatives \emph{before} performing the sums over spins and angular momenta in \eqref{eq:V_TT_ansatz} and \eqref{eq:OOO}-\eqref{eq:OOO_rearrangements}, our logic will hold. But what if we perform the sums first? Might they lead to $\hat\phi_1$ being effectively evaluated inside $\Omega_1$, and thus ``noticing'' its difference from the original $\phi_1$? This seems especially pertinent given that the worldlines $\gamma_i$ can lie arbitrarily close to each other, and thus to the boundaries between the regions $\Omega_i$. 

Our claim is that such a problem will not occur. This is because our construction only involves infinite towers of \emph{traceless} derivatives:
\begin{itemize}
	\item In the Sleight-Taronna vertex $V_{\text{ST}}$ from \eqref{eq:V_ST}, the derivatives are traceless in $EAdS_4$, since their indices are always contracted with traceless HS fields.
	\item In the new vertex $V_{\text{new,TT}}$ in transverse-traceless gauge, our ansatz \eqref{eq:geodesic_data_even}-\eqref{eq:V_TT_ansatz} implies that all derivatives are traceless in the 3d space transverse to the worldline.
	\item The vertex corrections $V_{\text{new}}-V_{\text{new,TT}}$ from section \ref{sec:locality:gauge} are constructed from gauge variations of $V_{\text{ST}}$ and $V_{\text{new,TT}}$, and thus also inherit their traceless-derivatives structure.
\end{itemize} 
Now, towers of traceless derivatives can define initial data for a source-free field, and they can generate translations along lightrays in Lorentzian signature. However, they \emph{can't} generate translations over any finite distance in Euclidean signature, nor can they ``tell the difference'' between a field that's source-free everywhere and one that is merely source-free in a finite neighborhood. Thus, we are safe from the vertices ``noticing'' the difference between e.g. $\phi_1$ and $\hat\phi_1$ outside of $\Omega_1$.

The last subtlety we'd like to address is the effect of the sums over spins in \eqref{eq:OOO} on the locality of the bulk diagrams. For fixed spins, we already know that $V_{\text{ST}}$ is strictly local, and that $V_{\text{new}}$ is non-local at most within $\sim 1$ AdS radius. How do the sums over spins affect these properties? For $V_{\text{ST}}$, the sum over spins introduces an infinite tower of derivatives, which indeed leads to some non-locality (we'll see this explicitly on an example in section \ref{sec:numeric}). However, we'll now argue that this non-locality is confined within $\sim 1$ AdS radius. This stems from a series of observations:
\begin{itemize}
	\item The powers of derivatives in the vertex formula \eqref{eq:V_ST} are just the spins $s_i$ themselves.
	\item Their coefficients are the coupling constants in \eqref{eq:V_ST}. As one or more spins grow large, these scale as $\sim \frac{(\sqrt{2})^{s_1+s_2+s_3}}{(s_1+s_2+s_3-1)!} = O\Big(\frac{(\sqrt{2})^{s_1+s_2+s_3}}{(s_1-1)!(s_2-1)!(s_3-1)!}\Big)$.
	\item The derivatives in \eqref{eq:V_ST} are contracted with DV fields \eqref{eq:phi}, whose spin-dependence amounts to an extra factor of $\sqrt{2}\,k^\mu$ whenever the spin is raised by 1. 
	\item The real and imaginary parts of $k^\mu = \frac{1}{2}(t^\mu + ir^\mu/R)$ both have norms between $0$ and $\frac{1}{2}$: $t^\mu$ goes from unit norm on the worldline to zero norm an infinity, while $r^\mu/R$ does the reverse. 
\end{itemize}
Overall, we see that the tower of derivatives in the $V_{\text{ST}}$ diagram is bounded by a product of Taylor series of the form $\prod_i\sum_{s_i} \frac{2^{s_i}}{(s_i-1)!}(k_i\cdot\nabla)^{s_i}$, which (up to a shift by one derivative) describe translations by the vectors $2k_i^\mu$, whose real and imaginary parts have norm bounded by $1$. The non-locality is therefore indeed confined to $\sim 1$ AdS radius.

Finally, for $V_{\text{new}}$, our claim is that the sum over spins in \eqref{eq:OOO} does not extend its non-locality beyond $\sim 1$ AdS radius. To see this, one can rerun the locality arguments from sections \ref{sec:locality:radial}-\ref{sec:locality:time}, with boundary bilocals $\calO^+_1,\calO^+_2$ in place of the currents $j_1^{(s_1)},j_2^{(s_2)}$, and with DV fields $\phi_1,\phi_2$ (involving all even spins) in place of the boundary-bulk propagators $\Pi_1,\Pi_2$.

\section{Example: locality in the (0,0,bilocal) correlator} \label{sec:numeric}

In this section, we perform a (partially numerical) study of the $\left< j_1^{(0)} j_2^{(0)}\calO_3 \right>$ correlator, between two spin-0 boundary ``currents'' and one bilocal. This will serve as a concrete example for several of the features discussed in section \ref{sec:locality}. 

\subsection{Bulk scalar modes}

As in section \ref{sec:locality:time}, we fix the bilocal's endpoints at $\ell_3^\mu = (\frac{1}{2},\frac{1}{2},\vec 0)$ and $\ell'^\mu_3 = (\frac{1}{2},-\frac{1}{2},\vec 0)$, and use these to induce a coordinate system \eqref{eq:x_cylindric}-\eqref{eq:ell_cylindric} on the bulk and boundary. We then use these coordinates' $\bbR\times SO(3)$ symmetry to arrange the scalar fields $h_1,h_2$ into modes with ``time'' frequency $\omega$ and angular momentum numbers $l,m$. Since the bilocal is invariant under the $\bbR\times SO(3)$, we can only have coupling between modes of $h_1,h_2$ with equal $l$, and equal \& opposite $\omega$ and $m$. Moreover, by $SO(3)$ symmetry, it's sufficient to study the $m=0$ modes. Thus, we are interested in modes of the form:
\begin{align}
 h_{\omega,l}(x) = e^{i\omega\tau}\psi_{\omega,l}(R)P_l(\mathbf{n\cdot n_0}) \ , \label{eq:modes}
\end{align}
where $P_l$ is a Legendre polynomial, and $\mathbf{n_0}$ is some fixed 3d unit vector. The modes' radial dependence $\psi_{\omega,l}(R)$ is found by solving the field equation $(\nabla\cdot\nabla + 2)h^{(0)} = 0$ in $EAdS_4$. This can be simplified by using the equation's conformal invariance, and the conformal relation between $EAdS_4$ and $\bbR\times(\text{half-}S_3)$:
\begin{align}
 dx\cdot dx = (1+R^2) d\tau^2 + \frac{dR^2}{1+R^2} + R^2 d\Omega^2 = (1+R^2)\left(d\tau^2 + d\alpha^2 + \sin^2\alpha\,d\Omega^2 \right) \ ,
\end{align}
where $d\Omega^2$ is the 2-sphere metric. The $S_3$ angle $\alpha$ is defined as $\alpha \equiv \arctan R$, and the asymptotic boundary $R=\infty$ becomes the $S_3$ equator $\alpha = \frac{\pi}{2}$. The problem now reduces to solving the Laplacian equation $(\nabla\cdot\nabla - \omega^2 - 1)\hat\psi = 0$ on the half-$S_3$. The solution that is regular at $R=0$ (i.e. at $\alpha=0$) is an $S_3$ spherical harmonic (see e.g. \cite{HyperSphericalHarmonics}) with complex angular momentum number (this is allowed because our $S_3$ doesn't continue beyond $\alpha=\frac{\pi}{2}$):
\begin{align}
 \hat\psi = \frac{1}{\sqrt{\sin\alpha}}\,P_{-\frac{1}{2}+i\omega}^{-\frac{1}{2}-l}(\cos\alpha)\,P_l(\mathbf{n\cdot n_0}) \ ,
\end{align}
where $P_l^m$ is the associated Legendre function. Note that despite the appearance of a complex parameter, $P_{-\frac{1}{2}+i\omega}^{-\frac{1}{2}-l} = P_{-\frac{1}{2}-i\omega}^{-\frac{1}{2}-l}$ is a real function. Converting back from $\alpha$ to $R$, and multiplying by the conformal factor $\frac{1}{\sqrt{1+R^2}}$, we obtain the radial dependence of our modes \eqref{eq:modes} as:
\begin{align}
 \psi_{\omega,l}(R) = \frac{1}{\sqrt{R\sqrt{1+R^2}}}\,P_{-\frac{1}{2}+i\omega}^{-\frac{1}{2}-l}\!\left(\frac{1}{\sqrt{1+R^2}}\right) \ . \label{eq:psi}
\end{align}
In the asymptotic analysis of the modes \eqref{eq:modes}, we can use $R^{-1}$ as the holographic coordinate $z$. Thus, the asymptotic data of the modes \eqref{eq:modes} with weights $\Delta=1,2$ can be extracted as the coefficients of $R^{-1}$ and $R^{-2}$ respectively in the boundary limit $x^\mu(\tau,R,\mathbf{n}) \rightarrow R\,\ell^\mu(\tau,\mathbf{n})$ at $R\rightarrow\infty$:
\begin{align}
 &h_{\omega,l}(x) \ \underset{x^\mu\to R\ell^\mu}{\longrightarrow} \ \frac{\varphi_{\omega,l}(\ell)}{R} + \frac{\pi_{\omega,l}(\ell)}{R^2} + O\!\left(\frac{1}{R^3}\right) \ ; \\
 &\varphi_{\omega,l}(\ell) = e^{i\omega\tau}P_l(\mathbf{n\cdot n_0})P_{-\frac{1}{2}+i\omega}^{-\frac{1}{2}-l}(0) \ ; \label{eq:weight_1_data} \\
 &\pi_{\omega,l}(\ell) = e^{i\omega\tau}P_l(\mathbf{n\cdot n_0})\left(P_{-\frac{1}{2}+i\omega}^{-\frac{1}{2}-l}\right)'(0) = -e^{i\omega\tau}P_l(\cos\theta)P_{-\frac{1}{2}+i\omega}^{\frac{1}{2}-l}(0) \ , \label{eq:weight_2_data}
\end{align}
where the value and derivative of the Legendre functions at zero can be found in e.g. \cite{LegendreSpecialValues}. The $\Delta=2$ boundary data \eqref{eq:weight_2_data} can be used to decompose our modes \eqref{eq:modes} in terms of the boundary-bulk propagators $\Pi^{(0)}(x;\ell)$ \eqref{eq:Pi}, whose own boundary data reads (see e.g. \cite{David:2020fea}):
\begin{align}
 \Pi^{(0)}(x;\hat\ell) = -\frac{1}{16\pi^2(x\cdot\hat\ell)} \ \underset{x^\mu\to R\ell^\mu}{\longrightarrow} \ -\frac{1}{16\pi^2 (\ell\cdot\hat\ell)R} - \frac{\delta^3(\ell,\hat\ell)}{4R^2} + O\!\left(\frac{1}{R^3}\right) \ . \label{eq:Pi_0_boundary_data}
\end{align}
Comparing \eqref{eq:weight_2_data} with \eqref{eq:Pi_0_boundary_data} and denoting the boundary coordinates of $\hat\ell$ as $(\hat\tau,\mathbf{\hat n})$, we get the decomposition:
\begin{align}
 h_{\omega,l}(x) = 4P_{-\frac{1}{2}+i\omega}^{\frac{1}{2}-l}(0) \int d\hat\tau\,e^{i\omega\hat\tau} \int d^2\mathbf{\hat n}\,P_l(\mathbf{\hat n\cdot n_0})\,\Pi^{(0)}(x;\hat\tau,\mathbf{\hat n}) \ . 
\end{align}
From this, we read off the boundary dual of the bulk modes \eqref{eq:modes} as a superposition of spin-0 ``currents'':
\begin{align}
 j^{(0)}_{\omega,l} = 4P_{-\frac{1}{2}+i\omega}^{\frac{1}{2}-l}(0) \int d\tau\,e^{i\omega\tau} \int d^2\mathbf{n}\,P_l(\mathbf{n\cdot n_0})\,j^{(0)}(\tau,\mathbf{n}) \label{eq:j_superposition}
\end{align}

\subsection{Ingredients of the correlator}

We are now ready to plug the above $(\omega,l)$ modes into the correlator formula \eqref{eq:jjO}. On the boundary side, this describes a correlator between $\calO(\ell_3,\ell'_3)\equiv\calO_3$ and two spin-0 operators of the form \eqref{eq:j_superposition}, i.e. $j^{(0)}_{\omega,l}$ and $j^{(0)}_{-\omega,l}$. In coordinate space, the CFT correlator \eqref{eq:correlators_bilocal} for this case reads:
\begin{align}
 \begin{split}
   \left<j^{(0)}(\ell_1) j^{(0)}(\ell_2) \calO(\ell_3,\ell'_3) \right> &= \frac{NG(\ell_1,\ell_2)}{G(\ell_3,\ell'_3)}\big(G(\ell_1,\ell_3)G(\ell_2,\ell'_3) + G(\ell_2,\ell_3)G(\ell_1,\ell'_3) \big) \\
     &= \frac{NG(\ell_1,\ell_2)}{2\pi}\cosh\frac{\tau_1-\tau_2}{2} \ . \label{eq:CFT_correlator_raw}
 \end{split}
\end{align}
In frequency space, the CFT propagator $G(\ell_1,\ell_2)$ becomes just the inverse of minus the conformal Laplacian:
\begin{align}
 G(\ell_1,\ell_2) = -\frac{1}{\Box_\ell} \longrightarrow \frac{1}{\omega^2 + (l+\frac{1}{2})^2} \ , \label{eq:G_omega_l}
\end{align}
while the factor of $\cosh\frac{\tau_1-\tau_2}{2}$ becomes a frequency shift $\omega\rightarrow \omega \pm \frac{i}{2}$. Overall, the CFT correlator $\left< j^{(0)}_{\omega,l}\,j^{(0)}_{-\omega,l}\,\calO_3 \right>$ reads:
\begin{align}
 \left< j^{(0)}_{\omega,l}\,j^{(0)}_{-\omega,l}\,\calO_3 \right> = \frac{32N}{2l+1}\left(\int_{-\infty}^\infty d\tau\right)\left(P_{-\frac{1}{2}+i\omega}^{\frac{1}{2}-l}(0)\right)^2 \Re\frac{1}{\omega(\omega+i) + l(l+1)} \ , \label{eq:CFT_correlator}
\end{align}
where the appearance of an infinite $\tau$ integral is a standard expression of ``time'' translation symmetry.

Let us now turn to the bulk side of the correlator formula \eqref{eq:jjO}, where the scalar bulk fields $h_{\omega,l}$ and $h_{\omega,-l}$ are interacting with the DV field $\phi_3$ and its worldline $\gamma_3$. We begin with the Sleight-Taronna diagram $\sum_{s} S_{0,0,s}[V_{\text{ST}};h_{\omega,l},h_{-\omega,l},\phi_3]$. Due to the singular behavior \eqref{eq:V_000} of $V_{\text{ST}}^{(0,0,0)}$, we must treat the cases $s=0$ and $s>0$ separately. The $s=0$ diagram can be evaluated using the $\Delta=1$ boundary data of the modes $h_{\pm\omega,l}$ and of the DV field $\phi_3^{(0)}$. The former is given by \eqref{eq:weight_1_data}, while the latter is just the coefficient of $\frac{1}{R}$ in \eqref{eq:phi}, i.e. $1/\pi$. Plugging these into \eqref{eq:V_000}, we get:
\begin{align}
 -NS_{0,0,0}[V_{\text{ST}};h_{\omega,l},h_{-\omega,l},\phi_3] = \frac{32N}{2l+1}\left(\int_{-\infty}^\infty d\tau\right)\left(P_{-\frac{1}{2}+i\omega}^{-\frac{1}{2}-l}(0)\right)^2 \ . \label{eq:ST_diagram_000}
\end{align}
We now turn to the Sleight-Taronna diagram with $s>0$. The relevant vertex \eqref{eq:V_ST} reads simply:
\begin{align}
  V_{\text{ST}}^{(0,0,s)}(\del_{x_1};\del_{x_2};\del_{u_3}) = \frac{8\!\left(i\sqrt{2}\right)^s}{(s-1)!}\big[(\del_{u_3}\cdot\del_{x_1})^s + (\del_{u_3}\cdot\del_{x_2})^s \big] \ .
\end{align}
Plugging in our scalar modes $h_{\pm\omega,l}(x)$ and the DV field \eqref{eq:phi}, this becomes (keeping in mind that the participating spins $s$ are even):
\begin{align}
 V_{\text{ST}}^{(0,0,s)}h_{\omega,l}\,h_{-\omega,l}\,\phi_3^{(s)} = \frac{16}{\pi R(s-1)!}\big[ h_{-\omega,l} (2k\cdot\del_x)^s h_{\omega,l} + h_{\omega,l} (2k\cdot\del_x)^s h_{-\omega,l} \big] \ , \label{eq:V_ST_00s}
\end{align}
where $k^\mu = k^\mu(x;\ell_3,\ell'_3)$ is the null vector \eqref{eq:k} generated by the $\gamma_3$ geodesic. In our coordinates \eqref{eq:x_cylindric}, the derivative $2k\cdot\del_x$ along $k^\mu$ reads:
\begin{align}
 2k\cdot\del_x = t\cdot\del_x + \frac{i}{R}\,r\cdot\del_x = \frac{1}{1+R^2}\frac{\del}{\del\tau} + i\frac{\del}{\del R} \ . \label{eq:k_derivative}
\end{align}
Since $k^\mu$ is null $k\cdot k = 0$ and affine $(k\cdot\nabla)k = 0$, the line $\{x^\mu + 2ak^\mu | a\in\bbR \}$ is a (complexified) lightray in both $\bbR^{1,4}$ and $EAdS_4$. Explicitly, this lightray takes the form:
\begin{align}
 x^\mu \to x^\mu + 2ak^\mu \ : \quad (\tau,R,\mathbf{n})\to \big(\tau - i\arctan(R+ia) + i\arctan R, R+ia,\mathbf{n} \big) \ . \label{eq:k_line}
\end{align}
Shifting the field $h_{\omega,l}(x)$ along this lightray, we get:
\begin{align}
 h_{\omega,l}(x+2ak) = e^{i\omega\tau} e^{\omega[\arctan(R+ia) - \arctan R]}\,\psi_{\omega,l}(R+ia) P_l(\mathbf{n\cdot n_0}) \ .
\end{align}
In terms of these shifted fields, the $(2k\cdot\del_x)^s$ derivatives in \eqref{eq:V_ST_00s} can be recast as $\frac{d^s}{da^s}$. Integrating the vertex \eqref{eq:V_ST_00s} over $EAdS_4$ with the measure $d^4x = R^2 dR\,d\tau d^2\mathbf{n}$, we get:
\begin{align}
 &{-N}S_{0,0,s}[V_{\text{ST}};h_{\omega,l},h_{-\omega,l},\phi_3] = \frac{128N}{(2l+1)(s-1)!}\left(\int_{-\infty}^\infty d\tau\right) \label{eq:ST_diagram_00s} \\
   &\qquad \times \int_0^\infty RdR\,\psi_{\omega,l}(R)\!\left.\frac{d^s}{da^s}\Big( \cosh\big[\omega(\arctan(R+ia) - \arctan R) \big] \psi_{\omega,l}(R+ia) \Big)\right|_{a=0} \ , \nonumber
\end{align}
where $\psi_{\omega,l}(R)$ is the radial dependence function \eqref{eq:psi}. Summing the diagrams \eqref{eq:ST_diagram_00s} over spin channels $s$, we get essentially a Taylor series, carrying the scalar fields from $a=0$ to $a=\pm 1$ along the complex lightray \eqref{eq:k_line}, i.e. from $x^\mu$ to $x^\mu\pm 2k^\mu$. Explicitly, the sum of \eqref{eq:ST_diagram_00s} over positive even $s$ reads:
\begin{align}
  &{-N}\sum_{\text{even s}>0}S_{0,0,s}[V_{\text{ST}};h_{\omega,l},h_{-\omega,l},\phi_3] = \frac{128N}{2l+1}\left(\int_{-\infty}^\infty d\tau\right) \label{eq:ST_diagram_sum} \\
  &\qquad \times \Re\int_0^\infty RdR\,\psi_{\omega,l}(R)\!\left.\frac{d}{da}\Big( \cosh\big[\omega(\arctan(R+ia) - \arctan R) \big] \psi_{\omega,l}(R+ia) \Big)\right|_{a=1} \ . \nonumber
\end{align}
We see here an example of a feature discussed in section \ref{sec:locality:general}: the sum over spins introduces some non-locality into the Sleight-Taronna diagram, by effectively shifting the fields from one point $x^\mu$ to another $x^\mu\pm 2k^\mu$. However, this non-locality is contained within $\sim 1$ AdS radius, since both the real and imaginary parts of $2k^\mu$ have norms between 0 and 1.

Let us now turn to the other bulk diagrams on the RHS of \eqref{eq:jjO}. The double-exchange diagram $S_{s_1}[\Pi_1,\gamma_3]\,S_{s_2}[\Pi_2,\gamma_3]$ will appear in our setup as a delta-function term $\sim\delta(\omega)$, since it does not depend on the ``time'' difference $\tau_1-\tau_2$ between the boundary source points $\ell_1,\ell_2$. Therefore, this diagram will not contribute at any nonzero frequency $\omega$. This leaves only the $V_{\text{new}}$ diagram, which must therefore account for any difference between the CFT correlator \eqref{eq:CFT_correlator} and the Sleight-Taronna diagrams \eqref{eq:ST_diagram_000},\eqref{eq:ST_diagram_sum}. Eliminating common factors, we can express this relationship as:
\begin{align}
 W_{\text{new}} ={}& W_{\text{CFT}} - W_{\text{ST}}^{(0)} - \sum_{\text{even s}>0}W_{\text{ST}}^{(s)} \ , \label{eq:W_new}
\end{align}
where the known pieces are given by:
\begin{align}
 W_{\text{CFT}} ={}& \left(P_{-\frac{1}{2}+i\omega}^{\frac{1}{2}-l}(0)\right)^2 \Re\frac{1}{\omega(\omega+i) + l(l+1)} \ ; \\
 W_{\text{ST}}^{(0)} ={}& \left(P_{-\frac{1}{2}+i\omega}^{-\frac{1}{2}-l}(0)\right)^2 \ ; \\
 W_{\text{ST}}^{(s)} ={}& 4\int_0^\infty RdR\,\psi_{\omega,l}(R)\left.\frac{d^s}{da^s}\Big( \cosh\big[\omega(\arctan(R+ia) - \arctan R) \big] \psi_{\omega,l}(R+ia) \Big)\right|_{a=0} \nonumber \\
 &\text{(for }s>0\text{)} \ ,
\end{align}
with the sum formula:
\begin{align}
 \begin{split}
   \sum_{\text{even s}>0}W_{\text{ST}}^{(s)} ={}& 4\Re\int_0^\infty RdR\,\psi_{\omega,l}(R) \\
     &\quad \times \left.\frac{d}{da}\Big( \cosh\big[\omega(\arctan(R+ia) - \arctan R) \big] \psi_{\omega,l}(R+ia) \Big)\right|_{a=1} \ .
 \end{split} \label{eq:W_ST_sum}
\end{align}

\subsection{Locality analysis}

Having been brought to the form \eqref{eq:W_new}-\eqref{eq:W_ST_sum}, the bulk and boundary diagrams can now be readily evaluated in Mathematica, for various values of the ``time'' frequency $\omega$ and angular momentum number $l$. In particular, we can examine the behavior of the new vertex's contribution $W_{\text{new}}$, and compare to the locality discussion in section \ref{sec:locality}. We begin with radial locality. By our criterion from section \ref{sec:locality:radial}, this requires $W_{\text{new}}$ to be regular at $\ell_1=\ell_2$. Thus, in frequency space, we expect $W_{\text{new}}$ to decay exponentially at large frequencies. In our present simple context of $\left< j^{(0)}_1 j^{(0)}_2 \calO_3 \right>$ correlators, we can make this expectation more detailed. 

Let us start in coordinate space. For the moment, let's consider the $EAdS_4$ boundary as the 3-sphere $\{\ell^\mu\in\bbR^{1,4}|\ell_\mu \ell^\mu = 0\,;\,\ell^0=1\}$. We then fix $\ell_3,\ell'_3$ at two opposite poles $(1,0,\pm n^a)$, with $n^a$ some 3d unit vector, and set $\ell_1,\ell_2$ nearly coincident at $(1,1,\pm\xi^a/2)$, for some infinitesimal 3d vector $\xi^a$ with norm $|\xi|$. The CFT correlator \eqref{eq:CFT_correlator_raw} then diverges as $G(\ell_1,\ell_2)\sim 1/|\xi|$. In Fourier space at large frequencies, this becomes (c.f. \eqref{eq:G_omega_l}):
\begin{align}
 W_{\text{CFT}} \sim \frac{1}{\Omega^2} \ . \label{eq:W_CFT_scaling}
\end{align}
Here, we introduce $\Omega$ as a generic notation for boundary frequencies, combining $\omega$ and $l$ as $\Omega\approx\sqrt{\omega^2+l^2}$ at large $\omega$ and/or $l$.

Now, let's consider the contributions $W_{\text{ST}}^{(s)}$ from the Sleight-Taronna diagram in various spin channels. To do this, it is helpful to apply a conformal transformation to the boundary 3-sphere, stretching the distance between $\ell_1,\ell_2$ by a factor of $\sim |\xi|^{-1}$ so as to bring them to opposite poles $(1,0,\pm\xi^a/|\xi|)$, while squeezing the distance between $\ell_3,\ell'_3$ and bringing them to $(1,-1,\pm|\xi|n^a)$; see figure \ref{fig:dilatation}. After this conformal transformation, the DV fields $\phi^{(s)}_3$ behave at leading order as spin-$s$ boundary-bulk propagators $\Pi_3^{(s)}$, with prefactors (i.e. boundary polarization tensors) of the form $\sim |\xi|^s(n^{a_1}\!\dots n^{a_s}-\text{traces})$. By rotational invariance, the Sleight-Taronna diagram then takes the form $\sim \frac{\xi^{a_1}}{|\xi|}\!\dots\frac{\xi^{a_s}}{|\xi|}|\xi|^s(n_{a_1}\!\dots n_{a_s}-\text{traces}) = n^{a_1}\!\dots n^{a_s}(\xi_{a_1}\!\dots\xi_{a_s}-\text{traces})$. We can now undo the conformal transformation, picking up a factor of $\sim 1/|\xi|$ due to the combination of weights w.r.t. $\ell_1,\ell_2,\ell_3,\ell'_3$:
\begin{figure}%
	\centering%
	\includegraphics[scale=0.6]{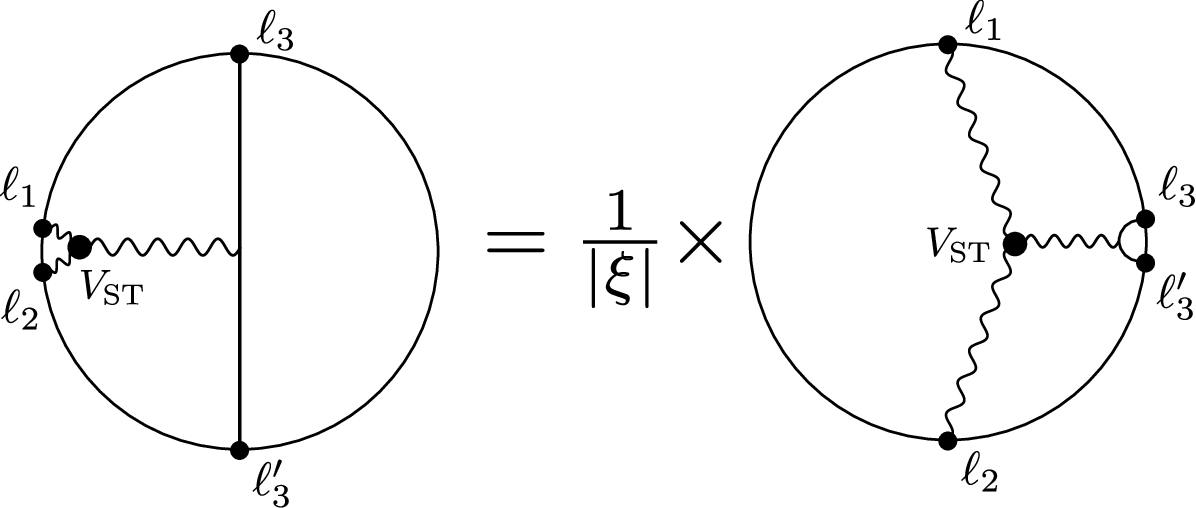} \\
	\caption{A conformal transformation on the boundary that clarifies the behavior of the $\left< j^{(0)}_1 j^{(0)}_2 \calO_3 \right>$ correlator in the $\ell_1=\ell_2$ limit.}
	\label{fig:dilatation} 
\end{figure}%
\begin{figure}%
	\centering%
	\subfigure[{\ $\omega=l+\frac{1}{2}$, $4\leq l\leq 8$. Slope is $-1.997$.}]%
	{\includegraphics[scale=0.85]{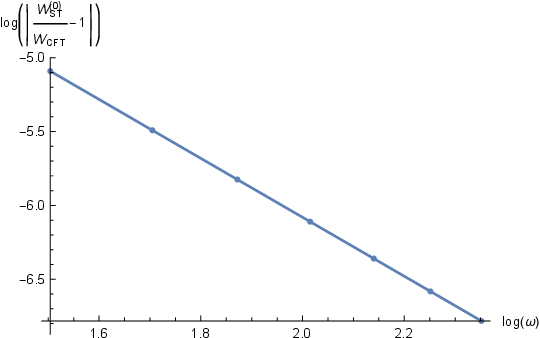}}
	\subfigure[{\ $l=0$, $4\leq\omega\leq 9$. Slope is $-1.99986$.}]%
	{\includegraphics[scale=0.85]{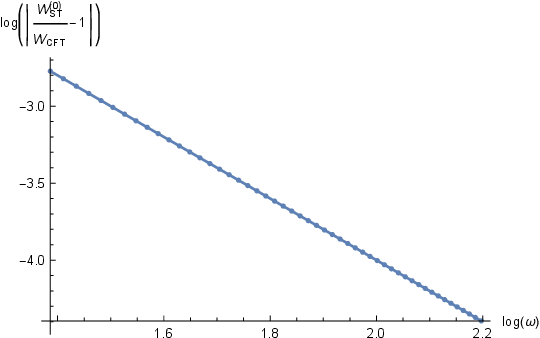}}
	\subfigure[{\ $\omega=0.01$, $4\leq l \leq 10$. Slope is $-1.978$.}]%
	{\includegraphics[scale=0.85]{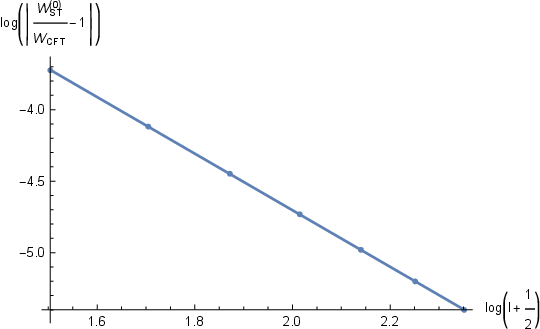}}
	\caption{Numerical log-log plots of the relative discrepancy between the CFT correlator and the spin-0 Sleight-Taronna contributions, at large boundary frequencies $\omega$ and/or $l$. As expected, the discrepancy decays with frequency as $\sim\Omega^{-2}$.}%
	\label{fig:discrep0}%
\end{figure}%
\begin{figure}%
	\centering%
	\subfigure[{\ $\omega=l+\frac{1}{2}$, $4\leq l\leq 10$. Slope is $-4.02$.}]%
	{\includegraphics[scale=0.85]{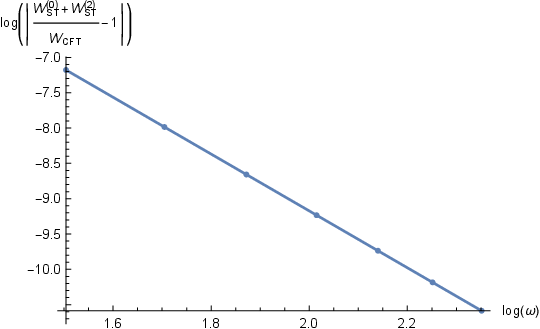}}
	\subfigure[{\ $l=0$, $4\leq\omega\leq 9$. Slope is $-4.25$.}]%
	{\includegraphics[scale=0.85]{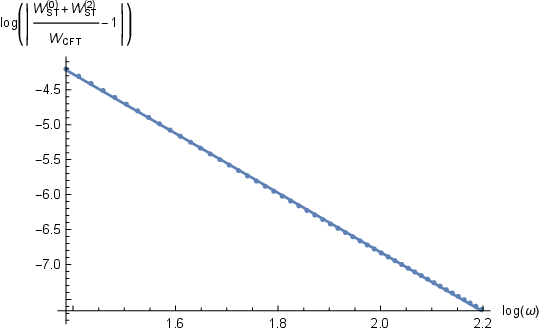}}
	\subfigure[{\ $\omega=0.01$, $4\leq l \leq 10$. Slope is $-3.82$.}]%
	{\includegraphics[scale=0.85]{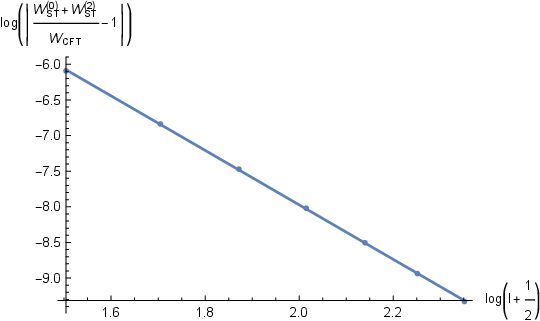}}
	\caption{Numerical log-log plots of the relative discrepancy between the CFT correlator and the (spin-0)+(spin-2) Sleight-Taronna contributions, at large boundary frequencies $\omega$ and/or $l$. As expected, the discrepancy decays with frequency as $\sim\Omega^{-4}$.}%
	\label{fig:discrep2}%
\end{figure}%
\begin{figure}%
	\centering%
	\subfigure[{\ $\omega=l+\frac{1}{2}$, $4\leq l\leq 10$. Slope is $-3.4$.}]%
	{\includegraphics[scale=0.85]{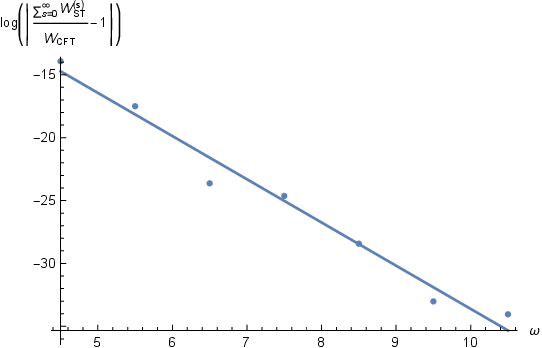}}
	\subfigure[{\ $l=0$, $4\leq\omega\leq 9$. Slope is $-1.55$.}]%
	{\includegraphics[scale=0.85]{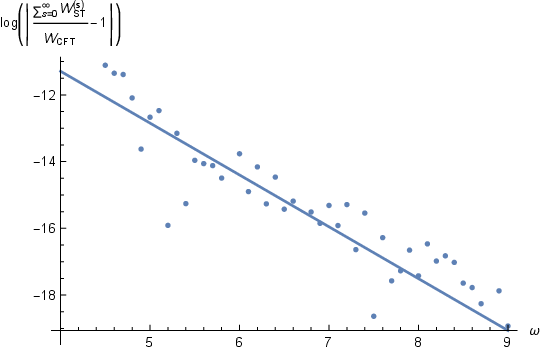}}
	\subfigure[{\ $l=1$, $0.1\leq\omega\leq 9$. Slope is $-2.5$.}]%
	{\includegraphics[scale=0.85]{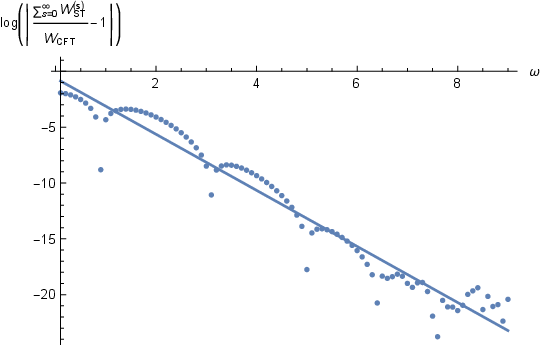}}
	\subfigure[{\ $l=5$, $0.1\leq\omega\leq 9$. Slope is $-1.54$.}]%
	{\includegraphics[scale=0.85]{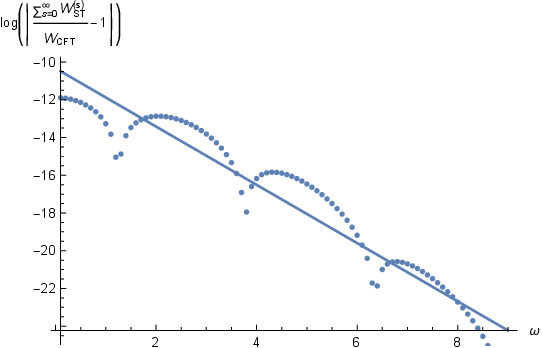}}
	\subfigure[{\ $\omega=0.01$, $4\leq l \leq 10$. Slope is $-2.5$.}]%
	{\includegraphics[scale=0.85]{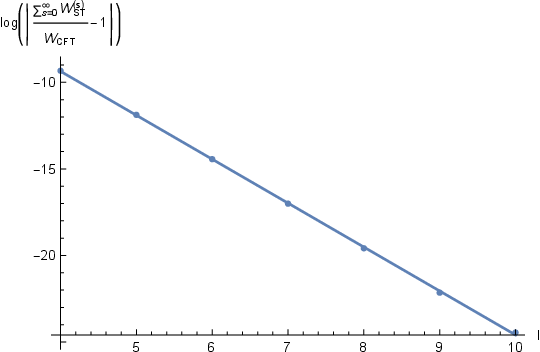}}
	\caption{Numerical log plots of the relative discrepancy between the CFT correlator and the all-spin Sleight-Taronna contributions, at large boundary frequencies $\omega$ and/or $l$. As expected, the discrepancy decays with frequency exponentially. This expresses the radial locality of the new bulk vertex.}%
	\label{fig:discrep}%
\end{figure}%
\begin{align}
 \Delta_1 + \Delta_2 - \Delta_3 - \Delta'_3 = 1 + 1 - \frac{1}{2} - \frac{1}{2} = 1 \ .
\end{align}
Thus, the small-$\xi$ behavior of $W_{\text{ST}}^{(s)}$ in the original conformal frame is $\sim (\xi^{a_1}\dots\xi^{a_s}-\text{traces})/|\xi|$. For $s=0$, this is divergent at $\xi^a=0$; for general spins, the $s$'th derivative with respect to $\xi^a$ is divergent. In frequency space, such singular short-distance behavior translates into power laws at large frequencies:
\begin{align}
 W_{\text{ST}}^{(s)} \sim \frac{1}{\Omega^{s+2}} \ .
\end{align}
Our radial-locality expectation can now be phrased in detail as follows. At large boundary frequencies $\Omega$, the spin-0 Sleight-Taronna diagram $W_{\text{ST}}^{(0)}$ should match the $\sim \Omega^{-2}$ behavior of the CFT correlator \eqref{eq:W_CFT_scaling}, leaving a $\sim \Omega^{-4}$ remainder; this remainder should be matched by the spin-2 diagram $W_{\text{ST}}^{(2)}$, leaving a $\sim \Omega^{-6}$ remainder, which should be matched by $W_{\text{ST}}^{(4)}$, and so forth. When all the spin-channels $W_{\text{ST}}^{(s)}$ have been taken into account, the remaining discrepancy, associated with $W_{\text{new}}$, should decay exponentially as $\sim e^{-\kappa\Omega}$, with some order-1 coefficient $\kappa$. 

These expectations are nicely confirmed by numerics. In fact, the numerics shows that it's sufficient for \emph{either} $\omega$ or $l$ to be large, and that values of $3\sim5$ already behave as ``large''. In figure \ref{fig:discrep0}, we display log-log plots of the relative discrepancy $\frac{W_{\text{ST}}^{(0)}}{W_{\text{CFT}}} - 1$ as a function of frequency, showing a $\sim \Omega^{-2}$ behavior, as predicted above. Similarly, in figure \ref{fig:discrep2}, we display log-log plots of $\frac{W_{\text{ST}}^{(0)} + W_{\text{ST}}^{(2)}}{W_{\text{CFT}}} - 1$, showing that it behaves as $\sim \Omega^{-4}$. Most importantly, in figure \ref{fig:discrep}, we display log plots of $\frac{\sum_s W_{\text{ST}}^{(s)}}{W_{\text{CFT}}} - 1 = -\frac{W_{\text{new}}}{W_{\text{CFT}}}$, showing its exponential decay. The exponential decay is particularly clean when $l$ grows with $\omega$ fixed at a small value, or when $\omega$ and $l$ grow together as $\omega = l+\frac{1}{2}$ (inspired by the boundary Laplacian formula \eqref{eq:G_omega_l}). In other setups, the exponential decay is noisier, sometimes with superposed periodic patterns. The exponent $\kappa$ varies widely between setups, but is always of order 1.

\begin{figure}%
	\centering%
	\subfigure[{\ $l=0$, CFT correlator}]%
	{\includegraphics[scale=0.75]{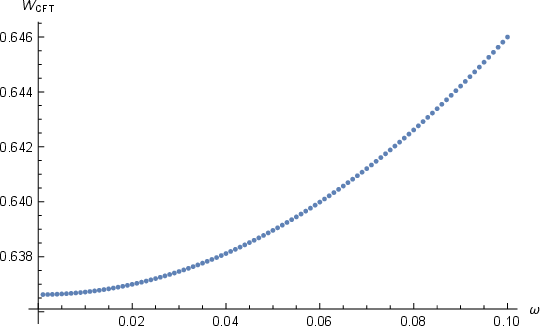}}
	\subfigure[{\ $l=5$, CFT correlator}]%
    {\includegraphics[scale=0.75]{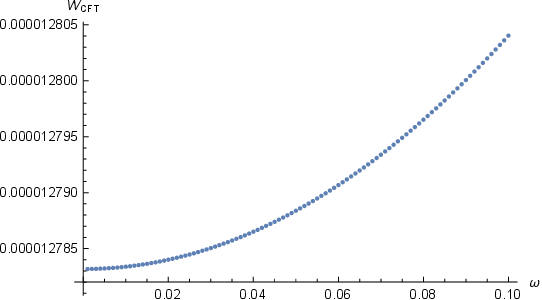}}
	\subfigure[{\ $l=0$, spin-0 Sleight-Taronna term}]%
	{\includegraphics[scale=0.75]{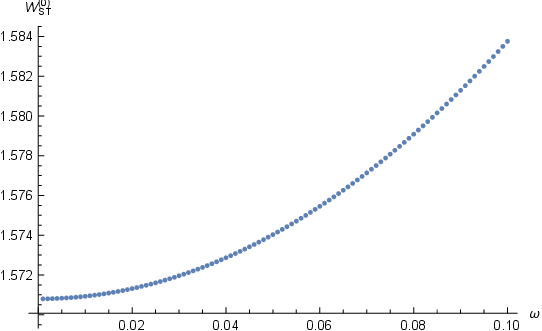}}
	\subfigure[{\ $l=5$, spin-0 Sleight-Taronna term}]%
    {\includegraphics[scale=0.75]{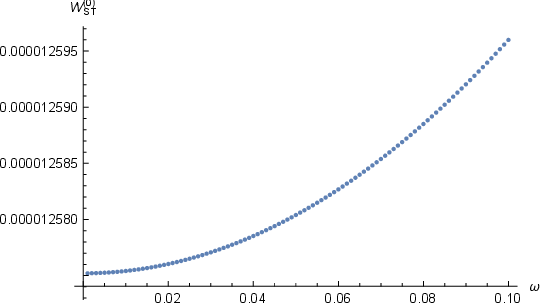}}
	\subfigure[{\ $l=0$, spin-2 Sleight-Taronna term}]%
	{\includegraphics[scale=0.75]{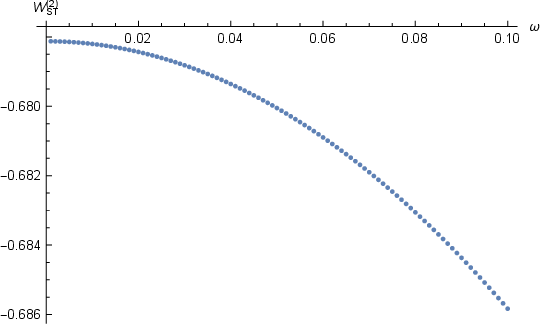}}
	\subfigure[{\ $l=5$, spin-2 Sleight-Taronna term}]%
    {\includegraphics[scale=0.75]{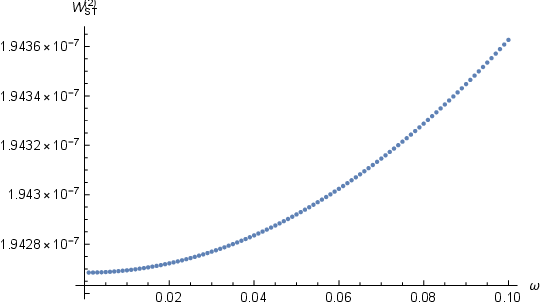}}
	\subfigure[{\ $l=0$, all-spins Sleight-Taronna term}]%
	{\includegraphics[scale=0.75]{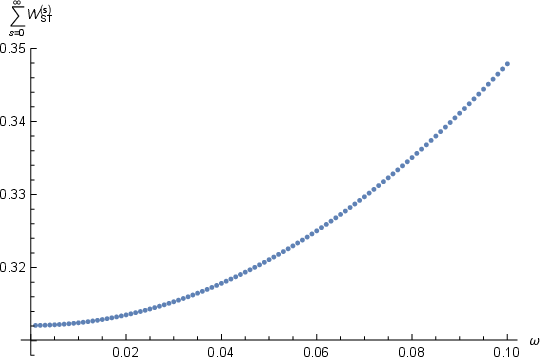}}
	\subfigure[{\ $l=5$, all-spins Sleight-Taronna term}]%
	{\includegraphics[scale=0.75]{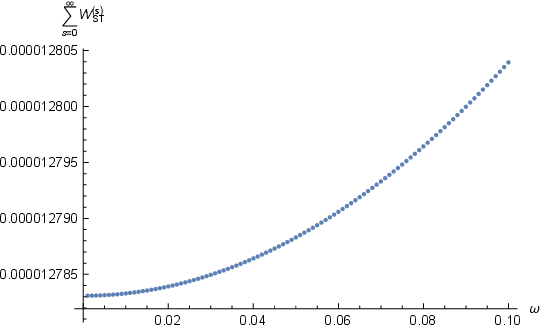}}
	\caption{Numerical plots of the CFT correlator and bulk Sleight-Taronna contributions in various channels, at $l=0,5$ and $0.001\leq\omega\leq0.1$. The regular behavior at $\omega\to 0$ expresses exponential decay at large ``time'' separations, which implies the ``time'' locality of the new vertex.}%
	\label{fig:small_omega}%
\end{figure}

Finally, let us turn to the issue of ``time'' locality. Here, our statement in section \ref{sec:locality:time} was that the CFT correlator and all bulk diagrams decay exponentially at large time separation $|\tau_1-\tau_2|$, with the exception of matching $O(1)$ terms \eqref{eq:cubic_for_time_locality}-\eqref{eq:double_exchange} in the CFT correlator and the double-exchange diagram. In frequency space, the $O(1)$ terms from \eqref{eq:cubic_for_time_locality}-\eqref{eq:double_exchange} become delta-function contributions $\sim\delta(\omega)$, which aren't visible in our analysis. This leaves the terms that should decay exponentially at large $|\tau_1-\tau_2|$, which, in frequency space, means the absence of singularities at small $\omega$. Thus, we expect $W_{\text{CFT}}$ and $W_{\text{ST}}^{(s)}$ to behave regularly as $\omega$ approaches zero. As we can see in figure \ref{fig:small_omega}, this expectation is also borne out by the numerics.

\section{Alternative approach to Sleight-Taronna diagram for (0,0,bilocal)} \label{sec:delta_gauge}

At an early stage of this work, we carried out exploratory calculations of the Sleight-Taronna contributions $S_{0,0,s}[V_{\text{ST}};\Pi_1,\Pi_2,\phi]$ to the $\left< j^{(0)}_1 j^{(0)}_2 \calO \right>$ correlator, using a different approach from that of the previous section. Though we didn't get far along this path, we report the details here for future reference. The idea is to calculate not the diagram itself, but its boundary Laplacians $\Box_{\ell}$ and/or $\Box_{\ell'}$ with respect to the endpoints of the bilocal $\calO(\ell,\ell')$. This has the advantage of reducing the bulk integral to just a 1d integral over the worldline $\gamma$, because the boundary Laplacians \eqref{eq:Box_Phi}-\eqref{eq:double_Box} of the DV field are delta-function-like distributions with support on $\gamma$. However, this is true not for the DV field in the original gauge \eqref{eq:phi}, but in the recently discovered gauges \eqref{eq:Phi}-\eqref{eq:Phi_symm}. Because these gauges are not traceless, we must pay the price of extending $V_{\text{ST}}$ beyond traceless gauge, where it is no longer given by the simple formula \eqref{eq:V_ST}. 

In this section, we describe one analytic calculation with the above technique. Using the DV field $\Phi^{(s)}_{\text{symm}}$ in the gauge \eqref{eq:Phi_symm}, we will demonstrate the vanishing of the double Laplacian $\Box_\ell\Box_{\ell'} S_{0,0,s}[V_{\text{ST}};\Pi_1,\Pi_2,\Phi_{\text{symm}}]$ for any spin $s\geq 2$, in the simple case where the scalar propagators $\Pi_1,\Pi_2$ have the \emph{same} boundary source point $\ell_1=\ell_2$. The reason for using the double Laplacian instead of a single Laplacian $\Box_\ell$ or $\Box_{\ell'}$ is its symmetry under $\ell\leftrightarrow\ell'$, which simplifies the worldline integral.

\subsection{Extending $V_{\text{ST}}^{(0,0,s)}$ beyond traceless gauge}

Our first task is to extend $V_{\text{ST}}$ beyond traceless gauge. For general spins $(s_1,s_2,s_3)$, this extension is not fully known (the somewhat incomplete state of the art for general cubic vertices in AdS is \cite{Francia:2016weg}; note that it uses the formalism of \cite{Joung:2011ww} rather than \cite{Sleight:2016dba}, i.e. scaling weights $2-s$ rather than $s+1$ with respect to the embedding-space coordinates $x^\mu$). However, in the special case $V^{(0,0,s)}_{\text{ST}}$, the extension is easy to work out. We begin by writing the original vertex in the form:
\begin{align}
 V^{(0,0,s)}_{\text{ST}} h_1 h_2 h^{(s)} \sim J^{\mu_1\dots\mu_s}h^{(s)}_{\mu_1\dots\mu_s} \ , \label{eq:V_J}
\end{align}
where $J^{\mu_1\dots\mu_s}$ is a bulk spin-$s$ current constructed from the scalar fields $h_1,h_2$, and we hide the coupling constants in the proportionality symbol ``$\sim$''. The fact that $V_{\text{ST}}^{(0,0,s)}$ is gauge-invariant within traceless gauge corresponds to the statement that $J^{\mu_1\dots\mu_s}$ is conserved in $EAdS_4$, up to trace terms and a gradient term:
\begin{align}
 \nabla_{\mu_1}J^{\mu_1\dots\mu_s} = \nabla^{(\mu_2}\tilde J^{\mu_3\dots\mu_s)} + \text{traces} \ . \label{eq:div_J}
\end{align}
Knowing $\tilde J^{\mu_1\dots\mu_{s-2}}$, we can construct a corrected current $\hat J^{\mu_1\dots\mu_s}$, which is conserved up to trace terms only:
\begin{align}
 \hat J^{\mu_1\dots\mu_s} = J^{\mu_1\dots\mu_s} - \frac{s}{2}\,g^{(\mu_1\mu_2}\tilde J^{\mu_3\dots\mu_s)} \ ; \quad \nabla_{\mu_1}\hat J^{\mu_1\dots\mu_s} = \text{traces} \ . \label{eq:corrected_J}
\end{align}
This then defines a vertex that is gauge-invariant without restriction to traceless gauge:
\begin{align}
 \hat V^{(0,0,s)}_{\text{ST}} h_1 h_2 h^{(s)} \sim \hat J^{\mu_1\dots\mu_s}h^{(s)}_{\mu_1\dots\mu_s} = J^{\mu_1\dots\mu_s}h^{(s)}_{\mu_1\dots\mu_s} - \frac{s}{2}\,\tilde J^{\mu_1\dots\mu_{s-2}}h^{(s)}_{\mu_1\dots\mu_{s-2}\nu}{}^\nu \ . 
\end{align}
Now, in the particular vertex formula \eqref{eq:V_ST}, the current $J^{\mu_1\dots\mu_s}$ reads:
\begin{align}
 J^{\mu_1\dots\mu_s} = P^{\mu_1}_{\nu_1}\!\dots P^{\mu_s}_{\nu_s} \big(h_1\del^{\nu_1\dots\nu_s}h_2 + h_2\del^{\nu_1\dots\nu_s}h_1\big) \ , \label{eq:J}
\end{align}
where $\del^{\nu_1\dots\nu_s}\equiv\del^{\nu_1}\!\dots\del^{\nu_s}$ are 5d partial derivatives with respect to $x^\mu\in\bbR^{1,4}$, and $P^\mu_\nu$ are the projectors \eqref{eq:P} from $\bbR^{1,4}$ onto the $EAdS_4$ hyperboloid. When contracting with the HS field $h^{(s)}_{\mu_1\dots\mu_s}$ in \eqref{eq:V_J}, these projectors can be omitted. However, they are important for calculating the covariant divergence in \eqref{eq:div_J}. Using the definition \eqref{eq:nabla} of the $EAdS_4$ covariant derivative, and setting $x\cdot x = -1$ at the end, we calculate the divergence as:
\begin{align}
 \nabla_{\mu_1}J^{\mu_1\dots\mu_s} = P^{\mu_2}_{\nu_2}\!\dots P^{\mu_s}_{\nu_s}\big(\del_\rho h_1\del^{\nu_2\dots\nu_s}\del^\rho h_2 - s h_1\del^{\nu_2\dots\nu_s}h_2 + (1\leftrightarrow 2) \big) + \text{traces} \ , \label{eq:div_J_explicit}
\end{align}
which should be equal (up to traces) to $\nabla^{(\mu_2}\tilde J^{\mu_3\dots\mu_s)} = P^{\mu_2}_{\nu_2}\dots P^{\mu_s}_{\nu_s}\del^{(\nu_2}\tilde J^{\nu_3\dots\nu_s)}$. It is now easy to guess and verify an expression for $\tilde J^{\mu_1\dots\mu_{s-2}}$:
\begin{align}
 \tilde J^{\mu_1\dots\mu_{s-2}} = P^{\mu_1}_{\nu_1}\!\dots P^{\mu_{s-2}}_{\nu_{s-2}}
   \sum_{n=0}^{s-2}(-1)^n \Big(\del^{(\nu_1\dots\nu_n}\del_\rho h_1\del^{\nu_{n+1}\dots\nu_{s-2})}\del^\rho h_2 - s \del^{(\nu_1\dots\nu_n} h_1\del^{\nu_{n+1}\dots\nu_{s-2})} h_2\Big) \ . \label{eq:J_tilde}
\end{align}
Putting everything together and reverting to the notation of section \ref{sec:preliminaries:ST}, the corrected vertex reads:
\begin{align}
 \begin{split}
   \hat V_{\text{ST}}^{(0,0,s)} \sim{}& (\del_{u_3}\cdot\del_{x_1})^s + (\del_{u_3}\cdot\del_{x_2})^s \\
     &- \frac{s}{2}(\del_{u_3}\cdot\del_{u_3})(\del_{x_1}\cdot\del_{x_2} - s)\sum_{n=0}^{s-2}(-1)^n (\del_{u_3}\cdot\del_{x_1})^n(\del_{u_3}\cdot\del_{x_2})^{s-2-n} \ .
 \end{split} \label{eq:V_corrected}
\end{align}

\subsection{Inserting the double Laplacian of the DV field}

The currents \eqref{eq:J},\eqref{eq:J_tilde} are to be integrated against the double Laplacian \eqref{eq:double_Box}-\eqref{eq:Q} of the DV field:
\begin{align}
 (\Phi^{(s)}_{\text{symm}})_{\mu_1\dots\mu_s} &\sim Q_{\mu_1\dots\mu_s} - \frac{1}{4}g_{(\mu_1\mu_2}Q_{\mu_3\dots\mu_s)} \ ; \\
 Q_{\mu_1\dots\mu_p} &= \calT_{\mu_1\dots\mu_p}\big(\nabla\cdot\nabla - p(p-1) \big)\delta^3(x;\ell,\ell') \ ; \\
 \calT_{\mu_1\dots\mu_p} &= t_{\mu_1}\dots t_{\mu_p} - \text{traces} \ ,
\end{align}
to form the cubic diagram:
\begin{align}
 &\Box_\ell\Box_{\ell'} S_{0,0,s}[V_{\text{ST}};h_1,h_2,\Phi_{\text{symm}}] \nonumber \\
 &\quad \sim \int_{EAdS_4} d^4x \left( Q_{\mu_1\dots\mu_s} - \frac{1}{4}g_{(\mu_1\mu_2}Q_{\mu_3\dots\mu_s)} \right)\left(J^{\mu_1\dots\mu_s} - \frac{s}{2}g^{(\mu_1\mu_2}\tilde J^{\mu_3\dots\mu_s)} \right) \label{eq:double_Box_integral_raw} \\
 &\quad = \int_{EAdS_4} d^4x \left( Q_{\mu_1\dots\mu_s}J^{\mu_1\dots\mu_s} + \frac{1}{2}Q_{\mu_1\dots\mu_{s-2}}\left(\frac{s}{s-1}\tilde J^{\mu_1\dots\mu_{s-2}} - \frac{1}{2}J^{\nu\mu_1\dots\mu_{s-2}}_\nu \right) \right) \ . \nonumber
\end{align}
Using the free field equation $(\del\cdot\del)h_i = 0$ and the scaling property $(x\cdot\del)h_i = -h_i$ for the two scalar fields $i=1,2$, we can evaluate the trace $J^{\nu\mu_1\dots\mu_{s-2}}_\nu$ as:
\begin{align}
 \begin{split}
   J^{\nu\mu_1\dots\mu_{s-2}}_\nu &= P^{\mu_1}_{\nu_1}\!\dots P^{\mu_{s-2}}_{\nu_{s-2}}\,h_1 g_{\sigma\rho} \del^{\rho\sigma\nu_1\dots\nu_{s-2}}h_2 + (1\leftarrow 2) \\
     &=  -\frac{1}{x\cdot x}\,P^{\mu_1}_{\nu_1}\!\dots P^{\mu_{s-2}}_{\nu_{s-2}}\,h_1 x_\sigma x_\rho \del^\rho\del^\sigma\del^{\nu_1\dots\nu_{s-2}}h_2 + (1\leftarrow 2) \\
     &= -\frac{s(s-1)}{x\cdot x}\,P^{\mu_1}_{\nu_1}\!\dots P^{\mu_{s-2}}_{\nu_{s-2}}\big(h_1 \del^{\nu_1\dots\nu_{s-2}}h_2 + h_2 \del^{\nu_1\dots\nu_{s-2}}h_1 \big) \ .
 \end{split} \label{eq:trace_J}
\end{align}
To simplify the integral \eqref{eq:double_Box_integral_raw}, we first move around the derivatives in the traceless structure \eqref{eq:Q} as:
\begin{align}
 Q_{\mu_1\dots\mu_p} = \big(\nabla\cdot\nabla - p(p+2) \big)\big(\calT_{\mu_1\dots\mu_p}\delta^3(x;\ell,\ell')\big) \ . \label{eq:Q_rearranged}
\end{align}
This follows from:
\begin{align}
 \big((\nabla\cdot\nabla)t^\mu\big)\delta^3(x;\ell,\ell') = -3t^\mu\delta^3(x;\ell,\ell') \ ; \quad (\nabla_\nu t^\mu)\nabla^\nu\delta^3(x;\ell,\ell') = 3t^\mu\delta^3(x;\ell,\ell') \ ,
\end{align}
which in turn follows from:
\begin{align}
 &\nabla_\mu t_\nu = -2t_{(\mu}r_{\nu)} \ ; \quad \nabla_\mu r_\nu = g_{\mu\nu} - t_\mu t_\nu - r_\mu r_\nu \ ; \\
 &r^\mu\delta^3(x;\ell,\ell') = 0 \ ; \quad (r\cdot\nabla)\delta^3(x;\ell,\ell') = -3\delta^3(x;\ell,\ell') \ .
\end{align}
We can now use \eqref{eq:Q_rearranged} to integrate \eqref{eq:double_Box_integral_raw} by parts, moving the Laplacians $\nabla\cdot\nabla$ onto the currents $J^{\mu_1\dots\mu_s},J^{\nu\mu_1\dots\mu_{s-2}}_\nu,\tilde J^{\mu_1\dots\mu_{s-2}}$. Then the delta functions, now free of derivatives, yield the following integral over the worldline $\gamma(\ell,\ell')$:
\begin{align}
 \begin{split}
   &\Box_\ell\Box_{\ell'} S_{0,0,s}[V_{\text{ST}};h_1,h_2,\Phi_{\text{symm}}] \sim \int_{-\infty}^\infty d\tau \left( \calT_{\mu_1\dots\mu_s}\big(\nabla\cdot\nabla - s(s+2) \big)J^{\mu_1\dots\mu_s} \vphantom{\frac{1}{2}} \right. \\
   &\left.\qquad {}+ \frac{1}{2}\calT_{\mu_1\dots\mu_{s-2}}\big(\nabla\cdot\nabla - s(s-2) \big)\left(\frac{s}{s-1}\tilde J^{\mu_1\dots\mu_{s-2}} - \frac{1}{2}J^{\nu\mu_1\dots\mu_{s-2}}_\nu \right) \right) \ .
 \end{split} \label{eq:worldline_integral_raw}
\end{align}
Now, recall that $J^{\mu_1\dots\mu_s},J^{\nu\mu_1\dots\mu_{s-2}}_\nu,\tilde J^{\mu_1\dots\mu_{s-2}}$ take the form \eqref{eq:J},\eqref{eq:J_tilde},\eqref{eq:trace_J} of $EAdS_4$ projections of simple (but not tangential to $EAdS_4$) embedding-space tensors. To evaluate $EAdS_4$ derivatives of such quantities, we use following identities, which hold for any $\bbR^{1,4}$ tensor $f_{\mu_1\dots\mu_p}$, and are straightforward to develop from the basic formula \eqref{eq:nabla}:
\begin{align}
 \nabla_\rho(P_{\mu_1}^{\nu_1}\!\dots P_{\mu_p}^{\nu_p} f_{\nu_1\dots\nu_p}) 
    &= P_{\mu_1}^{\nu_1}\!\dots P_{\mu_p}^{\nu_p}\left(P_\rho^\sigma \del_\sigma f_{\nu_1\dots\nu_p} - \frac{p}{x\cdot x}\,g_{\rho(\nu_1}f_{\nu_2\dots\nu_p)\sigma}x^\sigma  \right) \\
 (\nabla\cdot\nabla)(P_{\mu_1}^{\nu_1}\!\dots P_{\mu_p}^{\nu_p}f_{\nu_1\dots\nu_p}) &= P_{\mu_1}^{\nu_1}\!\dots P_{\mu_p}^{\nu_p}\left( \left[\del\cdot\del - \frac{1}{x\cdot x}\left((x\cdot\del)^2 + 3(x\cdot\del) - p\right)\right] f_{\nu_1\dots\nu_p} \right. 
 \nonumber \\
 &\left.{} - \frac{2p}{x\cdot x}\,\del_{(\nu_1}\big(f_{\nu_2\dots\nu_p)\rho}x^\rho\big) + \frac{p(p-1)}{(x\cdot x)^2}\,g_{(\nu_1\nu_2}f_{\nu_3\dots\nu_s)\rho\sigma}x^\rho x^\sigma \right) \ . \label{eq:Laplacian_with_projectors_1}
\end{align}
With some further manipulation, we can bring \eqref{eq:Laplacian_with_projectors_1} into the alternative form:
\begin{align}
 &(\nabla\cdot\nabla)(P_{\mu_1}^{\nu_1}\!\dots P_{\mu_p}^{\nu_p}f_{\nu_1\dots\nu_p}) = P_{\mu_1}^{\nu_1}\!\dots P_{\mu_p}^{\nu_p}\left( \left[\del\cdot\del - \frac{1}{x\cdot x}\left((x\cdot\del)^2 + 3(x\cdot\del) + p\right)\right] f_{\nu_1\dots\nu_p} \right. 
  \nonumber \\
 &\qquad\qquad \left.{} - \frac{2p}{x\cdot x}\,x^\rho\del_{(\nu_1}f_{\nu_2\dots\nu_p)\rho} + \frac{p(p-1)}{(x\cdot x)^2}\,g_{(\nu_1\nu_2}f_{\nu_3\dots\nu_s)\rho\sigma}x^\rho x^\sigma \right) \nonumber \\
 &= P_{\mu_1}^{\nu_1}\!\dots P_{\mu_p}^{\nu_p}\left( \left[\del\cdot\del - \frac{1}{x\cdot x}\left((x\cdot\del)^2 + x\cdot\del + p\right)\right] f_{\nu_1\dots\nu_p} \right. \nonumber \\
 &\qquad\qquad \left.{} - \frac{2(p+1)}{x\cdot x}\,x^\rho\del_{(\nu_1}f_{\nu_2\dots\nu_p\rho)} + \frac{p(p-1)}{(x\cdot x)^2}\,g_{(\nu_1\nu_2}f_{\nu_3\dots\nu_s)\rho\sigma}x^\rho x^\sigma \right) \ . \label{eq:Laplacian_with_projectors_2}
\end{align}
For $J^{\mu_1\dots\mu_s}$ and its trace, it's convenient to apply \eqref{eq:Laplacian_with_projectors_1}. However, for $\tilde J^{\mu_1\dots\mu_{s-2}}$, it's more convenient to apply \eqref{eq:Laplacian_with_projectors_2}, since by construction, a symmetrized gradient reduces it to the divergence \eqref{eq:div_J_explicit} of $J^{\mu_1\dots\mu_s}$. Using $(\del\cdot\del)h_i = 0$ and $(x\cdot\del)h_i = -h_i$ for the scalar fields $i=1,2$, setting $x\cdot x = -1$, and working up to trace terms, we get:
\begin{align}
 \begin{split}
 &(\nabla\cdot\nabla)J^{\mu_1\dots\mu_s} = P^{\mu_1}_{\nu_1}\!\dots P^{\mu_s}_{\nu_s}\big(2\del_\rho h_1\del^{\nu_1\dots\nu_s}\del^\rho h_2 - (s^2+2)h_1\del^{\nu_1\dots\nu_s}h_2 \\
   &\quad - 2s^2\del^{(\nu_1}h_1\del^{\nu_2\dots\nu_s)}h_2 \big) + (1\leftrightarrow 2) + \text{traces} \ ; 
 \end{split} \\
 \begin{split}
 &(\nabla\cdot\nabla)J^{\nu\mu_1\dots\mu_{s-2}}_\nu = s(s-1)P^{\mu_1}_{\nu_1}\!\dots P^{\mu_{s-2}}_{\nu_{s-2}}\big(2\del_\rho h_1\del^{\nu_1\dots\nu_{s-2}}\del^\rho h_2 \\
   &\quad - (s^2-4s+6)h_1\del^{\nu_1\dots\nu_{s-2}}h_2 - 2(s-2)^2\del^{(\nu_1}h_1\del^{\nu_2\dots\nu_{s-2})}h_2 \big) + (1\leftrightarrow 2) + \text{traces} \ ;
 \end{split}
\end{align}
\begin{align}
 \begin{split}
   &(\nabla\cdot\nabla)\tilde J^{\mu_1\dots\mu_{s-2}} = 2s(s-1)P^{\mu_1}_{\nu_1}\!\dots P^{\mu_{s-2}}_{\nu_{s-2}}\big( {-\del_\rho h_1}\del^{\nu_1\dots\nu_{s-2}}\del^\rho h_2 + (s-1)h_1\del^{\nu_1\dots\nu_{s-2}} h_2 \\
   &\quad + (1\leftrightarrow 2)\big) + P^{\mu_1}_{\nu_1}\!\dots P^{\mu_{s-2}}_{\nu_{s-2}}\sum_{n=0}^{s-2}(-1)^n\big(2\del^{(\nu_1\dots\nu_n}\del_{\rho\sigma}h_1\del^{\nu_{n+1}\dots\nu_{s-2})}\del^{\rho\sigma}h_2 \\
   &\qquad\qquad\qquad\qquad + s(s+2)\del^{(\nu_1\dots\nu_n}\del_\rho h_1\del^{\nu_{n+1}\dots\nu_{s-2})}\del^\rho h_2 \\
   &\qquad\qquad\qquad\qquad - s(s^2-2)\del^{(\nu_1\dots\nu_n}h_1\del^{\nu_{n+1}\dots\nu_{s-2})}h_2 \big) + \text{traces} \ .
 \end{split}
\end{align}
Plugging this back into the worldline integral \eqref{eq:worldline_integral_raw}, and pulling out an overall factor of 2, we arrive at the following expression for the diagram:
\begin{align}
 &\Box_\ell\Box_{\ell'} S_{0,0,s}[V_{\text{ST}};h_1,h_2,\Phi_{\text{symm}}] \nonumber \\
 &\sim \int_{-\infty}^\infty d\tau\,\Bigg(\calT_{\mu_1\dots\mu_s}\bigg[\del_\nu h_1\del_{\mu_1\dots\mu_s}\del^\nu h_2 
   - (s^2+s+1)h_1\del^{\mu_1\dots\mu_s}h_2 - s^2\del^{(\mu_1}h_1\del^{\mu_2\dots\mu_s)}h_2 + (1\leftrightarrow 2) \bigg] \nonumber \\
 &\qquad -\frac{1}{4} \calT_{\mu_1\dots\mu_{s-2}}\bigg[ s(3s-1)\del_\nu h_1\del^{\mu_1\dots\mu_{s-2}}\del^\nu h_2 - s(s-1)(s^2-s+3)h_1\del^{\mu_1\dots\mu_{s-2}}h_2 \nonumber \\
 &\qquad\qquad - s(s-1)(s-2)^2\del^{(\mu_1} h_1\del^{\mu_2\dots\mu_{s-2})}h_2 + (1\leftrightarrow 2) \label{eq:00s_worldline_integral} \\
 &\qquad\qquad - \frac{2s}{s-1}\sum_{n=0}^{s-2}(-1)^n\del^{\mu_1\dots\mu_n}\del_{\nu\rho}h_1\del^{\mu_{n+1}\dots\mu_{s-2}}\del^{\nu\rho}h_2 \nonumber \\
 &\qquad\qquad - \frac{4s^2}{s-1}\sum_{n=0}^{s-2}(-1)^n\del^{\mu_1\dots\mu_n}\del_\nu h_1\del^{\mu_{n+1}\dots\mu_{s-2}}\del^\nu h_2 + 2s^2\sum_{n=0}^{s-2}(-1)^n\del^{\mu_1\dots\mu_n} h_1\del^{\mu_{n+1}\dots\mu_{s-2}} h_2 \bigg] \Bigg) \ . \nonumber
\end{align}

\subsection{Evaluating the $\ell_1=\ell_2$ case}

We now specialize to the case where the scalar fields $h_1,h_2$ are both proportional to the boundary-bulk propagator from the \emph{same} boundary source point $\ell_1=\ell_2\equiv L$:
\begin{align}
 h_1(x) = h_2(x) = -\frac{1}{L\cdot x} \ .
\end{align}
The embedding-space derivatives of $h_1,h_2$ are then given by:
\begin{align}
 \del^{\mu_1\dots\mu_n}h_1 = \del^{\mu_1\dots\mu_n}h_2 = n!\left(-\frac{1}{L\cdot x}\right)^{n+1} \ .
\end{align}
Plugging this into \eqref{eq:00s_worldline_integral} and pulling out an overall factor of $2(-1)^{s+1}$, we get (note that all terms with contracted derivatives vanish, since $L^\mu$ is null):
\begin{align}
 \begin{split}
   \Box_\ell\Box_{\ell'} S_{0,0,s} \sim \int_{-\infty}^\infty &d\tau\,\Bigg(\frac{\calT_{\mu_1\dots\mu_s}L^{\mu_1}\dots L^{\mu_s}}{(L\cdot x)^{s+2}}\left[(s^2+s+1) s! + s^2(s-1)! \right] \\
   &-\frac{1}{4} \frac{\calT_{\mu_1\dots\mu_{s-2}}L^{\mu_1}\dots L^{\mu_{s-2}}}{(L\cdot x)^s}\bigg[s(s-1)(s^2-s+3)(s-2)! \\
   &\qquad  + s(s-1)(s-2)^2(s-3)! - s^2\sum_{n=0}^{s-2}(-1)^n n!(s-2-n)! \bigg] \Bigg) \ .
 \end{split} \label{eq:00s_L_integral_raw}
\end{align}
Using the identity:
\begin{align}
 \sum_{n=0}^p(-1)^n n!(p-n)! = \frac{2(p+1)!}{p+2} \ ,
\end{align}
which holds for all even $p$, we simplify \eqref{eq:00s_L_integral_raw} as (pulling out an overall factor of $(s+1)!$):
\begin{align}
 \Box_\ell\Box_{\ell'} S_{0,0,s} \sim \int_{-\infty}^\infty d\tau\,\left((s+1)\frac{\calT_{\mu_1\dots\mu_s}L^{\mu_1}\dots L^{\mu_s}}{(L\cdot x)^{s+2}} - \frac{s-1}{4} \frac{\calT_{\mu_1\dots\mu_{s-2}}L^{\mu_1}\dots L^{\mu_{s-2}}}{(L\cdot x)^s} \right) \ .
 \label{eq:00s_L_integral}
\end{align}
We thus need to evaluate quantities of the form (c.f. \eqref{eq:T}, and note that $p=s,s-2$ is even):
\begin{align}
 \calT_{\mu_1\dots\mu_p}L^{\mu_1}\dots L^{\mu_p} = p!\,\calT^{(p)}(x,t,L) = \frac{1}{2^p}\sum_{n=0}^{p/2} \binom{p+1}{2n+1} ({-q_{\mu\nu}}L^\mu L^\nu )^n (t\cdot L)^{p-2n} \ . \label{eq:T_L}
\end{align}
Here, $x^\mu$ is a point on the $\gamma(\ell,\ell')$ geodesic, $t^\mu$ is the unit tangent to the geodesic at $x^\mu$, and $q_{\mu\nu} = \eta_{\mu\nu} + x_\mu x_\nu - t_\mu t_\nu$ is the metric of the 3d space perpendicular to both. Without loss of generality, we can choose the bilocal's endpoints $\ell^\mu,\ell'^\mu$ and the boundary source point $L^\mu$ of the scalar fields as:
\begin{align}
 \ell^\mu = \left(\frac{1}{2},\frac{1}{2},0,0,0\right) \ ; \quad \ell'^\mu = \left(\frac{1}{2},-\frac{1}{2},0,0,0\right) \ ; \quad L^\mu = (1,0,1,0,0) \ .
\end{align}
This sets the geodesic at $x^\mu = x^\mu(\tau;\ell,\ell') = (\cosh\tau,\sinh\tau,\vec 0)$, with unit tangent $t^\mu = (\sinh\tau,\cosh\tau,\vec 0)$. We thus have:
\begin{align}
 x\cdot L = -\cosh\tau \ ; \quad t\cdot L = -\sinh\tau \ ; \quad q_{\mu\nu}L^\mu L^\nu = 1 \ .
\end{align}
This allows us to evaluate \eqref{eq:T_L} as:
\begin{align}
 \calT_{\mu_1\dots\mu_p}L^{\mu_1}\dots L^{\mu_p} = \frac{1}{2^p}\sum_{n=0}^{p/2} \binom{p+1}{2n+1}(-1)^n \sinh^{p-2n}\tau = \frac{1}{2^p}\Im(\sinh\tau+i)^{p+1} \ .
\end{align}
Dividing by $(L\cdot x)^{p+2} = \cosh^{p+2}\tau$ and integrating over $\tau$, we get:
\begin{align}
 \begin{split}
   &\int_{-\infty}^\infty d\tau\,\frac{\calT_{\mu_1\dots\mu_p}L^{\mu_1}\dots L^{\mu_p}}{(L\cdot x)^{p+2}} = \frac{1}{2^p}\Im\int_{-\infty}^\infty \frac{d\tau}{\cosh\tau} \left(\tanh\tau + \frac{i}{\cosh\tau}\right)^{p+1} \\
   &\quad = \frac{1}{2^p}\Im\int_0^\pi d\beta\,(\cos\beta + i\sin\beta)^{p+1} = \frac{1}{2^p}\int_0^\pi d\beta\,\sin[(p+1)\beta] = \frac{1}{2^{p-1}(p+1)} \ .
 \end{split} \label{eq:T_integral}
\end{align}
where we substituted $\tanh\tau\equiv\cos\beta$. Plugging \eqref{eq:T_integral} back into \eqref{eq:00s_L_integral}, we see that the two terms in \eqref{eq:00s_L_integral} cancel. Thus, we managed to show that in this simple case, the diagram $\Box_\ell\Box_{\ell'}S_{0,0,s}[V_{\text{ST}};\Pi,\Pi,\Phi_{\text{symm}}]$ vanishes.

\section{Discussion} \label{sec:discuss}

In this paper, we showed that the boundary correlator of three bilocals in HS holography can be reproduced by physically sensible bulk structures, which extend the Sleight-Taronna cubic vertex without sacrificing the principle of locality. We also showed that the  Sleight-Taronna vertex itself satisfies nice gauge-invariance properties outside its intended range of applicability. 

The most important direction for future work is to write down explicitly the new field-field-worldline vertex $V_{\text{new}}$ from section \ref{sec:locality}. As we've seen, this requires calculating Sleight-Taronna diagrams for two boundary-bulk propagators and one DV solution. The worldline localization technique of section \ref{sec:delta_gauge} may prove helpful, but it comes with the difficulty of extending the Sleight-Taronna vertex beyond traceless gauge in one of its three ``legs''.

Our paper was carefully phrased to refer to the \emph{minimal} type-A bulk theory, dual to the $O(N)$ model on the boundary. However, we repeatedly found it convenient to talk about e.g. the un-symmetrized bilocals $\calO(\ell,\ell')$ of the $U(N/2)$ model, rather than the symmetrized ones $\calO_+(\ell,\ell')$ of the $O(N)$ model. It is thus tempting to extend the entire discussion to the $U(N/2)$ model, by allowing all integer spins in the bulk. In fact, most of our results and arguments can be immediately generalized in this way (note that the calculations in sections \ref{sec:numeric}-\ref{sec:delta_gauge} in any case involve only even spins, and would not be affected). 

The only unknown is whether, with the inclusion of odd spins, the Sleight-Taronna vertex \eqref{eq:V_ST} continues to reproduce the boundary correlators $\left< j_1^{(s_1)} j_2^{(s_2)} j_3^{(s_3)} \right>$ as in \eqref{eq:on_shell_correlator}. Since $\left< j_1^{(s_1)} j_2^{(s_2)} j_3^{(s_3)} \right>$ vanishes for odd $s_1+s_2+s_3$, it is sensible to define $V_{\text{ST}}^{(s_1,s_2,s_3)} = 0$ for this case; in fact, in transverse-traceless gauge, the definition \eqref{eq:V_ST} already has this property, due to section \ref{sec:free_invariance}'s Lemma 2. Thus, the remaining question is whether the vertex \eqref{eq:V_ST} reproduces the correlators for (even,odd,odd) combinations of spins $(s_1,s_2,s_3)$. We expect that the answer is yes, but we haven't worked it out one way or the other. Section \ref{sec:free_invariance}'s gauge-invariance results for the vertex \eqref{eq:V_ST} hold just as well in the (even,odd,odd) case. Section \ref{sec:locality}'s locality arguments also survive the extension to odd spins. Specifically, if the Sleight-Taronna vertex correctly describes the (even,odd,odd) $\left< j_1^{(s_1)} j_2^{(s_2)} j_3^{(s_3)} \right>$ correlator, then the statements of section \ref{sec:locality} simply carry through. If not, then the (even,odd,odd) correlator will still be described by \emph{some} local vertex; unlike the Sleight-Taronna vertex, this may require some gauge corrections when generalized from transverse-traceless to general traceless gauge, but these will again be local. With such corrections taken into account, the main statements of section \ref{sec:locality} vis. the locality of the new field-field-worldline vertex $V_{\text{new}}$ will continue to hold.

As noted in the Introduction, our larger ambition is to use the cubic structure explored in this paper as a building block for constructing all the correlators of HS theory, in a way that repackages all non-locality into the structure and interactions of DV solutions and their geodesic ``worldlines''. This idea will be explored in detail in a separate publication \cite{FeynmanRules}.

\section*{Acknowledgements}

We are grateful to Sudip Ghosh and Mirian Tsulaia for discussions. This work was supported by the Quantum Gravity Unit of the Okinawa Institute of Science and Technology Graduate University (OIST).

\end{document}